\documentclass{JHEP3}
\usepackage{amsmath,amssymb}
\usepackage{graphicx}
\usepackage{slashed}
\usepackage{multirow}
\usepackage{booktabs}

\newcommand{\abs}[1]{\bigl\vert #1 \bigr\vert}
\newcommand{\absb}[1]{\left\vert#1\right\vert}

\newcommand{\beq}{\begin{eqnarray}}% can be used as {equation} or {eqnarray}
\newcommand{\eeq}{\end{eqnarray}}

\newcommand{\ul}{\underline}

\def\vol{{\mathrm{vol}}}
\def\w{\mathrm{w}}

\def\tr{{\mathrm{tr}}}

\def\R{{\mathbb{R}}}
\def\Id{{\mathbb{I}}}

\def\cF{{\mathcal{F}}}

\def\cS{{\mathcal{S}}}

\def\cK{{\mathcal{K}}}

\def\cM{{\mathcal{M}}}

\def\ud{{\mathrm{d}}}
\def\uD{{\mathrm{D}}}
\def\ue{{\mathrm{e}}}
\def\ui{i}

\def\pa{\partial}

\def\al{\alpha}
\def\sig{\sigma}
\def\ga{\gamma}
\def\G{\Gamma}

\def\lam{\lambda}
\def\Lam{\Lambda}
\def\ep{{\epsilon}}

\def\r2{{\sqrt{2}}}
\def\h{{\eta}}

\def\h0{\hat{h}}
\def\Vr0{\hat{V}_{r}}
\def\Vp0{\hat{V}_{\phi}}

\def\r2{\sqrt{2}}
\def\be{\begin{equation}}
\def\ee{\end{equation}}
\def\bes{\begin{subequations}}
\def\ees{\end{subequations}}
\def\bea{\begin{eqnarray}}
\def\eea{\end{eqnarray}}

\def\cF{\mathcal{F}}

\def\si1{\sin\theta_1}
\def\si2{\sin\theta_2}
\def\cs1{\cos\theta_1}
\def\cs2{\cos\theta_2}
\def\T21{\tan^2\theta_1/2}
\def\T22{\tan^2\theta_2/2}
\def\CT21{\cot^2\theta_1/2}
\def\CT21{\cot^2\theta_1/2}

\def\bT{{\bf{T}}}
\def\CY{{\text{\tiny{CY}}}}
\def\raw{\rightarrow}

\def\re{\mbox{Re }}

\def\tr{\mbox{Tr }}

\title{Open String Wavefunctions in Warped
Compactifications} 

\author{Fernando Marchesano,$^1$ Paul McGuirk,$^{2,3}$ Gary Shiu$^{1,2,3}$ \\
$^1$ PH-TH Division, CERN, CH-1211 Geneva 23, Switzerland
\vskip 3pt
$^2$ Department of Physics, University of Wisconsin, Madison,
WI 53706, USA\\
$^3$ Department of Physics and SLAC, Stanford University, Stanford, CA 94305, USA
}

\date{\today}

\abstract{We analyze the wavefunctions for open strings in warped compactifications, and compute the warped K\"ahler potential for the light modes of a probe D-brane. This analysis not only applies to the dynamics of D-branes in warped backgrounds, but also allows to deduce warping corrections to the closed string K\"ahler metrics via their couplings to open strings. We consider in particular the spectrum of D7-branes in warped Calabi-Yau orientifolds, which provide a string theory realizations of the Randall-Sundrum scenario. We find that certain background fluxes, necessary in the presence of warping, couple to the fermionic wavefunctions and qualitatively change their behavior. This modified dependence of the wavefunctions are needed for consistency with supersymmetry, though it is present in non-supersymmetric vacua as well.  We discuss the deviations of our setup from the RS scenario and, as an application of our results, compute the warping corrections to Yukawa couplings in a simple model. Our analysis is performed both with and without the presence of D-brane world-volume flux, as well as for the case of backgrounds with varying dilaton.}


\preprint{CERN-PH-TH/2008-227\\
MAD-TH-08-14\\
SU-ITP-08/34}

\begin{document}

\section{Introduction}

Scenarios with warped extra dimensions provide us with a rich framework to address long-standing puzzles in physics Beyond the Standard Model. In the presence of warping the energies of localized states are suppressed by the gravitational redshift and so, as pointed out in \cite{RS}, this may offer a geometric explanation of the electroweak-gravity hierarchy.

While this feature has been mainly exploited in the context of 5D models as the original Randall-Sundrum (RS) scenarios and extensions thereof, it does clearly apply to more general warped backgrounds. In particular, it is also manifest in warped compactifications of string theory \cite{Douglas:2006es,verlinde99,Dasgupta:1999ss,Greene:2000gh,Becker:2000rz,gkp01}, especially for those strongly warped regions that can be asymptotically described as $AdS_5 \times X_5$ for some compact manifold $X_5$, and which provide a natural extension of the RS scenario to a UV complete theory. As a result, these so-called `warped throats' have become a powerful tool to construct phenomenologically attractive
models of particle physics and cosmology from string theory, and are nowadays an essential ingredient in explicit constructions of string inflationary models \cite{stringcosmologyreviews}.

Given the above, it is natural to wonder how the dynamics governing warped compactifications can be understood from a string theory/supergravity perspective. In particular, in order to draw precise predictions from string warped models it is necessary to understand the low energy effective action that arises upon dimensional reduction. The derivation of such warped effective theory has proven to be a subtle problem even if one restricts to the closed string/gravity sector of the theory \cite{Giddings:2005ff,fm06,bcdgmqs06,stud,Douglas:2008jx}, although simple expressions can be given for certain subsectors  \cite{Frey:2008xw}. While these results represent significant progress in the derivation of warped effective theories, in order to accommodate constructions where the Standard Model can be realized closed strings are not enough,\footnote{At least in the context of type II string compactifications, where such developments have taken place.} and one should include D-branes in the picture. Hence, it is crucial to go beyond the previous analyses and study the effective theory for the associated open string degrees of freedom in warped backgrounds.

In this work we take an initial foray in this direction by studying open string wavefunctions in warped compactifications. In order to extract the 4D effective action for the open string degrees of freedom, we first need to compute their internal wavefunctions and then carry out a dimensional reduction. As is well known in phenomenological studies of warped extra dimensions \cite{WED}, warping has the effect of localizing massive modes to regions of strong warping because of the gravitational potential. As we shall see, warped compactifications in string theory have new added features. Other than the background  geometry which has been accounted for in the aforementioned studies, string theory contains background field strengths that, due to the equations of motion, are necessarily non-vanishing in the presence of warping. Not only do these field strengths couple to open string fermionic degrees of freedom, but they couple differently depending on the extra-dimensional chirality of such fields, which results in different warp factor dependence for their internal wavefunctions. For warped backgrounds that preserve supersymmetry, our results allow us to determine the warped corrected K\"ahler metrics for open strings, and to show that this different warp factor dependence is crucial for the kinetic terms of 4D fields in the same supermultiplet to match.\footnote{Let us stress that our analysis does not directly invoke 4D supersymmetry, since we analyze the open string wavefunctions for bosonic and fermionic fields separately. Therefore, the method of obtaining open string wavefunctions discussed here can be applied to non-supersymmetric warped backgrounds as well.} We will in addition find that open string wavefunctions act as probes of the warped geometry; their kinetic terms allowing us to deduce the K\"ahler metrics of the closed strings that couple to them and hence the {\it combined} warped K\"ahler potential. The closed string K\"ahler metrics obtained in this way indeed reproduce the recent results of \cite{stud,Frey:2008xw}. We however expect our method to have more general applicability, including situations where the direct closed string derivations have not yet been carried out.

In particular, we will focus on deriving the open string wavefunctions of D7-branes in warped type IIB/F-theory backgrounds. As pointed out in the literature (see e.g. \cite{gg06,abv06}), this setup provides a string theory realization of those 5D Warped Extra Dimension (WED) models where the SM gauge fields and fermions are located in the $AdS_5$ bulk \cite{WED}, and which have been suggested as a possible solution of the flavor puzzle. Indeed, in this 5D scenario the hierarchy between the various SM masses and mixing angles (i.e., the flavor hierarchy) results from the different localization of fermions in the extra dimensions, since the varying degrees of overlap of their wavefunctions with that of the Higgs field lead to hierarchical Yukawa couplings. In the string theory setup that we consider, the D7-branes and their intersections give rise to non-Abelian gauge symmetries and chiral matter. In particular, in a warped throat background of the form $AdS_5 \times X_5$ we can consider a D7-brane whose embedding is locally described as $AdS_5 \times X_3$, and so its open string wavefunctions are extended along the $AdS_5$ warped extra dimension. 

With a concrete realization of the {\it bulk} Randall-Sundum scenario, one can investigate whether the assumptions made in the phenonomenological studies of warped extra dimensions are justified or modified, and whether the $p$-form field strengths in string theory could lead to new variations of this basic idea. Furthermore, the open string wavefunctions obtained here enable us to calculate the physical Yukawa couplings for explicit chiral models, as we shall demonstrate in an explicit example.

More generally, the present work can be considered as an initial step towards the construction of the `Warped String Standard Model'. Besides the phenomenological appeals mentioned above, these warped models are interesting because they can be understood, by way of the AdS/CFT correspondence, as holographic duals of technicolor-like theories. Constructing these warped models from a UV complete theory allows us to go beyond a qualitative rephrasing of the strong coupling dynamics in terms of a putative gravity dual.  In addition, embedding such technicolor models in string theory may also suggest new model building possibilities.\footnote{See \cite{csu05} (and also \cite{frv08}) for the realization of this idea in the context of D3-brane at singularities.} Note that our analysis was carried out with all the essential ingredients, such as worldvolume fluxes. Therefore, our results can be applied to specific models once concrete constructions of such technicolor duals are found.

This paper is organized as follows. In Section \ref{sec:unmagnetized}, we study the D7-brane wavefunctions in the situation where the D7-brane worldvolume magnetic flux ${\cF}$ is absent. We begin with the simplest warped background which is conformally flat space and compute the wavefunctions of the bosonic and fermionic modes separately. Our treatment of the fermions follows from the $\kappa$-symmetric fermionic action in \cite{mrvv05} (see also \cite{bs06}), which takes into account the coupling of fermions to the background RR $p$-form field strengths in a manifested manner. Many of our results carry over directly to the more general case of a warped Calabi-Yau space, as discussed in subsection \ref{warpcy}, and to turning on background 3-form fluxes in such background, as shown in subsection \ref{subsec:backgroundfluxes}. In addition, in subsection \ref{F-theory} we also consider D7-branes in backgrounds with varying dilaton, which become relevant when these constructions are lifted to F-theory. The open string wavefunctions obtained in the earlier sections can be used to extract information about the warp factor dependence of the open string K\"ahler potential, discussed in subsection \ref{kahler}, and to analyze a simple chiral model in subsection \ref{model}. Finally, in Section \ref{sec:magnetized} we extend the above analysis to the more generic case of D7-branes with a non-vanishing magnetic flux $\cF$, which is an essential ingredient to obtain chirality in generic situations. We draw our conclusions in Section \ref{sec:conclusions}, and our conventions are spelled out in Appendix \ref{conv}.

\section{Unmagnetized D7-branes}\label{sec:unmagnetized}

\subsection{Warped backgrounds in string theory}
\label{subsec:warped}

As discussed in \cite{verlinde99,gkp01}, one can realize the Randall-Sundrum scenario by considering type IIB string theory on a (string frame) metric 
background of the form
\be
\ud s_{10}^{2}\, =\,  \Delta^{-1/2}\eta_{\mu\nu}\ud x^{\mu}\ud x^{\nu} + \Delta^{1/2}e^{\Phi} \hat{g}_{mn}\ud y^{m}\ud y^{n}
\label{10Dmetricst}
\ee
where $\Delta \equiv \Delta(y)$ is a warp factor that only depends on the extra six-dimensional space $X_6$ of metric $\hat{g}$. In the limit where the dilaton field $\Phi \equiv \Phi(y)$ is constant, the equations of motion constrain $\hat{g}$ to describe a Calabi-Yau metric. On the other hand, when $\Phi$ is non-constant $X_6$ will be a non-Ricci-flat K\"ahler three-fold manifold, which nevertheless serves as a base for an elliptically fibered Calabi-Yau four-fold $X_8$, as usual in F-theory constructions.

The above warp factor may be sourced by either localized sources like D3-branes and O3-planes or by the background field strengths $F_3$, $H_3$ present in the type IIB closed string sector.
 In both cases, consistency of the construction demands that the background field strength $F_5$ is also sourced. More precisely, the equations of motion require that $F_5$ is related with the warp factor and the dilaton as
\be
F_5\, =\, (1 + *_{10}) F^{\text{int}}_{5}\quad \quad \quad{F}^{\text{int}}_{5}\, =\,  \hat{*}_6 \ud \left(\Delta e^\Phi\right)
\label{F5st}
\ee
where $*_{10}$ stands for the Hodge star operator in the full 10D metric (\ref{10Dmetricst}) and  $\hat{*}_6$ in the unwarped 6D metric $\hat{g}$. Finally, together with a non-trivial dilaton profile a non-trivial RR scalar $C_0$ must be present, both of them related by the equation
\be
 \bar{\partial} \tau\, =\, 0
\label{cpxdil}
\ee
where $\tau = C_0 + i e^{-\Phi}$ is the usual type IIB axio-dilaton.

In order to introduce a Standard Model-like sector in this setup, one needs to consider open string degrees of freedom. These can be simply added to the above setup via embedding probe D-branes in this background. Such D-branes will not only give rise to 4D gauge theories upon dimensional reduction, but also to chiral matter fields charged under them. The simplest example of this is given by a D3-brane filling $\mathbb{R}^{1,3}$ and placed at some particular point $y_0 \in X_6$. While most quantities of the D3-brane gauge theory will be affected by the warp factor via the particular value of $1/\Delta(y_0)$, the internal wavefunctions for the D3-brane fields will have a trivial $\delta$-function profile.

A more non-trivial set of wavefunctions is given by the open string fields of a D7-brane wrapping a 4-cycle ${\cal S}_4 \subset X_6$. As now the wavefunctions can extend along a 4D subspace of $X_6$ they can feel non-trivially the effect of the warp factor, reproducing one of the essential ingredients of the WED models with SM fields localized on the bulk \cite{WED}. If we focus on a single D7-brane, then we will start from an 8D $U(1)$ gauge theory whose bosonic degrees of freedom are described by the so-called Dirac-Born-Infeld and Chern-Simons actions
\begin{subequations}
\begin{align}
  S_{\uD 7}^{\left.\mathrm{bos}\right.}=& \,S_{\uD 7}^{\left.\mathrm{DBI}\right.}
  + S_{\uD 7}^{\left.\mathrm{CS}\right.} \\
  S_{\uD 7}^{\left.\mathrm{DBI}\right.}=&
  -\tau_{\uD 7}\int_{\mathbb{R}^{1,3} \times {\cal S}_4}\ud^{8}\xi\, \ue^{-\Phi}
  \sqrt{\abs{\det\bigl(P[G] +\mathcal{{F}}\bigr)}}
  \\
  S_{\uD p}^{\left.\mathrm{CS}\right.}=& \,
  \tau_{\uD 7}\int_{\mathbb{R}^{1,3} \times {\cal S}_4} P[\mathcal{C}]\wedge \ue^{\mathcal{{F}}}
\end{align}
\end{subequations}
where $\tau_{\uD 7}^{-1} = (2\pi)^3 (2\pi\al')^{4}$
is the tension of the D7 brane,
and where $P[\dots]$ indicates that the 10D metric $G$ and the sum of RR potentials $\mathcal{C} = \sum_{p=0}^4 C_{2p}$ are pulled-back onto the D7-brane worldvolume. The same applies to the
NS-NS
$B$-field, which enters the action via the generalized two-form field strength
${\cal F} = P[B] + 2\pi \al' F$.
In the remainder of this section we will simplify our discussion by setting $B=0$ and $F$ to be exact. That is, we will set $F = dA$, where $A$ is the 8D gauge boson of the D7-brane worldvolume theory. In practice, this implies that $\cF=0$ up to fluctuations of $A$, a situation which will be denoted by $\langle \cF \rangle = 0$. With these simplifications, one can express the fermionic part of the D7-brane action as \cite{mrvv05}
\begin{equation}
\label{D7fermst}
  S_{\uD 7}^{\left.\mathrm{fer}\right.}\, =\, \tau_{\uD 7}
  \int\ud^{8}\xi\, \ue^{-\Phi}\sqrt{\abs{\det\, P[G]}}\,
  \bar{\Theta} P^{\uD 7}_-
  \biggl({\Gamma}^{\al}
  {\cal D}_{\al}-\frac{1}{2}{\cal O}\biggr)\,{\Theta}
\end{equation}
where ${\cal D}_\al$ is the operator appearing in the gravitino variation, its index $\al$ pulled-back into the D7-brane worldvolume, and ${\cal O}$ is the operator of the dilatino variation. The explicit expression of these operators are given in Appendix \ref{conv}, see eq.(\ref{ap:SUSYst}). As explained there, these two operators act in a 10D Majorana-Weyl bispinor 
\begin{equation}
\label{bispin}
  {\Theta} = \begin{pmatrix} {\theta}_{1} \\ 
    {\theta}_{2}\end{pmatrix}
\end{equation}
where both components have positive 10D chirality  $\Gamma_{(10)} \theta_i = \theta_i$. The fermionic degrees of freedom contained in (\ref{bispin}) are twice of what we would expect from an 8D supersymmetric theory, but they are halved by the presence of $P_-^{D7}$, which is a projector related with the $\kappa$-symmetry of the fermionic action.\footnote{Roughly speaking, (\ref{D7fermst}) is invariant under the transformation $\Theta \rightarrow \Theta + P_-^{D7} \kappa$, with $\kappa$ an arbitrary 10D MW bispinor. One can then use this symmetry to remove half of the degrees of freedom in $\Theta$.} For $\langle {\cal F}\rangle =0$ this projector is given by
\be
P_\pm^{\uD 7} \, =\, \frac{1}{2} \left(\Id \mp  \Gamma_{(8)} \otimes \sigma_2 \right)
\label{projD7F0}
\ee
where $\Gamma_{(8)}$ is the 8D chirality operator on the D7-brane worldvolume,\footnote{In our conventions the chirality matrix for a D$\left(2k+2\right)$-brane in $\R^{1,2k+1}$ is $\G_{(2k+2)} = i^{k} \G^{\ul{0\dots2k+1}}$, where $\G^{\ul{i}}$ are flat $\G$-matrices. For instance, a D7-brane extended along the directions $0\dots 7$ has  $\Gamma_{(8)} = -i \Gamma^{\ul{01234567}}$. \label{chiralf}} and $\sigma_2$ acts on the bispinor indices.

In order to dimensionally reduce the above construction to a 4D effective theory with canonically normalized kinetic terms, one first needs to convert the above quantities from the string to the Einstein frame. This basically amounts to using, instead of the metric $G_{MN}$ in (\ref{10Dmetricst}), the rescaled metric $G_{MN}^E \equiv e^{-\Phi/2} G_{MN}$. That is, in the Einstein frame we have the 10D metric background
\begin{equation}
\ud s_{10}^{2}=Z^{-1/2}\eta_{\mu\nu}\ud x^{\mu}\ud x^{\nu} + Z^{1/2}\hat{g}_{mn}\ud y^{m}\ud y^{n}
\label{10metricE}
\end{equation}
where $Z  = \Delta e^{\Phi}$ is the Einstein frame warp factor. Note that eqs.(\ref{F5st}) and (\ref{cpxdil}) are unchanged by this rescaling, and that in terms of $Z$ we have $F_5^{\mathrm{int}} = \hat{*}_{6}\ud Z$. While the D7-brane CS action does not depend on metric and hence is also not affected by such rescaling, the DBI action does change. The bosonic action now reads
\begin{equation}
  \label{D7DBIE}
  S_{\uD 7}^{\left.\mathrm{bos}\right.}
  \, =\, -\tau_{\uD 7}\int\ud^{8}\xi\, \ue^{\Phi}
  \sqrt{\abs{\det\left(P[G^E] + e^{-\Phi/2}  \mathcal{F}\right)}}
 \, +\,  \tau_{\uD 7}\int P\left[\mathcal{C}\right]\wedge \ue^{\mathcal{F}}
\end{equation}
where now $G^E$ refers to the metric tensor in (\ref{10metricE}). Finally, the fermionic D7-brane action also varies by going to the Einstein frame (see Appendix A) reading
\begin{equation}
\label{D7fermE}
  S_{\uD 7}^{\left.\mathrm{fer}\right.}\, =\, \tau_{\uD 7}
  \int\ud^{8}\xi\, \ue^{\Phi}\sqrt{\abs{\det\, P[G^E]}}\,
  \bar{\Theta} P^{\uD7}_-
  \biggl({\Gamma}^{\al}
  {\cal D}_{\al}^E+\frac12{\cal O}^E\biggr)\,{\Theta}
\end{equation}
where ${\cal O}^E$ and ${\cal D}_\al^E$ now refer to the dilatino and gravitino variations in the Einstein frame, as defined in (\ref{ap:SUSYE}). In the remainder of this paper we will always work with Einsten frame quantities, without indicating so with the superscript $E$.

\subsection{Warped flat space}\label{flat}

The simplest case of a warped background of the form (\ref{10metricE}) is constructed by taking the 6D metric $\hat{g}$ to be flat. This situation is easily obtained in string theory, by simply considering the backreaction of $N$ D3-branes in 10D flat space. While in such simple solution the internal space ${X_6} = {\R^6}$ is non-compact, one may turn to a compact setup by simply setting $X_6 = {\bf T}^6$, and adding the appropriate number of D3-branes and O3-planes such that the theory is consistent. In the latter construction the global form of the warp factor $Z$ will be a complicated function of the D3-brane positions, but close to a stack of D3-branes it will produce the well-known $\mathrm{AdS}_{5}\times S^{5}$ geometry that mimics the Randall-Sundrum scenario \cite{verlinde99}.

In the following we will derive the open string wavefunctions of a D7-brane in such conformally flat background. We will particularly focus on the warp factor dependence developed by the wavefunctions of both fermionic and bosonic zero modes, to be analyzed separately. This setup will not only be useful to make contact with the WED literature, but also to emphasize some simple features that remain true in the more general situations considered below. Finally, we will discuss some subtle issues that arise when considering D-brane fermionic actions of the form (\ref{D7fermst}), as well as an alternative derivation of the fermionic zero mode wavefunctions more suitable for further generalizations.

\subsubsection{Fermions}\label{flatferm}

Let us then consider a background of the form (\ref{10metricE}) with $\hat{g} = \hat{g}_{{\bf T}_6}$
(which implies a constant axio-dilaton $\tau=C_0 + \ui\ue^{-\Phi_{0}}$)
and a D7-brane spanning four internal dimensions of such a background. In particular, we will consider that the internal worldvolume of the D7-brane wraps a 4-cycle ${\cal S}_4 = {\bf T}^4 \subset {\bf T}^6$, so that we also have a conformally flat metric on the D7-brane worldvolume
\begin{equation}
\ud s_{\uD 7}^{2}=Z^{-1/2}\eta_{\mu\nu}\ud x^{\mu}\ud x^{\nu} + Z^{1/2}\sum_{a,b=1}^4 (\hat{g}_{\bT^4})_{ab}\, \ud y^{a}\ud y^{b}
\label{flat10metricE}
\end{equation}
where $\hat{g}_{\bT^4}$ is a flat $\bT^4$ metric.

Then, if in addition we do not consider any background fluxes $H_3$ or $F_3$, we have that the operators entering the D7-brane fermionic action (\ref{D7fermE}) are
\bes
\label{flatD7op}
\begin{align}
{\cal O}\, =& \, 0 \\
{\cal D}_\mu \, =& \, \nabla_\mu + \frac18 \slashed{F}^{\text{int}}_5 \Gamma_\mu i \sigma_2\, =\, \pa_\mu - \frac{1}{4} \G_\mu \slashed{\pa} \ln Z P_+^{O3}\\
{\cal D}_m \, =& \, \nabla_m + \frac18 \slashed{F}^{\text{int}}_5 \Gamma_m i \sigma_2\, =\, \pa_m  + \frac{1}{8}\pa_m \ln Z   - \frac{1}{4} \slashed{\pa}\ln Z \G_m  P_+^{O3}
\end{align}
\ees
where we have used the definitions (\ref{ap:SUSYE}) and the relation (\ref{F5st}). Here $\mu$ stands for $\R^{1,3}$ coordinates, $m$ labels the internal $\bT^6$ coordinates and the slash-notation stands for a contraction over bulk indices as in (\ref{ap:slashnot}).  Finally, we have defined the projectors
\be
P_\pm^{O3}\, =\, \frac12 \left(\Id \pm \G_{(6)} \otimes \sigma_2 \right) 
\label{O3proj}
\ee
where as in  (\ref{ap:proj}) $\G_{(6)}$ is the 6D chirality operator in $\bT^6$. These projectors separate the space of bispinors $\Theta$ into two sectors: those modes $\Theta$ annihilated by $P_-^{O3}$ and those annihilated by $P_+^{O3}$. Pulling-back the above operators\footnote{This amounts to pulling-back the index $M$ of ${\cal D}_M$, and not indexless quantities like $\slashed{\pa} \ln Z$ or  ${\cal O}$.} onto the D7-brane worldvolume we obtain that the term in parentheses in (\ref{D7fermE}) reads
\be
\label{flatD7Dirac}
\Gamma^\mu {\cal D}_\mu + \Gamma^a {\cal D}_a + \frac12{\cal O}\, =\, \slashed{\pa}_4^{\text{ext}} + {\slashed{\pa}}_4^{\text{int}} + \left({\slashed{\partial}}_4^{\text{int}}  \ln Z\right) \left(\frac{1}{8} - \frac{1}{2}  P_+^{O3}\right)
\ee
where $a$ runs over the internal D7-brane coordinates, $\slashed{\partial}_4^{\text{ext}} \equiv \Gamma^\mu \partial_\mu$ and $\slashed{\partial}_4^{\text{int}} \equiv \Gamma^a \partial_a$. Note that both of these operators contain a warp factor: $\slashed{\partial}_4^{\text{ext}} = Z^{1/4} \slashed{\partial}_{\R^{1,3}}$ and $\slashed{\partial}_4^{\text{int}} = Z^{-1/4} \slashed{\partial}_{\bT^{4}}$.

Plugging (\ref{flatD7Dirac}) into (\ref{D7fermE}), one can proceed with the dimensional reduction of the D7-brane fermionic action. First, we halve the degrees of freedom in (\ref{bispin}) by considering a bispinor of the form
\beq
\label{bispin1}
  {\Theta} = \begin{pmatrix} {\theta} \\ 0
\end{pmatrix}
\eeq
which is an allowed choice for fixing the $\kappa$-symmetry of the action. We can then express the D7-brane action as 
\be
\label{flatD7fermE1}
  S_{\uD 7}^{\left.\mathrm{fer}\right.}\, =\, \tau_{\uD 7}\, \ue^{\Phi_0}
  \int_{\R^{1,3}} \ud^{4}x \int_{\bT^4}  \ud\hat{\text{vol}}_{\bT^4} 
\,  \bar{\theta} \slashed{D}^w
  \,{\theta}
\ee
where $\theta$ stands for a conventional 10D MW spinor, $\ud\hat{\text{vol}}_{\bT^4}$ for the unwarped volume element of $\bT^4$ and the warped Dirac operator is given by
\beq
\label{wDirac}
\slashed{D}^w\, =\, {\slashed{\pa}}_4^{\text{ext}} + {\slashed{\pa}}_4^{\text{int}} - \frac18 \left({\slashed{\partial}}_4^{\text{int}} \ln Z\right) \left(1 + 2\G_{\text{Extra}}\right)
\eeq
$\G_{\text{Extra}} = \ud\slashed{\text{vol}}_{\bT^4}$ being the chirality operator for the internal dimensions of the D-brane. For instance, if we considered a D7-brane extended along the directions $0\dots 7$ then we would have $\G_{\text{Extra}} = \G^{\ul{4567}}$, with $\G^{\ul{i}}$ defined in (\ref{ulG:ap}).

Second, we split the 10D Majorana-Weyl spinor $\theta$ as
\beq
\theta\, =\, \chi + B^* \chi^* \quad \quad \chi\, =\,  \theta_{4D} \otimes \theta_{6D}
\label{splitgaug1}
\eeq
where $\theta_{4D}$ are four and $\theta_{6D}$ six-dimensional Weyl spinors, both of negative chirality, and $B = B_4 \otimes B_6$ is the Majorana matrix (\ref{ap:Maj}). 

Finally, one must decompose (\ref{splitgaug1}) as a sum of eigenstates under the (unwarped) 4D Dirac operator. More precisely, we consider the KK ansatz
\be
\label{KKflatferm}
\theta\, =\, \sum_\omega \theta^{\left.\omega\right.}\, =\, \sum_\omega \theta_{4D}^{\left.\omega\right.}(x) \otimes \theta_{6D}^{\left.\omega\right.}(y) + \sum_\omega \left(B_4\theta_{4D}^{\left.\omega\right.}(x)\right)^* \otimes \left(B_6\theta_{6D}^{\left.\omega\right.}(y)\right)^*
\ee
and we impose that $\G_{(4)}\slashed{\pa}_{\R^{1,3}} (B_4 \theta_{4D}^\omega)^*= - m_\omega\, \theta_{4D}^\omega$ where $\G_{(4)}$ is the 4D chirality operator. This indeed implies that each component $\theta^\omega$ of the sum above is  an eigenvector of $\G_{(4)}\slashed{\pa}_{\R^{1,3}}$, with a 4D mass  eigenvalue $|m_\omega|$.\footnote{As recalled in the appendix, we consider the eigenvalues of $\{\G_{(4)}\slashed{\pa}_{\R^{1,3}}, \G_{(4)}\slashed{\pa}_{\bT^4}\}$ instead of $\{\slashed{\pa}_{\R^{1,3}}, \slashed{\pa}_{\bT^4}\}$ because the former set of operators do commute and can hence be simultaneously diagonalized.} Imposing the 10D on-shell condition $\slashed{D}^w \theta = 0$ we arrive at the following 6D equation for the internal wavefunction of such
eigenvector\footnote{Na\"ively, this equation looks like it ignores the
Majorana-Weyl nature of $\theta$.  However, as discussed in Sec
\ref{gaugefix}, this is the equation of motion that we should use.}
\be
\label{flatD7eigen1}
\G_{(4)}\left[{\slashed{\pa}}_{\bT^4} - \frac18 \left({\slashed{\partial}}_{\bT^4}  \ln Z\right) \left(1 + 2\G_{\text{Extra}}\right) \right] \theta_{6D}^{\left.\omega\right.} \, =\, Z^{1/2} m_\omega (B_6 \theta_{6D}^{\left.\omega\right.})^*
\ee
It is then easy to see that the 4D zero modes of the action (\ref{D7fermE}) are given by
\bes
\label{flatD7zero1}
\begin{align}
\label{flatwilsonino1}
\theta_{6D}^0 = {Z^{-1/8}}\eta_- 
\quad & \text{for} \quad \G_{\text{Extra}}\, \eta_-\, =\, - \eta_-\\
\theta_{6D}^0 = {Z^{3/8}}\eta_+ 
\quad & \text{for} \quad \G_{\text{Extra}}\, \eta_+\, =\, \eta_+
\label{flatphotino1}
\end{align}
\ees
where $\eta_\pm$ are constant 6D spinor modes with $\pm$ chirality in the D7-brane extra dimensions. In particular, if we consider a D7-brane extended along $01234578$, then $\G_{\text{Extra}} = \G^{\ul{4578}}$ and the fermionic zero modes will have the following internal wavefunctions
\be
\theta_{6D}^{0,0}\, =\,
 Z^{3/8} \, \eta_{---}
\quad \quad
\theta_{6D}^{0,3}\, =\,
Z^{3/8} \, \eta_{++-}
\ee
and
\be
\label{eq:wilsoniniwaves}
\theta_{6D}^{0,1}\, =\,
Z^{-1/8} \, \eta_{-++}
\quad \quad
\theta_{6D}^{0,2}\, =\, 
Z^{-1/8} \, \eta_{+-+}
\ee
where the 6D fermionic basis $\{\eta_{---}, \eta_{++-}\dots\}$ has been defined in Appendix \ref{conv}.

Hence, we find that the warp factor dependence of the open string fermionic wavefunction depends on the chirality of such fermion in the D-brane extra dimensions. Note that this is because of the presence of $F_5$ in the D7-brane Dirac action. Indeed, had we considered an 8D Super Yang-Mills action instead of (\ref{D7fermE}), no projector $P_+^{O3}$ would have appeared in (\ref{flatD7Dirac}) nor any $\G_{\text{Extra}}$ operator in (\ref{wDirac}). Hence, the zero mode solution would have been $\theta_6^0 = Z^{1/8} \eta$ regardless of the eigenvalue of $\eta$ under $\G_{\text{extra}}$, as found in \cite{abv06}.

Note that (\ref{flatD7zero1}) implies a specific warp factor dependence on the 4D kinetic terms of the D7-brane zero modes. These are obtained by inserting them into (\ref{flatD7fermE1}). For (\ref{flatwilsonino1}) we find
\be
\label{ktermwil}
S_{\uD 7}^{\left.\mathrm{fer}\right.}\, =\, \tau_{\uD 7} \,\ue^{\Phi_0}
\int_{\R^{1,3}} \ud^{4}x\, \bar{\theta}_{4D} \slashed{\pa}_{\R^{1,3}} \theta_{4D} \int_{\bT^4}  \ud\hat{\text{vol}}_{\bT^4}\, \eta_-^\dag \eta_- 
\ee
so we have to divide by $\tau_{\uD 7} e^{\Phi_0}\hat{\text{vol}} (\bT^4)$ to obtain a canonically normalized kinetic term. Hence, for these zero modes nothing changes with respect to the unwarped case. On the other hand, for (\ref{flatphotino1}) we find
\be
\label{ktermphot}
S_{\uD 7}^{\left.\mathrm{fer}\right.}\, =\, \tau_{\uD 7}\, \ue^{\Phi_0}
\int_{\R^{1,3}} \ud^{4}x\, \bar{\theta}_{4D} \slashed{\pa}_{\R^{1,3}} \theta_{4D} \int_{\bT^4}  \ud\hat{\text{vol}}_{\bT^4}\, Z \eta_+^\dag \eta_+
\ee
which involves the warped volume $\text{vol} (\bT^4)$. In the following we will see that both kinetic terms are precisely the ones required to match those of the bosonic modes, as required by supersymmetry.

\subsubsection{Bosons}\label{bosonsflat}

In order to compute the D7-brane bosonic wavefunctions in a flat warped background, let us first analyze the degrees of freedom contained in the bosonic action (\ref{D7DBIE}). First we have the 8D gauge boson $A_\al$, that enters the bosonic action via its field strength $F = dA$ in ${\cal F} = P[B] + 2\pi \al' F$. Second, we have the transverse oscillations of the D7-brane worldvolume, that look like scalars from the 8D point of view, and that enter the bosonic action via the pull-back of $G$, $B$ and ${\cal C}$. Indeed, let us consider a D7-brane extended along the directions $01234578$. One can describe a deformation of this worldvolume on the transverse directions $69$ via two scalars $Y^6$ and $Y^9$, that depend on the worldvolume coordinates $x^\mu$ $\mu = 0, 1, 2, 3$ and $y^a$ $a= 4, 5, 7,8$. The pull-back 
of the metric in the deformed D7-brane is given by 
\be
\label{pbmetric}
\begin{array}{rcl}
P[G]_{\al\beta}& =& G_{\al\beta} + G_{ij} \pa_\al Y^i\pa_\beta Y^j + \pa_\al Y^i G_{i\beta} +  \pa_\beta Y^i G_{i\al} \\
& = & G_{\al\beta} + k^2 G_{ij} \pa_\al \sigma^i\pa_\beta \sigma^j
\end{array}
\ee
where $\al, \beta \in \{01234578\}$ are worldvolume coordinates and $i,j \in \{6, 9\}$ are transverse coordinates. In the second line we have used the fact that in our background $G_{i\alpha} = 0$ and redefined $Y^i = 2\pi \al' \sigma^i = k \sigma^i$ for later convenience. Clearly, the same expression applies for any flat D7-brane in flat space.

In general, a similar expansion applies for the pull-back of the B-field, although as before we are taking $B=0$ and a constant dilaton $\Phi = \Phi_0$. With these simplifications the DBI action for the D7-brane reads
\begin{eqnarray}\label{expD7DBI}
 S_{\uD 7}^{\left.\mathrm{DBI}\right.}& = &
  -\tau_{\uD 7}\int \ud^{8}\xi\, \ue^{\Phi}
  \sqrt{\abs{\det\bigl(P[G] +e^{-\Phi/2}\mathcal{{F}}\bigr)}}\\ \nonumber
  & = &-\tau_{D7} \int_{\mathbb{R}^{1,3}} \ud^4x \int_{\bT^4}  \ud\hat{\text{vol}}_{\bT^4}\,  e^{\Phi_0} \left\{ 1 + \frac12 k^2 G_{ij} G^{\al\beta} \pa_\al \sigma^i \pa_\beta \sigma^j + e^{-\Phi_0}\frac14 k^2 F_{\al\beta}F^{\al\beta} + \dots\right\} \\ \nonumber
  & = &  \left[S_{\uD 7}^{\left.\mathrm{DBI}\right.}\right]_0 - 
\left(8\pi^{3}k^{2}\right)^{-1} \int_{\mathbb{R}^{1,3}}   \ud^4x \int_{\bT^4}  \ud\hat{\text{vol}}_{\bT^4}\, \left(\frac12 e^{\Phi_0} G_{ij} G^{\al\beta} \pa_\al \sigma^i \pa_\beta \sigma^j + \frac14 F_{\al\beta}F^{\al\beta} + \dots \right)
\end{eqnarray}
where we have used the formula
\begin{equation}
  \label{eq:dettr}
  \det\left(1+M\right)=1 + \tr\left(M\right)
  +\frac{1}{2}\bigl[\tr\left(M\right)\bigr]^{2}
  -\frac{1}{2}\tr\bigl(M^{2}\bigr) + \cdots
\end{equation}
and dropped the terms containing more than two derivatives. Also, in the last line of (\ref{expD7DBI}) we have separated between a zero energy contribution to the D7-brane action and the contribution coming from derivative terms, the latter being the relevant part when computing the open string bosonic wavefunctions.

Besides the DBI action, the open string bosons enter the CS action of the D7-brane, which for the background at hand reads
\be
S_{\uD 7}^{\left.\mathrm{CS}\right.}\, = \,
\frac{\tau_{\uD 7}}{2} \int P[C_4]\wedge {\mathcal{F}} \wedge {\mathcal{F}}\, =\, \frac12  (2\pi k^2)^{-1} \int \left(C_4^{\text{ext}} + C_4^{\text{int}} \right) \wedge F \wedge F
\ee
as all the other RR potentials besides $C_4$ are turned off. We have also separated $C_4$ into internal and external components, with $C_4^{\text{ext}}$ containing $C_{0123}$ and $C_4^{\text{int}}$ the component $C_{abcd}$ whose indices lie all along the extra dimensions.\footnote{Note that a  background $C_4$ component of the form $C_{\mu\nu ab}$ would break 4D Poincar\'e invariance.} Finally, since the term $F\wedge F$ already contains two derivatives, we have neglected any term of the form $\pa_\al \sigma^i$ arising from expanding the pull-back of $C_4$ as in (\ref{pbmetric}).

As a result one can see that, up to two-derivative terms, the Chern-Simons action does not contain the D7-brane geometric deformations $\sigma^i$. The 8D action of such scalar fields then arises from the DBI expansion (\ref{expD7DBI}), and amounts to
\be
\label{flatD7scal}
S^{\left.\text{scal}\right.}_{\uD 7}\, =\, - \frac12
\left(8\pi^{3}k^{2}\right)^{-1}
e^{\Phi_0} \int_{\R^{1,3}} \ud^{4} x \int_{\bT^4} \ud \hat{\text{vol}}_{\bT^4}\, \hat{g}_{ij} \left(Z \eta^{\mu\nu}   \pa_\mu \sigma^i \pa_\nu \sigma^j + \hat{g}_{\bT^4}^{ab} \pa_a \sigma^i \pa_b \sigma^j \right)
\ee
and so we obtain the following 8D equation of motion
\be
\Box_{\R^{1,3}}\sigma^i + Z^{-1} \Box_{\bT^4} \sigma^i\, =\, 0
\ee
where $\Box_{\R^{1,3}} = \eta^{\mu\nu}\pa_\mu\pa_\nu$ and $\Box_{\bT^4} = \hat{g}_{\bT^4}^{ab}\pa_a\pa_b$. Performing a KK expansion
\begin{equation}
  \label{kkmod}
  \sigma^i \left(x^\mu,y^a\right)=\sum_{\omega}\zeta^i_{\omega}\bigl(x^\mu\bigr)
  s^i_{\omega}\bigl(y^a\bigr)
\end{equation}
and imposing the 4D Klein-Gordon equation $\Box_{\R^{1,3}} \zeta^i_\omega= m_\omega^2 \zeta^i_\omega$ we arrive at the eigenmode equation
\be
\label{KGint}
\Box_{\bT^4} s^i_\omega\, =\, - Z m^2_\omega\, s^i_\omega
\ee
that again contains a warp factor dependence. Such warp factor is however irrelevant when setting $m_\omega = 0$ and so we obtain that zero modes $s^i_0$ may either have a constant or linear dependence on the $\bT^4$ coordinates $y^a$. By demanding that $s^i_0$ is well-defined in $\bT^4$, that is by imposing the periodicity conditions on  $s^i_0(y^a +1) = s^i_0(y^a)$, the linear solution is discarded and we are left with a constant zero mode, that describes an overall translation of the D7-brane in the $i^{th}$ transverse coordinate.

Note that a trivial warp factor for scalar zero modes does not contradict our previous results for fermions, where we obtained warped wavefunctions. Indeed, in a supersymmetric setup like ours, the bosonic and fermionic wavefunctions should not necessarily match because of the presence of the (warped) vielbein in the
SUSY transformations.  However, the 4D effective kinetic terms should match. These are obtained by plugging $s^i_0 = const.$ in (\ref{flatD7scal}), after which we obtain
\be
\label{ktermscal}
S^{\left.\text{scal}\right.}_{\uD 7}\, =\, - \frac12 
\left(8\pi^{3}k^{2}\right)^{-1}
e^{\Phi_0} \int_{\R^{1,3}} \ud^{4}x\, \hat{g}_{ij} \eta^{\mu\nu} \pa_\mu \zeta_0^{i} \pa_\nu \zeta_0^j  \int_{\bT^4} \ud \hat{\text{vol}}_{\bT^4}\,  Z s_0^{i} s_0^j
\ee
which again involves a warped volume, like in (\ref{ktermphot}). Hence we find that the geometric zero modes of a D7-brane are related by supersymmetry with fermionic zero modes of the form (\ref{flatphotino1}).

Finally, by inserting the whole KK expansion (\ref{kkmod}) into the 8D action (\ref{flatD7scal}) and imposing (\ref{KGint}) one obtains the following 4D effective action
\be
\label{4Dscal}
S^{\left.\text{scal}\right.}_{\uD 7}\, =\, - \frac12
\left(8\pi^{3}k^{2}\right)^{-1} e^{\Phi_0} \sum_\omega  \int_{\R^{1,3}} \ud^{4}
x\,\hat{g}_{ij}  \left( \eta^{\mu\nu} \pa_\mu \zeta_{\omega}^i \pa_\nu \zeta_\omega^j  + m^2_\omega \zeta_{\omega}^i\zeta_\omega^j \right) \int_{\bT^4} \ud \hat{\text{vol}}_{\bT^4}\,  Z s_{\omega}^i s_\omega^j
\ee
where we have used that those wavefunctions with different 4D mass eigenvalue are orthogonal, in the sense that
\begin{equation}
  \int_{\bT^4}  \ud \hat{\text{vol}}_{\bT^4}\,  Z \hat{g}_{ij} s_\omega^i s_\chi^j\, =\, 0  \quad \quad  \text{if} \quad \quad m^2_\omega \neq m^2_\chi
  \end{equation}
as implied by the Sturm-Liouville problem eq.(\ref{KGint}).  Our primary concern is toward the zero modes and henceforth, we will will not consider the KK
modes.

Regarding the 8D gauge boson $A_\al$, the 8D action up to two derivatives reads
\be\nonumber
S_{\uD 7}^{\mathrm{gauge}} = - \frac14
\left(8\pi^{3}k^{2}\right)^{-1}
\int%_{\mathbb{R}^{1,3}\times \bT^4}  \hspace*{-.8cm} 
\ud^4x\, \frac{\ud\hat{\text{vol}}_{\bT^4}}{\sqrt{\hat{g}_{\bT^4}}}\, \left[\sqrt{\hat{g}_{\bT^4}}  F_{\al\beta}F^{\al\beta} - \frac12 \left( C_4^{\text{int}}  \epsilon^{\mu\nu\rho\sigma} F_{\mu\nu}F_{\rho\sigma} + C_4^{\text{ext}} \epsilon^{abcd} F_{ab}F_{cd} \right) \right]
\ee
where $\epsilon$ is a tensor density taking the values $\pm1$. As before $\al, \beta$ run over all D7-brane indices, $\mu,\nu,\rho,\sigma$ over the external $\R^{1,3}$ indices and $a,b,c,d$ over the internal $\bT^4$ indices of the D7-brane.
The gauge boson can be split in terms of 4D Lorentz indices as
$A_\al = (A_\mu, A_a)$ where the components $A_{\mu}$ give a 4D gauge boson while
the components $A_{a}$ give scalars in 4D.  The action contains a term that
mixes the scalars with the 4D photon
\begin{equation}
  \left(8\pi^{3}k^{2}\right)^{-1}
  \int_{\mathbb{R}^{1,3}}\ud^{4}x\,\int_{\mathbf{T}^{4}}\ud\hat{\vol}_{\mathbf{T}^{4}}
  \partial^{a}A_{a}\partial^{\mu}A_{\mu}
\end{equation}
which comes from the $F_{\mu a}F^{\mu a}$ term after integrating by parts twice.
In analogy with what is sometimes done in RS (see e.g. \cite{Randall:2001gb}),
this term can be gauged away by the addition of an $R_{\Xi}$ gauge-fixing term
to
the action,
\begin{equation}
  \label{eq:gaugefixterm}
  S_{\uD 7}^{\left.\Xi\right.}=-\left(8\pi^{3}k^{2}\right)^{-1}
  \int_{\mathbb{R}^{1,3}}\ud^{4}x\, \int_{\mathbf{T}^{4}}
  \ud\hat{\vol}_{\mathbf{T}^{4}}\, \frac{1}{2\Xi}\bigl(
  \partial^{\mu}A_{\mu}+\Xi\partial^{a}A_{a}\bigr)^{2}
\end{equation}
The form of this term is chosen to cancel the mixing term while preserving
Lorentz invariance. With this gauge choice, the $A_{\mu}$ and $A_{a}$ components
decouple. The action for $A_{\mu}$ in the $R_{\Xi}$ gauge is
\begin{align}
  S_{\uD 7}^{\left.\mathrm{photon}\right.}
  =-\bigl(8\pi^{3}k^{2}\bigr)^{-1}
  \int_{\mathbb{R}^{1,3}}
  \ud^{4}x
  \int_{\mathbf{T}^{4}}
  \frac{\ud\hat{\vol}_{\mathbf{T}^{4}}}{\sqrt{\hat{g}_{\mathbf{T}^{4}}}}
  &\biggl[\sqrt{\hat{g}_{\mathbf{T}^{4}}}\biggl(\frac{1}{4}F_{\mu\nu}F^{\mu\nu}
  +\frac{1}{2\Xi}\left(\partial^{\mu}A_{\mu}\right)^{2}\biggr)  \\ \notag
  & + \frac{1}{2}\sqrt{\hat{g}_{\mathbf{T}^{4}}}\eta^{\mu\nu}
  \hat{g}_{\mathbf{T}^{4}}^{ab}
  \partial_{a}A_{\mu}\partial_{b}A_{\nu}
  -\frac{1}{8}C_{4}^{\mathrm{int}}\epsilon^{\mu\nu\rho\sigma}
  F_{\mu\nu}F_{\rho\sigma}\biggr]
\end{align}
which results in the equation of motion
\begin{equation}
  \Box_{\mathbb{R}^{1,3}}A_{\nu}
  -\bigl(1-\frac{1}{\Xi}\bigr)
  \eta^{\mu\sigma}\partial_{\nu}\partial_{\mu}A_{\sigma}
  + Z^{-1}\Box_{\mathbf{T}^{4}}A_{\nu}=0
  \label{flatD7eomgauge}
\end{equation}
where again, $\Box_{\mathbb{R}^{1,3}}$ and $\Box_{\mathbf{T}^{4}}$ are the
unwarped Laplacians on $\mathbb{R}^{1,3}$ and $\mathbf{T}^{4}$
respectively.  Here
 we have used that $\hat{g}_{\bT^4}$ is constant, that $Z, C_4$ are
$\mathbb{R}^{1,3}$-independent, and that $F_{\rho\sigma} = \pa_\rho A_\sigma - \pa_\sigma A_\rho$ is an exact two-form.  Similarly, for the 4D Lorentz scalars $A_{a}$, we obtain the
action
\begin{align}
  S_{\uD 7}^{\left.\mathrm{wl}\right.}
  =-\bigl(8\pi^{3}k^{2}\bigr)^{-1}
  \int_{\mathbb{R}^{1,3}}\ud^{4}x\int_{\mathbf{T}^{4}}
  \frac{\ud\hat{\vol}_{\mathbf{T}^{4}}}{\sqrt{\hat{g}_{\mathbf{T}^{4}}}}
  \biggl[&\sqrt{\hat{g}_{\mathbf{T}^{4}}}\biggl(
  \frac{1}{4}F_{ab}F^{ab}+\frac{\Xi}{2}\bigl(\partial^{a}A_{a}\bigr)^{2}\biggr) 
  \notag \\
  &+\frac{1}{2}\sqrt{\hat{g}_{\mathbf{T}^{4}}}\partial_{\mu}A_{n}\partial^{\mu}A^{n}
  -\frac{1}{8}C_{4}^{\mathrm{ext}}\epsilon^{abcd}F_{ab}F_{cd}\biggr]
\end{align}
from which we get the equation of motion in the $R_{\Xi}$ gauge
\begin{equation}
  \Box_{\mathbb{R}^{1,3}}A^{a}
  +Z^{-1/2}\partial_{b}F^{ba}+\Xi\partial^{a}\bigl(Z^{-1/2}\partial^{b}A_{b}\bigr)
  +\frac{Z^{-1/2}}{\sqrt{\hat{g}_{\mathbf{T}^{4}}}}\epsilon^{abcd}
  \partial_{b}\bigl(Z^{-1}F_{cd}\bigr)=0
\label{flatD7eomwl}
\end{equation}
where we have made use of $C_4^{\text{ext}} = Z^{-1} + const.$, as implied by our bulk supergravity ansatz, and more precisely by (\ref{10Dmetricst}) and (\ref{F5st}).  

Let us now consider the following KK decomposition for the 4D gauge boson
\be
A_{\mu}\bigl(x,y\bigr)\,=\,\sum_{\omega}A_{\mu}^\omega \bigl(x^\mu\bigr)\al^{\omega}(y^a)
\ee
with the 4D wavefunction satisfying the massive Maxwell equation in the
$R_{\Xi}$ gauge
\begin{equation}
  \label{eq:flatmassivemaxwell}
  \Box_{\mathbb{R}^{1,3}}A_{\mu}^{\omega}
  -\bigl(1-\frac{1}{\Xi}\bigr)
  \eta^{\nu\sigma}\partial_{\mu}\partial_{\nu}A_{\sigma}^{\omega}
  =m_{\omega}^{2}A_{\mu}^{\omega}
\end{equation}
So that in an specific $R_{\Xi}$ gauge,
(\ref{flatD7eomgauge}) amounts
to 
\be
\label{KGgauge}
\Box_{\bT^4} \al^\omega \, =\, - Z m^2_\omega\, \al^\omega
\ee
Hence, we recover the same spectrum of internal KK wavefunctions as for the transverse scalar (\ref{kkmod}).
In particular, we recover a constant zero mode $\al^0$ and an effective  kinetic term given by the real part of the 4D gauge kinetic function
\be
f_{D7}\, =\, 
\bigl(8\pi^{3}k^{2}\bigr)^{-1}\int_{\bT^4}
 \frac{\ud\hat{\text{vol}}_{\bT^4}}{\sqrt{\hat{g}_{\bT^4}}}
\left(Z \sqrt{\hat{g}_{\bT^4}} + i C_4^{\text{int}}  \right) (\al^0)^2
\label{gkfD7}
\ee
whose holomorphicity has been studied in \cite{bdkmmm06}. Notice that the kinetic terms again involve a warped volume, so we conclude that the D7-brane 4D gaugino is also given by a fermionic zero mode of the form  (\ref{flatphotino1}).

Similarly, one can decompose the $\R^{1,3}$ scalars arising from $A_\al$ as
\be
A_{a}\bigl(x,y\bigr)\,=\,\sum_{\omega}w^{\omega}_a \bigl(x^\mu\bigr) W^{\omega}_a \bigl(y^a\bigr).
\ee
and impose the 4D on-shell condition $\Box_{\R^{1,3}} w_{a}^\omega= m_\omega^2 w_{a}^\omega$. Then the 8D eom (\ref{flatD7eomwl})  becomes
\begin{equation}
  \partial_{b}F^{\omega\, ba}+
  Z^{1/2}\Xi\partial^{a}\bigl(Z^{-1/2}\partial^{b}W_{b}^{\omega}\bigr)
  +\frac{1}{\sqrt{\hat{g}_{\mathbf{T}^{4}}}}\epsilon^{abcd}
  \partial_{b}\bigl(Z^{-1}F^{\omega}_{cd}\bigr)=
  -Z^{1/2}m_{\omega}^{2}W^{\omega\, a}
\label{flatD7eigenwl}
\end{equation}
where we have defined $F^\omega_{ab} \equiv \pa_a W_b^\omega - \pa_b W_a^\omega = dW^\omega$.
Note that if we chose the 4D Lorenz gauge $\Xi=0$,
in the case of the zero modes $m_{\omega = 0} = 0$ the above equation is equivalent to
\be
d\left[Z^{-1}\left(1 - *_{\bT^4}\right) F^{\, 0}\right]\, =\, 0
\label{wlzeromode}
\ee
where $F^0 = \frac12 F^0_{ab} dy^a \wedge dy^b$ is the zero-mode two-form. This implies that $(1 - *_{\bT^4}) F^0 = Z \omega_2$, where $\omega_2$ is a harmonic, anti-self-dual two-form in $\bT^4$. Because $F^0$ is exact, the integral of $Z\omega_2$ over any two-cycle of $\bT^4$ has to vanish, and so we deduce that $\omega_2 = 0$. Hence $F^0 = *_{\bT^4} F^0$ is a self-dual form and, again using the exactness of $F^0$, we deduce that $F^0 = 0$. This is solved by taking $W_a^0 = const.$, like for the previous bosonic wavefunctions. Finally, inserting such $W_a^0$ in the 8D bosonic action we obtain the 4D effective action in the 4D Lorenz gauge $\Xi = 0$
\be
\label{ktermwl}
S^{\left.\text{wl}\right.}_{\uD 7}\, =\, - \frac12 
\left(8\pi^{3}k^{2}\right)^{-1}
 \int_{\R^{1,3}} \ud^{4}x\, \hat{g}^{ab}_{\bT^4} \eta^{\mu\nu} \pa_\mu w^0_a \pa_\nu w_b^0  \int_{\bT^4} \ud \hat{\text{vol}}_{\bT^4}\,  W_a^0 W_b^0
\ee
which only involves the unwarped $\bT^4$ volume. This matches with the 4D kinetic terms of their fermionic superpartners (\ref{flatwilsonino1}). 
Note that in imposing the 4D Lorenz gauge, language there is still
a residual gauge symmetry which in 8D language is
$A_{\alpha}\to A_{\alpha}-\partial_{\alpha}\Lambda$ where
$\partial_{\mu}\Lambda=0$.  It is easy to see that this residual gauge symmetry
is respected by the entire 4D effective action and we can use it to set
$W_{a}^{0}$ to be constant.

Although the equations were solved in the 4D Lorenz gauge, $W_{a}^{0}=const.$
and $m_{0}=0$ is a solution to
(\ref{flatD7eigenwl})
for any choice of $\Xi$.  However, for the KK modes, some of the masses
will depend on the choice of gauge.  This is related to the fact that, except
for the zero mode, each of the vectors $A_{\mu}^{\omega}$ has a mass and so
corresponds to the gauge boson of a spontaneously broken gauge symmetry in
the effective 4D language.  The modes with $\Xi$-dependent masses correspond to
Goldstone bosons that are eaten by KK vectors which then become massive.
Similarly, $\alpha_{0}=\mathrm{const.}$ is a zero mode of
(\ref{eq:flatmassivemaxwell}) for any choice of $\Xi$.  Finally, one can again
show that the KK modes are orthogonal with the zero modes as they were for the
position modulus.

\subsubsection{Summary and comparison to RS}\label{summary}

In the previous subsections we have analyzed the zero modes of a 
D7 brane wrapping a 4-cycle in a warped compactification.
One could see this as a step towards a string theory realization of an extended
supersymmetric RS scenario~\cite{Gherghetta:2000qt}. In the standard WED setup, 
4D fields result from the dimensional reduction of the zero modes of 5D fields propagating 
in the bulk of $\mathrm{AdS}_{5}$.\footnote{These bulk RS models also involve an orbifold
$\mathbf{S}^{1}/\mathbb{Z}_{2}$. The effect of the orbifold is however  to project out
certain zero modes and does not effect the dependence on the warp factor
of the surviving modes.}  Unlike for flat space, the
supersymmetry algebra in $\mathrm{AdS}_{5}$ implies that component fields have
different 5D masses \cite{Shuster:1999zf}. In particular, the 4D gauge boson
and gaugino come from a 5D $\mathcal{N}=1$ vector supermultiplet.  Gauge
invariance requires that the 5D vector component is massless, while SUSY
requires that the 5D gaugino has mass $\frac{1}{2}K$ where $K=1/R$ is the
AdS curvature. %\footnote{There is a slight complication when the 5$^{th}$ dimension
%is orbifolded, but again this does not effect the dependence of the zero modes on
%the warp factor which is what we are focusing on here.}  
Similarly, the matter fields result from the
reduction of a 5D hypermultiplet, the component fields of which each have
a different mass.

\TABLE[t]{
\begin{tabular}{lccclcc}
\toprule
\multicolumn{3}{c}{RS} & \hspace{4pt} & \multicolumn{3}{c}{D7} \\
\cline{1-3}\cline{5-7}
4D Field & $p$ & $q$ & & 4D Field &$p$ &$q$ \\
\cline{1-3}\cline{5-7}
gauge boson & $0$ & \multirow{2}{*}{$1/4$} & & 
gauge boson/modulus & $0$ & \multirow{2}{*}{$1$} \\
gaugino & $3/8$ & & & gaugino/modulino & $3/8$ & \\
\cline{1-3}\cline{5-7}
matter scalar & $(3-2c)/8$ & \multirow{2}{*}{$(1-c)/2$} & &
Wilson line & $0$ & \multirow{2}{*}{$0$} \\
matter fermion & $(2-c)/4$ & & & Wilsonino & $-1/8$ & \\
\bottomrule
\end{tabular}
\label{table:warppowers}
\caption{Warp factor dependence for internal wavefunctions ($p$) and
K\"ahler metric ($q$) in the RS scenario
and the D-brane construction consdered here.  In RS, the gauge boson and
gaugino come from a 5D vector multiplet while the matter scalar and fermion
come from a 5D hypermultiplet.  The 5D mass of the fermion in the hypermultiplet
is $cK$ with $K$ the AdS curvature.  The additional degrees of freedom from
these supermultiplets are projected out by the orbifold action is RS.
The wavefunctions in SUSY RS are worked out
in~\cite{Gherghetta:2000qt} (our conventions differ slightly from theirs in
that we take the
ansatz for the 5D fermion to be $\Psi_{\mathrm{L,R}}\left(x,y\right)
=\psi_{\mathrm{L,R}}\left(x\right)\chi_{\mathrm{L,R}}\left(y\right)$ while
\cite{Gherghetta:2000qt} uses a power of the warp factor in the
decomposition.)
}}

The D7-brane construction here differs not only because 
of the existence of additional spatial dimensions, but also because of the presence of
additional background fields, namely the RR potential $C_{4}$ that couples to open string modes 
via the D7-brane CS and fermionic action. This results into a different behavior of the internal wavefunctions when compared to the analogous RS zero modes, as shown in Table \ref{table:warppowers}. For each field, the wavefunction can be
written as $Z^{p}\eta$ where $\eta$ is a constant function with the appropriate
Lorentz structure.  The kinetic terms for each 4D field can then be written
schematically as
\begin{equation}
\label{genkterm}
  \int_{\mathbb{R}^{1,3}}\ud^{4}x
  \bar{\phi}D\phi\int_{\mathrm{int}}\ud\hat{\vol}_{\mathrm{int}}\,
  Z^{q}\bar{\eta}\eta
\end{equation}
where $\phi$ is a 4D field with kinetic operator $D$, $\eta$ is the
corresponding constant
internal wavefunction
and `$\mathrm{int}$' denotes the unwarped internal space 
($\mathbf{S}^{1}/\mathbb{Z}_{2}$ for
RS or $\mathbf{T}^{4}$ here).  Since both
the D-brane construction considered here and the extended SUSY RS model
are supersymmetric, the 4D fields can be arranged into supermultiplets with
the same value of $q$ for each component field.  These are also given in
Table \ref{table:warppowers}.

\subsubsection{More on the equation of motion}\label{gaugefix}

When deducing the fermionic equation of motion $\slashed{D}^w\theta =0$ from the $\kappa$-fixed action (\ref{flatD7fermE1}), we have apparently ignored the Majorana-Weyl nature of $\theta$.\footnote{We would like to thank D. Simi\'c and L. Martucci for discussions related to this subsection.}  Indeed, the MW condition implies that in deriving the equation of motion, $\theta$ and $\bar{\theta}$ cannot be varied independently. As a consequence, if given the two actions
\begin{equation}
  \tau_{\uD 7}\int\ud^{8}\xi\, 
  \bar{\theta}\Gamma^{\alpha}\partial_{\alpha}\theta
  \quad\quad \text{and} \quad \quad
   \tau_{\uD 7}\int\ud^{8}\xi\, 
  \bar{\theta}\Gamma^{\alpha}\bigl(\partial_{\alpha} - \partial_{\alpha}\ln f\bigr)\theta
  \label{actionMW}
\end{equation}
with $f$ an arbitrary function, then the resulting equation of motion is simply $\Gamma^{\alpha}\partial_{\alpha}\theta=0$ in both cases, solved by $\theta=\eta$ with $\eta$ a constant MW spinor. This is in clear contrast to the case where $\theta$ in (\ref{actionMW}) is a Weyl spinor, since then for the second action the eom solution is given by $\theta = f\eta$. This could have been anticipated from the fact that the 10D MW nature of $\theta$ implies that $\bar{\theta}\Gamma^{a_{1}\cdots a_{n}}\theta$ is non-vanishing only for $n=3,7$. Hence, we have that $\bar{\theta} (\slashed{\partial} \ln f) \theta \equiv 0$ and so, in the MW case, both actions in (\ref{actionMW}) are the same.

Going back to the fermionic action (\ref{flatD7fermE1}), we have that
\be
\bar{\theta}_\pm \slashed{D}^w \theta_\pm\, \equiv\, \bar{\theta}_\pm \left(\slashed{\partial}_4^{\text{ext}} + {\slashed{\pa}}_4^{\text{int}}\right) \theta_\pm
\ee
where $\slashed{D}^w$ is given by (\ref{wDirac}) and $\theta_\pm$ are 10D MW spinors with $\pm1$ eigenvalue under $\G_{\text{Extra}}$, just like those constructed from (\ref{flatD7zero1}). Hence, by analogy with (\ref{actionMW}) one could na\"ively conclude that the actual zero mode equation is given by ${\slashed{\pa}}_4^{\text{int}}\theta^0_{6D} = 0$, instead of $\slashed{D}^w \theta^0_{6D} = 0$.

A more careful analysis shows that this is not the case. Indeed, 
\be
\delta  S_{\uD 7}^{\left.\mathrm{fer}\right.}\,  =\,  \tau_{\uD 7}\, \ue^{\Phi_0}
%\int_{\uD 7}% 
\int\ud^{8}\xi\, \overline{\delta \theta} \slashed{D}^w \theta + \bar{\theta}  \slashed{D}^w \delta \theta 
\, =\,  2 \tau_{\uD 7}\, \ue^{\Phi_0}
%\int_{\uD 7}
\int\ud^{8}\xi\,   \overline{\delta \theta} \slashed{D}^w \theta
\label{variation}
\ee
where we have used that
\be
\int\ud^{8}\xi\, Z^{-1/4} \bar{\theta} {\slashed{\pa}}_{\bT^{4}} {\delta\theta}
 =\int\ud^{8}\xi\,  Z^{-1/4} \overline{\delta \theta} \left({\slashed{\pa}}_{\bT^{4}} -\frac{1}{4} {\slashed{\pa}}_{\bT^{4}} \ln Z \right) {\theta}
\ee
and that $\overline{\theta} {\slashed{\pa}}_{\bT^{4}} \ln Z {\delta\theta} = -\overline{\delta \theta}{\slashed{\pa}}_{\bT^{4}} \ln Z {\theta}$. Hence, from (\ref{variation}) we read that the equation of motion is indeed $\slashed{D}^w \theta = 0$. Note that we would have obtained the same result if we had treated $\theta$ and $\bar\theta$ as independent fields.

While in principle one could apply the same kind of computation to deduce the equation of motion for the more general backgrounds to be discussed below, let us instead follow the results of \cite{bs06}. There, using the action presented in \cite{Bandos:1997rq} (similar to that in \cite{mrvv05} to quadratic order in fermions) the following equation of motion was deduced for an unmagnetized D7-brane
\begin{equation}
  P^{\uD7}_-
  \biggl({\Gamma}^{\al}
  {\cal D}_{\al}^E+\frac12{\cal O}^E\biggr)\,{\Theta}=0
\label{bs06eom}
\end{equation}
which is again the equation found from (\ref{D7fermE}) if we na\"ively ignore the MW nature of $\Theta$. %An analogous statement applies for magnetized D7-branes.

A subtle point in deriving (\ref{bs06eom}) is that a particular gauge choice in the fermionic sector must be made. Indeed, in \cite{bs06} the background superdiffeomorphisms were used to choose a supercoordinate system in which the $\uD 7$-brane does not extend in the Grassmann-odd directions of superspace. One may then wonder whether such fermionic gauge fixing is compatible with the gauge fixing choices taken in the bosonic sector. One can check this by comparing the SUSY transformations in
10D with those in 4D.  In the absence of NS-NS flux, the $\kappa$-fixed SUSY transformations for the bosonic modes are
\cite{mrvv05}
\begin{subequations}
\label{eq:10dsusyvar}
\begin{align}
  \delta_{\epsilon}Y^{i} =&\,\bar{\epsilon}\, \Gamma^{i}\theta \\
  \delta_{\epsilon}A_{\alpha} =&\, \bar{\epsilon}\, \Gamma_{\alpha}\theta
\end{align}
\end{subequations}
where $\epsilon$ is  the 10D Killing spinor.  We can compare these against the SUSY
transformations in 4D for a
chiral multiplet $\left(\phi,\psi\right)$ and a vector multiplet
$\left(\lambda,A\right)$,
\begin{subequations}
\begin{align}
  \delta_{\varepsilon}\phi=&\, \bar{\varepsilon}\, \psi \\
  \delta_{\varepsilon}A_{\mu}=&\, \bar{\varepsilon}\, \lambda
\end{align}
\label{4dsusytr}
\end{subequations}
where $\varepsilon$ is a constant 4D spinor and hence independent of the warp factor. 
This implies that when we dimensionally reduce (\ref{eq:10dsusyvar}), we will only recover the
standard 4D transformations (\ref{4dsusytr}) if the warp factor dependence of bosons and 
fermions follows a particular relation. Indeed, if we take the zero modes $A_\al$ and $Y^i$ to
have no warp factor dependence as in subsection \ref{bosonsflat}, and if we notice that $\G^i \sim \G_\mu \sim Z^{-1/4}$, $\G_a \sim Z^{1/4}$, $\bar{\ep} \sim Z^{-1/8}$, then it is easy to see that precisely the fermionic wavefunctions of subsection \ref{flatferm} are those needed to cancel the warp factor dependence in the r.h.s. of (\ref{eq:10dsusyvar}).

%To summarize, the equation of motion resulting from (\ref{D7fermE}) is
%ambiguous due to the Majorana-Weyl nature of the spinors $\theta$.  This
%ambiguity can be fixed following \cite{bs06} which is consistent with
%the gauge choices made in the bosonic sector, in the sense that we
%recover the standard 4D SUSY transformations (\ref{4dsusytr}) upon dimensional
%reduction. Basically, the gauge choice in the fermionic amounts to ignore the Majorana-Weyl
%nature of $\theta$ when deriving its Dirac equation from (\ref{D7fermE}). Finally, while in order
%to cross-check the consistency of our framework we have made us  of 4D supersymmetry,
% we expect the general result to hold in a non-supersymmetric setup as well.

\subsubsection{Alternative $\kappa$-fixing}\label{altk}

When analyzing the D7-brane fermionic action, the $\kappa$-fixing choice (\ref{bispin1}) has the clear advantage of expressing everything in terms of a conventional 10D spinor $\theta$, in contrast to the less familiar bispinor $\Theta$ that would appear in general. Taking other choices of $\kappa$-fixing may, however, provide their own vantage point. Indeed, we will show below that taking a different $\kappa$-fixing choice not only allows to rederive the results above, but also to better understand the structure of D7-brane zero modes in a warped background.

More precisely, let us as before consider the action (\ref{D7fermE}) in waped flat space, but now we choose $\Theta$ such that $P_-^{\uD7}\Theta = 0$. The action (\ref{D7fermE}) then reads
\begin{equation}
\label{flatD7fermEalt}
  S_{\uD 7}^{\mathrm{fer}}\, =\, \tau_{\uD 7} \ue^{\Phi_0}
  \int_{\R^{1,3}} \ud^{4}x \int_{\bT^4}  \ud\hat{\text{vol}}_{\bT^4} 
\,  \bar{\Theta} \slashed{D}^w
  \,{\Theta}
\end{equation}
where $\slashed{D}^w$ is now given by (\ref{flatD7Dirac}). Following a similar strategy as in subsection \ref{flatferm}, we split the 10D Majorana-Weyl spinors $\theta_i$ in (\ref{bispin}) as
\beq
\theta_i\, =\, \chi_i + B^* \chi_i^* \quad \quad \chi_i\, =\,  \theta_{i,4} \otimes \theta_{i,6}
\label{splitgaugalt}
\eeq
where $\theta_{i,4}$ are 4D and $\theta_{i,6}$  6D Weyl spinors, all of negative chirality, and $B = B_4 \otimes B_6$ is again the Majorana matrix (\ref{ap:Maj}). Because of the condition $P_-^{\uD7}\Theta = 0$ one can set $\theta_{1,4} = \theta_{2,4} = \theta_{4D}$, so  that we have
\be
\label{decomp46alt}
\Theta\, =\, \theta_{4D} \otimes \Theta_{6D} + B_4^*\theta_{4D}^* \otimes B_6^* \Theta_{6D}^*
\quad \quad
\Theta_{6D} \, =\, 
\left(
\begin{array}{c}
\theta_{1,6}\\ \theta_{2,6}
\end{array}
\right)
\ee
where $\Theta_{6D}$ satisfies $P_+^{\text{Extra}} \Theta_{6D} = 0$, with
\be
P_\pm^{\text{Extra}} \, =\, \frac12 \left( \Id \pm \G_{\text{Extra}} \otimes \sig_2 \right)
\ee

Decomposing  (\ref{decomp46alt}) as a sum of eigenstates under the (unwarped) 4D Dirac operator, and  imposing $\G_{(4)}\slashed{\pa}_{\R^{1,3}} (B_4 \theta_{4D}^\omega)^*= - m_\omega\, \theta_{4D}^\omega$ and $\slashed{D}^w \Theta = 0$ leads to the 6D bispinor equation
\be
\label{flatD7eigenalt}
\G_{(4)}\left[{\slashed{\pa}}_{\bT^4} - \frac18 \left({\slashed{\partial}}_{\bT^4}  \ln Z\right) \left(1 + 2\G_{\text{Extra}}\otimes\sigma_{2}\right) \right] \Theta_{6D}^\omega \, =\, Z^{1/2} m_\omega (B_6 \Theta_{6D}^\omega)^*
\ee
which is analogous to (\ref{flatD7eigen1}). Finally, instead of (\ref{flatD7zero1}) we obtain 
\bes
\label{flatD7zeroalt}
\begin{align}
\label{flatwilsoninoalt}
\Theta_{6D}^0 = \frac{Z^{-1/8}}{\sqrt{2}} 
\left(
\begin{array}{c}
\eta_{-}\\ i\eta_{-}
\end{array}
\right) 
\quad & \text{for} \quad \G_{\text{Extra}} \eta_-\, =\, - \eta_-\quad \quad \text{Wilsonini}\\
\Theta_{6D}^0 = \frac{Z^{3/8}}{\sqrt{2}} 
\left(
\begin{array}{c}
i\eta_{+}\\ \eta_{+}
\end{array}
\right)
\quad & \text{for} \quad \G_{\text{Extra}} \eta_+\, =\, \eta_+\quad \quad \text{gaugino + modulino}
\label{flatphotinoalt}
\end{align}
\ees
and so we recover the same warp factor dependence in terms of the extra-dimensional chirality of the spinor. It is also easy to see that upon inserting such solutions into the D7-brane action we recover the same 4D kinetic terms as in (\ref{ktermwil}) and (\ref{ktermphot}).

Interestingly, the above set of zero modes have a simple interpretation in the context of 10D type IIB supergravity. Indeed, note that for this choice of $\kappa$-fixing the D7-brane zero modes can be rewritten as
\bes
\label{flatD7zero2}
\begin{align}
\label{flatD7zerog}
\Theta =  Z^{-1/8} \Xi_- &\quad \quad \text{with} \quad \quad P_+^{D3} \Xi_- \, =\, P_-^{D7} \Xi_- \,=\, 0\\
\Theta =  Z^{3/8} \Xi_+ &\quad \quad \text{with} \quad \quad P_-^{D3} \Xi_+  \, =\, P_-^{D7} \Xi_+  \,=\, 0
\end{align}
\ees
and $\Xi_\pm$ constant bispinors. This last expression can be easily deduced from (\ref{flatD7Dirac}) and the fact that $P_\pm^{O3}$ and
\be
P_\pm^{D3}\, =\, \frac12 \left(\Id \pm \G_{(4)}  \otimes \sig_2 \right)
\label{projD3}
\ee
are equivalent when acting on type IIB Weyl spinors. As explained in the appendix \ref{conv}, $P_-^{D3}$ is the projector that has to be inserted in the D3-brane fermionic action, in the same sense that  $P_-^{D7}$ is inserted in (\ref{D7fermE}). This implies that 10D bispinors satisfying $P_-^{D3} \Theta = 0$ will enter the D3-brane action, while those satisfying $P_+^{D3} \Theta = 0$ will be projected out. For instance, a D3-brane in flat 10D space will have precisely four 4D fermion zero modes of the form $\Theta = const.$, $P_-^{D3} \Theta = 0$. Such a D3-brane, which is a 1/2 BPS object, breaks the amount of 4D supersymmetry as ${\cal N} = 8 \rightarrow {\cal N} = 4$, so these four zero modes can be interpreted as the four goldstini of the theory. Conversely, the constant bispinors satisfying $P_+^{D3} \Theta = 0$ can be identified with the four generators of the ${\cal N} =4$ superalgebra surviving the presence of the D3-brane.

If we now consider a warped background created by a backreacted D3-brane, we have four Killing (bi)spinors generating the corresponding ${\cal N} =4$ SUSY. Those Killing bispinors $\ep$ must satisfy ${\cal O}\ep = {\cal D}_\mu \ep = {\cal D}_m \ep =0$, where ${\cal O}$ and ${\cal D}_M$ are given by (\ref{flatD7op}). It is easy to see that the solution are of the form $\ep = Z^{-1/8} \Theta$ where $\Theta$ is constant and, as argued above, satisfies $P_+^{D3}\Theta = 0$. Introducing a D7-brane in this background will break the bulk supersymmetry as ${\cal N} =4 \rightarrow {\cal N} =2$, so the D7-brane should develop two goldstino zero modes. Now, by taking the $\kappa$-fixing choice $P_-^{D7} \Theta = 0$ the Dirac action takes the simple form (\ref{flatD7fermEalt}), and so such goldstini amount to the pull-back of the above Killing spinors into the D7-brane\footnote{Recall that $\slashed{D}^w$ is a linear combination of gravitino and dilatino operators, pulled-back into the D7-brane worldvolume.} or, more precisely, those which are not projected out by the condition $P_-^{D7} \Theta = 0$. These are precisely the zero modes in (\ref{flatD7zerog}), whose warp factor dependence is thus to be expected.

Hence, we again see by supersymmetry arguments that such modes could never have a warp factor dependence of the form $Z^{1/8}$, which would only be allowed if we turned off the RR flux $F_5$ from our background. Indeed, in that case the background would not satisfy the equations of motion, so no supersymmetry would be preserved and the arguments above do not apply.

\subsection{Warped Calabi-Yau}\label{warpcy}

Let us now extend the above analysis to include warped backgrounds (\ref{10metricE})  with a non-flat internal space $X_6$. We will however still consider a constant axio-dilaton field $\tau = C_0 + ie^{-\Phi_0}$, which constrains $X_6$ to be a Calabi-Yau manifold. This basically means that the holonomy group of $X_6$ must be contained in SU(3), which in turn guarantees that there is a globally defined 6D spinor $\eta^{\CY}$, invariant under the SU(3) holonomy group and satisfying the equation
\beq
\nabla^\CY_m\, \eta^\CY_-\, =\, 0
\label{ccspinor}
\eeq
where $\nabla^\CY$ is the spinor covariant derivative constructed from the unwarped, Calabi-Yau metric of $X_6$, and where we have taken $\eta^\CY$ to be of negative chirality. If we choose $X_6$ to be of proper SU(3) holonomy, meaning that its holonomy group is contained in SU(3) but not in any SU(2) subgroup of the latter, then the solution to (\ref{ccspinor}) is unique, and the only other covariantly constant spinor besides $\eta_-^\CY$ is its conjugate $\eta^\CY_+ = (B_6 \eta_-^\CY)^*$.

As emphasized in the literature, these facts are crucial in specifying the supersymmetry generators of not only unwarped, but also warped Calabi-Yau backgrounds. Indeed, it is easy to see that for a warped Calabi-Yau the 10D gravitino and dilatino variation operators are given by
\bes
\label{CYD7op}
\begin{align}
{\cal O}\, =& \, 0 \\
{\cal D}_\mu \, =& \, \pa_\mu - \frac{1}{4} \G_\mu \slashed{\pa} \ln Z P_+^{O3}\\
{\cal D}_m \, =& \, \nabla_m^\CY + \frac{1}{8}\pa_m \ln Z   - \frac{1}{4} \slashed{\pa}\ln Z \G_m  P_+^{O3}
\end{align}
\ees
where $P_+^{O3}$ is again defined by (\ref{O3proj}). In terms of these operators the background supersymmetry conditions read ${\cal O} \ep = {\cal D}_\mu\ep = {\cal D}_m\ep = 0$, where $\ep$ a type IIB bispinor like (\ref{bispin}). If we now take the ansatz
\beq
\label{ansatzCY}
\ep\, =\, 
\left(
\begin{array}{c}
\ep_1 \\ \ep_2
\end{array}
\right)
\quad \quad
\ep_i\, =\, \lam_i + B^*\lam_i^*
\quad \quad
\lam_i\, =\, \ep_{i,4D}(x) \otimes \ep_{i,6D}(y)
\eeq
with $\ep_{i,4D}$ and $\ep_{i,6D}$ of negative chirality, it is easy to see that ${\cal D}_\mu \ep = 0$ imposes $P_+^{O3}\ep = 0$ and $\pa_\mu \ep = 0$, while ${\cal D}_m\ep = 0$ in addition sets $\ep_{i,6D}$ proportional to $Z^{-1/8}\eta^\CY_-$. That is, our warped Killing bispinor is of the form
\beq
\ep\, = \, \ep_{4D} \otimes 
Z^{-1/8} \left(
\begin{array}{c}
\eta^\CY_- \\ i \eta^\CY_-
\end{array}
\right) - i 
B_4^* \ep_{4D}^* \otimes 
Z^{-1/8} \left(
\begin{array}{c}
i \eta^\CY_+ \\  \eta^\CY_+
\end{array}
\right)
\label{wkilling}
\eeq
where $\ep_{4D}$ is a constant 4D spinor that, upon compactification, will be identified with the generator of ${\cal N} =1$ supersymmetry in $\R^{1,3}$. Note that in (\ref{wkilling}) we have set $\ep_{1,4D} = \ep_{2,4D} = \ep_{4D}$ because such identification is enforced by the condition $P_+^{O3} \ep = 0$. On the other hand, if we take the unwarped limit $Z \rightarrow 1$ then $P_+^{O3} \ep = 0$ no longer needs to be imposed, and so $\ep_{1,4D}$ and $\ep_{2,4D}$ are independent spinors that generate a 4D ${\cal  N} =2$ superalgebra. Thus we recover the fact that any source of warp factor breaks the Calabi-Yau ${\cal N} = 2$ supersymmetry down to ${\cal N} =1$.

Let us now consider a D7-brane in this background. For simplicity, we will first take the limit of constant warp factor $Z \rightarrow 1$, while nevertheless imposing the condition $P_+^{O3} \ep = 0$ on the background Killing spinor. The worldvolume of such a D7-brane is then of the form $\R^{1,3} \times \cS_4$, where $\cS_4$ is a four-cycle inside $X_6$. Being a dynamical object, our D7-brane will tend to minimize its energy which, since we are assuming $\langle \cF \rangle = 0$ and constant dilaton, amounts to minimizing the volume of $\cS_4$. In the context of Calabi-Yau manifolds there is a well-known class of volume-minimizing objects, known as calibrated submanifolds, that are easily characterized in terms of the globally defined 2 and 3-forms $J$ and $\Omega$ present in any Calabi-Yau. In particular, for a four-cycle $\cS_4$ the calibration condition reads $-\frac12 P[J\wedge J] = \ud\text{vol}_{\cS_4}$, where $P[\cdot]$ again stand for the pull-back into $\cS_4$. Finally, this is equivalent to asking that $\cS_4$ is a complex submanifold of $X_6$, which is the assumption that we will take in the following.\footnote{In fact, a complex four-cycle $\cS_4$ satisfies either $P[J^2]  = 2\ud\text{vol}_{\cS_4}$ or $-P[J^2]  = 2\ud\text{vol}_{\cS_4}$, and both conditions define volume-minimizing objects in a Calabi-Yau. However, given our conventions in the D7-brane action only $P[J^2]  = - 2\ud\text{vol}_{\cS_4}$ will survive as a (generalized) calibration condition when we reintroduce a warp factor satisfying $F_5^{\mathrm{int}} = \hat{*}_{6}\ud Z$. This choice of calibration in warped backgrounds matches the conventions of \cite{mrvv05} and \cite{ms05}, while the opposite choice $P[J^2]  = 2\ud\text{vol}_{\cS_4}$ is taken in \cite{gmm05,jockers05}. Changing from one choice to the other amounts to interchange the definitions of D7-brane vs. anti-D7-brane or, in terms of the fermionic action, redefining $P_\pm^{D7} \leftrightarrow P_\mp^{D7}$. This also explains why, in the next section, we consider a self-dual worldvolume flux $\cF = *_{\cS_4} \cF$ for a BPS D7-brane, instead of the anti-self-dual choice taken in \cite{gmm05}.\label{ft:sign}}

Given this setup, one may analyze which are the bosonic degrees of freedom of our D7-brane and, in particular, which are the massless bosonic modes from a 4D perspective. The answer turns out to be quite simple, and only depends on topological quantities of the four-cycle $\cS_4$. First, from the 8D gauge boson $A_M = (A_\mu, A_a)$ we obtain a 4D gauge boson $A_\mu$ and several 4D scalars $A_a$ whose internal wavefunctions $W_a$ can be used to build up a 1-form $W = W_a \ud\xi^a$ in $\cS_4$. Using that $F^W = \ud W = 0$ by assumption as well as the gauge freedom of $A_a$, one can identify the set of zero modes with the number of independent harmonic 1-forms in $\cS_4$. We then obtain $b_1(\cS_4)$ real scalar fields from dimensionally reducing $A_M$, or in other words $h^{(1,0)}(\cS_4) = b_1(\cS_4)/2$ complex Wilson lines. This result applies in particular to a flat D7-brane in flat space, where we have that $b_1(\bT^4) = 4$.

In addition, 4D scalar zero modes may arise from infinitesimal geometric deformations of the D7-brane internal dimensions $\cS_4 \rightarrow \cS_4'$ inside the Calabi-Yau $X_6$. Such deformations will be zero modes if the volume of the 4-cycle does not change, or otherwise said if $\cS_4'$ is still a complex submanifold. It can be shown that, if we describe such deformation via a vector $\sigma^a$ transverse to $\cS_4$, then $\cS_4'$ is complex only if $\sigma^a\Omega_{abc} \ud\xi^b\wedge \ud\xi^c$ is a harmonic (2,0)-form in ${\cal S}_4$. The number of complex scalar geometric moduli is then given by the number of independent harmonic (2,0)-forms of $\cS_4$, namely the topological number $h^{(2,0)}(\cS_4)$. For a flat D7-brane we have that $h^{(2,0)}(\bT^4) = 1$, and that the complex zero mode is the transverse translations of $\bT^4$ inside $\bT^6$.

Regarding the fermionic zero modes, one should obtain the same degrees of freedom as for bosonic zero modes, so that the 4D effective theory can be supersymmetric. This is because the calibration condition  $- \frac12 P[J\wedge J] = \ud\text{vol}_{\cS_4}$ used above is equivalent to $P_+^{D7} \ep =0$, where $\ep$ is taken as in (\ref{wkilling}) with $Z=1$, and which is the equation that a D7-brane needs to satisfy in order to be a supersymmetric, BPS object in a Calabi-Yau.

Let us describe how these zero modes look like, again taking the unwarped limit $Z \rightarrow 1$. As in subsection \ref{altk}, to remove the spurious degrees of freedom we will take the $\kappa$-fixing choice $P_-^{D7}\Theta = 0$ in (\ref{D7fermE}), which will simplify our discussion below. Then, the zero modes of this action must satisfy $P_-^{D7}\Theta = 0$ and $\slashed{\pa}_{\R^{1,3}} \theta_i = \G^a\nabla_a^\CY \theta_i = 0$, $a\in \cS_4$. An obvious choice for a zero mode would be to take $\Theta = \ep$,\footnote{Strictly speaking, here $\ep$ stands for the restriction of the spinor $\ep$, defined all over $\R^{1,3} \times X_6$ to the 8D slice $\R^{1,3}\times \cS_4$ where the D7-brane is localized. As these worldvolume restrictions for spinors can be understood from the context, we will not indicate them explicitly.} since $\nabla_a^\CY\eta^\CY_\pm =0$. However, the BPS condition $P_+^{D7} \ep =0$ is equivalent to $P_-^{D7}\ep = \ep$, and so this would-be fermionic zero mode is projected out by $\kappa$-fixing. Instead, following \cite{lmmt08} we can consider
\beq
\Theta\, = \, \theta_{4D} \otimes 
\frac{1}{\sqrt{2}}
\left(
\begin{array}{c}
i \eta^\CY_- \\  \eta^\CY_-
\end{array}
\right) - i 
B_4^* \theta_{4D}^* \otimes 
\frac{1}{\sqrt{2}}
\left(
\begin{array}{c}
\eta^\CY_+ \\ i\eta^\CY_+
\end{array}
\right)
\label{gauginoCY}
\eeq
with $\theta_{4D}$ constant and of negative 4D chirality. This bispinor is not only a D7-brane zero mode but also an universal one, since it is present for any BPS D7-brane. As pointed out in \cite{lmmt08}, upon dimensional reduction we can identify such zero mode with the 4D gaugino.

The rest of fermionic zero modes can be constructed from (\ref{gauginoCY}) (see e.g. \cite{jockers05,bhv08}). Indeed, by the basic properties of a Calabi-Yau, the covariantly constant spinor $\eta_-^\CY$ is annihilated by any holomorphic $\G$-matrix defined on $X_6$, namely $\G_{z^i} \eta_-^\CY = \G^{\bar{z}^i} \eta_-^\CY = 0$. Since $\cS_4$ is a complex manifold, the same is also true for the $\Gamma$-matrices living on $\cS_4$. Hence all the spinors that can be created from $\eta_-^\CY$ are of the form
\beq
\eta_W\, =\, W_a \G^{z^a}\eta_-^\CY \quad \quad \text{and} \quad \quad \eta_m\, =\, m_{ab} \G^{z^az^b}\eta_-^\CY
\eeq
with $a,b \in \cS_4$. Finally, one can show that $\G^a\nabla_a^\CY$ annihilates these spinors if and only if $W_a\ud z^a$ and $m_{ab} \ud z^a \wedge \ud z^b$ are harmonic (1,0) and (2,0)-forms in $\cS_4$, respectively.\footnote{Notice that $\G^a\nabla_a^\CY \neq \slashed{\nabla}_{\cS_4}$, since $\nabla_a^\CY$ is constructed from the metric in $X_6$ and not that in $\cS_4$. See \cite{bs06} for their precise relation. In the language of \cite{bhv08}, going from $\slashed{\nabla}_{\cS_4}$ to $\G^a\nabla_a^\CY$ involves introducing a twist in the Dirac operator.} This clearly matches the scalar degrees of freedom obtained above and, in particular, we can identify $\theta_W$ with internal wavefunction for the Wilsonini and $\theta_m$ with that for the modulini of the theory. More precisely, since we need to impose that $P_-^{D7}\Theta = 0$, we have that such fermion zero modes are
\bes
\label{zerofCY}
\begin{align}\nonumber
\Theta\, =  \,  \theta_{4D}& \otimes \Theta_{6D} +  B_4^*\theta_{4D}^* \otimes B_6^* \Theta_{6D}^*\\
\label{wilsoninoCY}
B_6^*\Theta_{6D}^* \, = & \,
\frac{1}{\sqrt{2}}
 \left(
\begin{array}{c}
i\eta_W\\ \eta_W
\end{array}
\right)
\quad \quad \text{for Wilsonini}\\
\label{modulinoCY}
\Theta_{6D} \, = & \, 
\frac{1}{\sqrt{2}}
\left(
\begin{array}{c}
i\eta_m\\ \eta_m
\end{array}
\right)
\quad \quad \text{for modulini + gaugino}
\end{align}
\ees

How do these zero modes change when we introduce back the warp factor? By taking the operators (\ref{CYD7op}), it is easy to see that the D7-brane fermionic action is again of the form (\ref{flatD7fermEalt}), now with
\be
\label{CYD7Dirac}
\slashed{D}^w\, =\, \slashed{\pa}_4^{\text{ext}} + \G^a\nabla_a^\CY + \left({\slashed{\partial}}_4^{\text{int}}  \ln Z\right) \left(\frac{1}{8} - \frac{1}{2}  P_+^{O3}\right)
\ee
Hence, the warped zero modes will again be given by (\ref{gauginoCY}) and (\ref{zerofCY}), but now multiplied with a certain power of the warp factor which depends on how $P_+^{O3}$ acts of them. In particular, it is easy to see that for (\ref{gauginoCY}) and (\ref{modulinoCY}) we have that $P_+^{O3} \Theta = \Theta$, so that the appropriate warp factor is given by $Z^{3/8}$. On the other hand, for (\ref{wilsoninoCY}) we have that $P_+^{O3} \Theta = 0$, and so Wilsonino zero modes need to be multiplied by a warp factor $Z^{-1/8}$. Finally, one can check that if we define $\G_{\text{Extra}} = \ud \slashed{\text{vol}}_{\cS_4}$ as the chirality operator of $\cS_4$ then $\G_{\text{Extra}} \eta_-^\CY= \eta_-^\CY$ and that the same is true for $\eta_m$, while the Wilsonini $\eta_W$ possess the opposite extra-dimensional chirality. Thus, we see that the result (\ref{flatD7zeroalt})  derived for warped flat space remains valid in warped Calabi-Yau compactifications. This will also imply that again both the gaugino and modulini will have a 4D kinetic term of the form (\ref{genkterm}) with $q=1$, while for the Wilsonini $q=0$ and nothing will change with respect to an unwarped compactification.

Considering the bosons, one can also see that the results from warped flat space apply to a warped Calabi-Yau, and so the wavefunctions for the gauge boson, Wilson lines and moduli do not carry the warp factor. Indeed, note that in this way the 4D kinetic terms of bosonic and fermionic superpartners will match, which is again a requirement of supersymmetry. One can also perform an explicit derivation via an explicit dimensional reduction for the D7-brane zero modes, along the lines of \cite{bdkmmm06} for the gauge boson and of \cite{bcdgmqs06} for the moduli.

\subsection{Adding background fluxes}\label{subsec:backgroundfluxes}

Let us now add background fluxes $H_3$, $F_3$ to our warped Calabi-Yau solution, while still considering D7-branes with ${\cal F} = 0$ in their worldvolume. We can do so by following the discussion in \cite{lmmt08}, adapted to our Einstein frame conventions of eq.(\ref{ap:SUSYE}). Indeed, one first imposes the constraint $G_3 = F_3 + i e^{-\Phi}H_3 = - i *_6G_3$, coming form the equations of motion \cite{gkp01}. This implies that the operators  ${\frak G}_3^\pm \equiv \slashed{F}_3\sigma_1\pm e^{-\Phi}\slashed{H}\sigma_3$ defined in (\ref{ap:SUSYE}) can be written as ${\frak G}_3^\pm = 2 e^{-\Phi}\slashed{H}\sigma_3 P_\mp^{O3}$, and so we have that the 10D gravitino and dilatino variations are
\bes
\label{fluxD7op}
\begin{align}
{\cal O}\, =& \, e^{-\Phi_0/2}\slashed{H}_3\sig_3 P_+^{O3} \\
{\cal D}_\mu \, =& \, \pa_\mu - \frac{1}{4} \G_\mu \slashed{\pa} \ln Z P_+^{O3} - \frac18  e^{-\frac{\Phi_0}{2}} \Gamma_\mu \slashed{H}_3\sig_3 P_-^{O3} \\
{\cal D}_m \, =& \, \nabla_m^\CY + \frac{1}{8}\pa_m \ln Z   + \frac{1}{4} \slashed{\pa}\ln Z \G_m  P_+^{O3} +\frac{e^{-\frac{\Phi_0}{2}}}{4} \left(\slashed{H}_3 \Gamma_m P_+^{O3}+ \frac12\Gamma_m \slashed{H}_3 P_-^{O3}\right)\sig_3 
\end{align}
\ees
from which we see that for a bispinor $\ep$ of the form (\ref{wkilling}) we have that ${\cal O} \ep = 0$ and
\be
{\cal D}_\mu \ep \, =\, {\cal D}_m \ep \, =\, 0 \quad \iff \quad  \slashed{H}_3 \sig_3 \ep\, =\, 0
\ee
which, as expected, happens if and only if $H_3$ is a $(2,1)+(1,2)$-form \cite{gp00}. Without imposing this latter condition, we can proceed to analyze the eigenmodes of the D7-brane fermionic action. Using the same conventions as for the warped Calabi-Yau case, we have that the Dirac operator is now given by
\be
\label{fluxD7Dirac}
\slashed{D}^w = \slashed{\pa}_4^{\text{ext}} + \G^a\nabla_a^\CY + \left({\slashed{\partial}}_4^{\text{int}}  \ln Z\right) \left(\frac{1}{8} - \frac{1}{2}  P_+^{O3}\right) + \frac12 e^{-\Phi_0/2} \left(\G^a(\slashed{H}_3)_a   - \slashed{H}_3 \right) P_+^{O3}\sig_3
\ee
and so we find that the new Dirac operator contains a piece which is exactly like the fluxless Dirac operator (\ref{CYD7Dirac}) plus a new piece proportional to the background flux $H_3$. From this piece is where the  flux-induced fermionic masses should arise from, following the microscopic analysis of \cite{Camara:2004jj}. From (\ref{fluxD7Dirac}) we see that in general the Wilsonini do not get any mass term, as already expected from the analysis in \cite{gmm05}. Regarding the gaugino and the modulini, they can get a mass term from $\G^a(\slashed{H}_3)_a   - \slashed{H}_3$, which projects out the components of $H_3$ that have just one index on the D7-brane worldvolume. As a component of $H_3$ with all three indices in $\cS_4$ is incompatible with our initial assumption $\langle{\cal F}\rangle = 0$, we are left with only those components of $H_3$ with two indices on ${\cal S}_4$, which we denote by $H_3^{(2)}$, contribute to fermionic mass terms. The Dirac operator can then be expressed as
\be
\label{fluxD7Diracb}
\slashed{D}^w\, =\, \slashed{\pa}_4^{\text{ext}} + \G^a\nabla_a^\CY + \left({\slashed{\partial}}_4^{\text{int}}  \ln Z\right) \left(\frac{1}{8} - \frac{1}{2}  P_+^{O3}\right) + \frac12 e^{-\Phi_0/2} \slashed{H}_3^{(2)} P_+^{O3}\sig_3
\ee
and so all those zero modes not lifted by the presence of the flux maintain the same warp factor dependence as in the fluxless case. The warp factor dependence of modes lifted by the flux is however more complicated, as the operator $\slashed{H}_3^{(2)}$ also depends on the warp factor. See \cite{bcdgmqs06} for a discussion on these issues in terms of bosonic modes.

\subsection{Extension to F-theory backgrounds}\label{F-theory}

The results above can be further extended to warped F-theory backgrounds, with metric (\ref{10metricE}) and a nonconstant dilaton field $\Phi$. Again, the 10D gravitino and dilatino variations can be deduced from (\ref{ap:SUSYE}). If for simplicity we assume no background 3-form fluxes they read
\bes
\label{FD7op}
\begin{align}
{\cal O}\, =& \, \slashed{\pa}\Phi - e^{\Phi} \slashed{F}_1i \sig_2 \\
{\cal D}_\mu \, =& \, \pa_\mu - \frac{1}{4} \G_\mu \slashed{\pa} \ln Z P_+^{O3}\\
{\cal D}_m \, =& \, \nabla_m^{X_6} + \frac{1}{4}e^{\Phi}(F_1)_m +  \frac{1}{8}\pa_m \ln Z   - \frac{1}{4} \slashed{\pa}\ln Z \G_m  P_+^{O3}
\end{align}
\ees
where  we have also allowed a non-trivial RR flux $F_1 = \re d\tau$, so that (\ref{cpxdil}) can be satisfied. Translating the discussion in \cite{gp01} to our formalism, one can look for Killing bispinors $\ep$ satisfying ${\cal D}_\mu\ep = {\cal D}_m\ep = 0$, again using the ansatz (\ref{ansatzCY}). We obtain a warped bispinor of the form
\be
\ep\, = \, \ep_{4D} \otimes 
Z^{-1/8} \left(
\begin{array}{c}
\eta^{X_6}_- \\ i \eta^{X_6}_-
\end{array}
\right) - i 
B_4^* \ep_{4D}^* \otimes 
Z^{-1/8} \left(
\begin{array}{c}
i \eta^{X_6}_+ \\ \eta^{X_6}_+
\end{array}
\right)
\label{wkillingF}
\ee
where again $\eta_-^{X_6}$ is a negative chirality 6D spinor, now
satisfying\footnote{This is the weak coupling and small
$C_{0}$ limit (that is, linearized) version of eq. (2.19) in
\cite{gp01}.}
\beq
\left(\nabla^{X_6}_m + \frac{1}{4}e^{\Phi}(F_1)_m\right)\, \eta^{X_6}_-\, =\, 0
\label{Fccspinor}
\eeq
instead of (\ref{ccspinor}). The fact that  $\eta^{X_6}_\pm$ are no longer covariantly constant implies that the holonomy group of $X_6$ cannot be in SU(3), and so $X_6$ cannot be a Calabi-Yau. However, from (\ref{Fccspinor}) one can see that the holonomy group is contained in U(3), which implies that $X_6$ is a complex, K\"ahler manifold. Hence, we can still introduce complex coordinates $z^i$ and holomorphic $\G$-matrices such that, as before, $\G_{z^i} \eta_-^{X_6} = \G^{\bar{z}^i} \eta_-^{X_6} = 0$. One can then check that the last supersymmetry condition ${\cal O}\ep = 0$ is equivalent to (\ref{cpxdil}).

As before, the BPS condition for a D7-brane $P_+^{D7} \ep = 0$ will restrict $\cS_4$ to be a complex submanifold of $X_6$ and, since $X_6$ is K\"ahler, this will mean that $\cS_4$ is minimizing its volume.\footnote{Notice that for a varying axio-dilaton $\tau$ the physically relevant question is whether the D7-brane is minimizing its energy, and more precisely its DBI + CS Lagrangian densities, rather than its volume. 
Of course, energy minimization turns also to be true for such D7-branes, as expected from their BPSness.} Taking the $\kappa$-fixing choice $P_-^{D7}\Theta = 0$ and the unwarped limit $Z \raw 1$, we will have again a D7-brane fermionic action of the form (\ref{flatD7fermEalt}), where now
\be
\label{FD7Dirac}
\slashed{D}^w\, =\, \slashed{\pa}_4^{\text{ext}} + \G^a\left(\nabla_a^{X_6} + \frac{1}{4} e^\Phi (F_1)_a\right)  - \frac{i}{2} e^\Phi\left( \slashed{F}_1\sig_2 - i \slashed{\pa} e^{-\Phi} \right)
\ee
Because of the holomorphicity of the dilaton, the zero modes of this Dirac operator will as before be of the form (\ref{gauginoCY}) and (\ref{zerofCY}), with the obvious replacement $\eta^\CY_- \raw \eta_-^{X_6}$.  While (\ref{gauginoCY}) will be a universal zero mode that corresponds to the D7-brane gaugino, the Wilsonino  and modulino zero modes will have to solve a differential equation, that will again relate them to the harmonic (1,0) and (2,0)-forms of $\cS_4$, respectively.\footnote{See \cite{bhv08} for a derivation of this spectrum using twisted Yang-Mills theory.}

Finally, we can restore the warp factor dependence on the D7-brane fermionic action, which amounts to add to (\ref{FD7Dirac}) a piece of the form
\be
\left({\slashed{\partial}}_4^{\text{int}}  \ln Z\right) \left(\frac{1}{8} - \frac{1}{2}  P_+^{O3}\right)
\ee
exactly like in warped flat and Calabi-Yau spaces. As a result, we will again have that the D7-brane gaugino and modulini depend on the warp factor as $Z^{3/8}$, while the Wilsonini do as $Z^{-1/8}$. The generalization to F-theory backgrounds with fluxes is then straightforward. 

\subsection{Effects on the K\"ahler potential}\label{kahler}

Just like for closed strings, one can interpret the effect of warping in the open string wavefunctions as a modification of the 4D K\"ahler potential and gauge kinetic functions. In order to properly interpret the effect of warping, we must convert our results to the 4D Einstein frame, which differs from the 10D Einstein frame by a Weyl transformation of the unwarped 4D metric
\begin{equation}
  \eta_{\mu\nu}\to\frac{\mathcal{V}^{0}}
  {\mathcal{V}_{\w}}\eta_{\mu\nu}
\end{equation}
where $\mathcal{V}_{\w}$ is the warped volume of the internal 6D space
\begin{equation}
  \mathcal{V}_{\w}=\int_{X_{6}}\ud\hat{\vol}_{X_{6}}Z
  \label{wvol}
\end{equation}
and $\mathcal{V}^{0}$ is the fiducial volume of the unwarped Calabi-Yau. This Weyl transformation gives a canonical 4D Einstein-Hilbert action with 4D gravitational constant
\begin{equation}
  \frac{1}{2\kappa_{4}^{2}}=\frac{\mathcal{V}^{0}}
  {2\kappa_{10}^{2}}
\end{equation}

Let us now analyze the different open string metrics. The D7-brane gauge kinetic function for the gauge boson was deduced for the toroidal case in (\ref{gkfD7}). From the results of Sec \ref{warpcy}, one can easily generalize this result to a D7-brane wrapping a 4-cycle $\cS_4$ in a warped Calabi-Yau as
\begin{equation}
  \label{eq:gkfd7cy}
  f_{\uD 7}=\left(8\pi^{3}k^{2}\right)^{-1}\int_{\mathcal{S}_{4}}
  \frac{\ud\hat{\vol}_{\mathcal{S}_{4}}}
  {\sqrt{\hat{g}_{\mathcal{S}_{4}}}}
  \bigl(Z\sqrt{\hat{g}_{\mathcal{S}_{4}}}+\ui C_{4}^{\mathrm{int}}
  \bigr)%\bigl(\alpha^{0}\bigr)^{2}
\end{equation}
where $\hat{g}_{\mathcal{S}_{4}}$ is the unwarped induced metric on $\mathcal{S}_{4}$, and $\ud\hat{\text{vol}}_{\cS_4}$ the corresponding volume element. Since the gauge kinetic function is Weyl invariant, this is not modified when moving to the 4D Einstein frame.  

The position moduli and modulini combine to form $\mathcal{N}=1$ chiral supermultiplets, the K\"ahler metric for which can be read from the kinetic term of the moduli, after converting it to the 4D Einstein frame.\footnote{The same philosophy has been applied in \cite{dlmp08} to compute (unwarped) open string K\"ahler metrics in the 10D SYM limit of type I theory, using the framework developed in \cite{cim04}.} Let us first consider the case where the D7 is wrapping $\mathbf{T}^{4}=\left(\mathbf{T}^{2}\right)_{i}\times\left(\mathbf{T}^{2}\right)_{j} \subset \bT^6$, where each torus has a complex structure defined by the holomorphic coordinate  
\begin{equation}
  z^{m}= y^{m+3}+\tau_{m} y^{m+6}
\end{equation}
Then, from (\ref{ktermscal}), the kinetic term in the 4D Einstein frame for the zero mode (dropping the KK index 0 on the 4D fields) in the warped toroidal
case is
\begin{equation}
  S_{\uD 7}^{\left.\mathrm{scal}\right.}
  =-\frac{k^{2}}{\kappa_{4}^{2}\mathcal{V}_{\w}}
  \int_{\mathbb{R}^{1,3}}
  \ud^{4}x\, \eta^{\mu\nu}\partial_{\mu}\zeta\partial_{\nu}\,\zeta^{\ast}
  \int_{\mathbf{T}^{4}}\ud\hat{\vol}_{\mathbf{T}^{4}}\, \ue^{\Phi_{0}}
  Zs_{0}s^{\ast}_{0}
  \bigl(\hat{g}_{\mathbf{T}^{4}}\bigr)_{k\bar{k}}
\end{equation}
where we have defined the complex field $\sigma=\left(\sigma_{3+k}+\tau_{k}\sigma_{6+k}\right)$ for $i\neq k \neq j$ and extracted the zero modes from the expansion (\ref{kkmod}). The K\"ahler metric is then
\begin{equation}
  \kappa_{4}^{2}\mathcal{K}_{\zeta\bar{\zeta}}=\frac{k^{2}}{\mathcal{V}_{\w}}
  \int_{\mathbf{T}^{4}}\ud\hat{\vol}_{\mathbf{T}^{4}}\, \ue^{\Phi_{0}}
  Zs_{0}s_{0}^{\ast}\left(\hat{g}_{\mathbf{T}^{4}}\right)_{k\bar{k}}
  \label{kinmod}
\end{equation}

If we now consider a D7-brane wrapping a 4-cycle $\cS_4$ in an unwarped Calabi-Yau, the D7-brane moduli can be expanded in a basis $\left\{s_{A}\right\}$ of complex deformations of $\cS_4$
\begin{equation}
  \sigma\bigl(x,y\bigr)=\zeta^{A}\left(x\right)s_{A}\left(y\right)
  + \bar{\zeta}^{\bar{A}}\bar{s}_{\bar{A}}\left(y\right)
\end{equation}
Following \cite{jl04}, the Einstein frame kinetic term can then be written as
\begin{equation}
  \ui\tau_{\uD 7}\int_{\mathbb{R}^{1,3}}\ue^{\Phi}\mathcal{L}_{A\bar{B}}\,
  \ud\zeta^{A}\wedge\ast_{4}\ud\bar{\zeta}^{\bar{B}}
  \label{ktermmod}
\end{equation}
where
\begin{equation}
  \mathcal{L}_{A\bar{B}}=
  \frac{\int_{\mathcal{S}_{4}}\, m_{A}\wedge
    m_{\bar{B}}}{\int_{X_{6}}\, \Omega^{\CY}\wedge
    \bar{\Omega}^{\CY}}
\end{equation}
and $\left\{m_{A}\right\}$ is a basis of harmonic $\left(2,0\right)$-forms related to $\left\{s_{A}\right\}$ via $m_A = \iota_{s_A} \Omega^\CY$. As we have seen, in the toroidal case the effect of warping introduces a warp factor in the integral over the internal wavefunctions and requires a Weyl rescaling with the warped volume rather than the unwarped one. The appropriate generalization for the warped Calabi-Yau case amounts then to
\begin{equation}
  \mathcal{L}_{A\bar{B}}\to\mathcal{L}^{\mathrm{w}}_{A\bar{B}}
  =\frac{\int_{\mathcal{S}_{4}} Z\, m_{A}\wedge
    m_{\bar{B}}}{\int_{X_{6}} Z\, \Omega^{\mathrm{CY}}\wedge
    \bar{\Omega}^{\mathrm{CY}}}
\end{equation}

Let us now try to combine these open string K\"ahler metrics with the kinetic terms in the closed string sector, studied in \cite{stud,Douglas:2008jx, Frey:2008xw}.  For the axio-dilaton, the result from \cite{stud} is
\begin{equation}
  -\int_{\mathbb{R}^{1,3}}\ud^{4}x\, \mathcal{K}_{\bar{t}t}\,\partial^{\mu}\bar{t}\,
  \partial_{\mu}t
\end{equation}
where $t$ is the axio-dilaton zero-mode, and the K\"ahler metric is given by
\begin{equation}
  \mathcal{K}_{\bar{t}t}
  =\frac{1}{8\left(\mathrm{Im}\tau\right)^{2}\mathcal{V}_{\mathrm{w}}}
  \int_{X_{6}}\ud^{6}y\, Z\, Y_{0}^{2}
\end{equation}
where $Y_{0}$ is the internal wavefunction for the zero mode. Since the equation of motion admits a constant zero mode, the integral is proportional to the warped volume which is canceled by the factor of
$\mathcal{V}_{\mathrm{w}}$ appearing in the denominator. That is, the kinetic term for the zero mode of the axio-dilaton is unaffected by the presence of warping. In the presence of D7 branes, the D7 geometric moduli and the axio-dilaton combine into a single K\"ahler coordinate $S$. In the unwarped Calabi-Yau  this combination is given by \cite{jl04}
\begin{equation}
  S = t-\kappa_{4}^{2}\tau_{\uD 7}\mathcal{L}_{A\bar{B}}
  \zeta^{A}\bar{\zeta}^{\bar{B}}
\end{equation}
and so the appropriate part of the K\"ahler potential is
\begin{equation}
  \mathcal{K}\ni \ln\bigl[-\ui\bigl(S-\bar{S}\bigr)
  -2\ui\kappa_{4}^{2}\tau_{\uD 7}\mathcal{L}_{A\bar{B}}\zeta^{A}
  \bar{\zeta}^{\bar{B}}\bigr]
\end{equation}
The kinetic term for $t$ is not modified by warping, which suggests that in
the presence of warping we should identify
\begin{equation}
  S^{\mathrm{w}} = t-\kappa_{4}^{2}\tau_{\uD 7}\mathcal{L}^{\mathrm{w}}_{A\bar{B}}
  \zeta^{A}\bar{\zeta}^{\bar{B}}
\end{equation}
and that the K\"ahler potential should be modified accordingly,
\begin{equation}
  \mathcal{K}\ni \ln\bigl[-\ui\bigl(S^{\mathrm{w}}-\bar{S}^{\mathrm{w}}\bigr)
  -2\ui\kappa_{4}^{2}\tau_{\uD 7}\mathcal{L}^{\mathrm{w}}_{A\bar{B}}\zeta^{A}
  \bar{\zeta}^{\bar{B}}\bigr]
  \label{Kal1}
\end{equation}
This correctly reproduces the quadratic-order kinetic terms for the
axio-dilaton and D7 deformation moduli.

Turning now to the Wilson line and Wilsonini,  their K\"ahler metric can be found from the Wilson line action. In the $\cS_4 =  \bT^2_i \times \bT_j^2$ case,  the components of the 1-form potential $A$ in complex coordinates are
\begin{equation}
  \label{eq:complexwl}
  A_{a}=\frac{\ui}{2\,\mathrm{Im}\left({\tau_{a}}\right)}
  \bigl(\tau_{a}^{\ast}A_{a+3}-A_{a+6}\bigr)
\end{equation}
for $a =i,j$. Converting (\ref{ktermwl}) to the Einstein frame, we find that the action for the massless modes is
\begin{equation}
  S_{\uD 7}^{\left.\mathrm{wl}\right.}
  =-\frac{k^{2}}{\kappa_{4}^{2}\mathcal{V}_{\w}}\int_{\mathbb{R}^{1,3}}
  \ud^{4}x\, \hat{g}_{\mathbf{T}^{4}}^{a\bar{b}}
  \eta^{\mu\nu}\partial_{\mu}w_{a}\partial_{\nu}w_{\bar{b}}^{\ast}
  \int_{\mathbf{T}^{4}}\ud\hat{\vol}_{\mathbf{T}^{4}}
  W_{a}^{\left(0\right)}W_{\bar{b}}^{\ast\left(0\right)}
\end{equation}
which finally gives the K\"ahler metric
\begin{equation}
  \label{eq:flatwilsonkahler}
  \kappa_{4}^{2}\mathcal{K}_{a\bar{b}}
  =\frac{k^{2}}{\mathcal{V}_{\w}}\int_{\mathbf{T}^{4}}
  \ud\hat{\vol}_{\mathbf{T}^{4}}W^{\left(0\right)}_{a}\, W^{\ast\left(0\right)}_{\bar{b}}
  \hat{g}^{a\bar{b}}_{\mathbf{T}^{4}}
\end{equation}
where the indices $a$ and $b$ are not summed over.

In the Calabi-Yau case, the Wilson lines of a D7 wrapping  $\mathcal{S}_{4}$ can be expanded as
\begin{equation}
  A_{a}\ud A^{a}=w_{I}\left(x\right)W^{I}\left(y\right) +
  \overline{w}_{\bar{I}}\left(x\right)\overline{W}^{\bar{I}}\left(y\right)
\end{equation}
where $\left\{W^{I}\right\}$ is a basis of harmonic $\left(1,0\right)$-forms on $\cS_4$. The kinetic term for the Wilson lines in the unwarped case is \cite{jl04}
\begin{equation}
  \ui\frac{2\tau_{\uD 7}k^{2}}{\mathcal{V}}\int_{\mathbb{R}^{1,3}}
  \mathcal{C}_{\alpha}^{I\bar{J}}
  v^{\alpha}\ud w_{I}\wedge\ast_{4}\ud\overline{w}_{\bar{J}}
\end{equation}
where ${\cal V}$ is the (unwarped) Calabi-Yau volume. If we now expand the K\"ahler form in a basis $\left\{\omega_{\alpha}\right\}$ of harmonic 2-forms
\begin{equation}
  \label{eq:expandedkahler}
  J^{\CY} = v^{\alpha}\omega_{\alpha}
\end{equation}
we can express $\mathcal{C}^{I\bar{J}}_\al$ as
\begin{equation}
  \mathcal{C}_{\alpha}^{I\bar{J}}=\int_{\mathcal{S}_{4}}
  P\left[\omega_{\alpha}\right]\wedge W^{I}\wedge
  \overline{W}^{\bar{J}}
\end{equation}
In the warped toroidal case, the effect of the warping on the Wilson line kinetic terms is to simply replace the volume with the warped volume.  Again, from Sec \ref{warpcy}, this result is independent of the shape of unwarped internal geometry so that in the warped Calabi-Yau case, the kinetic term for the Wilson lines is
\begin{equation}
  \label{eq:warpedcykterm}
  \ui\frac{2\tau_{\uD 7}k^{2}}{\mathcal{V_{\mathrm{w}}}}\int_{\mathbb{R}^{1,3}}
  \mathcal{C}_{\alpha}^{I\bar{J}}
  v^{\alpha}\ud w_{I}\wedge\ast_{4}\ud\overline{w}_{\bar{J}}
\end{equation}
where now  the warped volume $\mathcal{V}_{\mathrm{w}}$ appears in the denominator.

One may again wonder how these open string modes combine with the closed string ones in the full K\"ahler potential. In analogy with the results for the unwarped Calabi-Yau case, we would now expect that Wilson lines combine with the K\"ahler moduli. However, as pointed out in \cite{jl04} it is not an easy problem to derive the K\"ahler metrics from the general form of the K\"ahler potential. Let us instead consider the particular case of $X_6 = \bT^6$, $\cS_4 = \left(\bT^2\right)_i \times \left(\bT^2\right)_j$. In the unwarped case, the K\"ahler potential can be written as
\bea
\label{KuwT6}
  \mathcal{K}\ni
  -\ln\bigl[T_{\Lambda}+\overline{T}_{\Lambda}\bigr]
 & -& \ln\bigl[T_{i}+\overline{T}_{i}-6\ui\kappa_{4}^{2}\tau_{\uD 7}k^{2}
  \mathcal{C}_{i}^{I\bar{J}}
  w_{I}\overline{w}_{\bar{J}}\bigr]\\ \nonumber
  & - & \ln\bigl[T_{j}+\overline{T}_{j}-6\ui\kappa_{4}^{2}\tau_{\uD 7}k^{2}
  \mathcal{C}_{j}^{I\bar{J}}
  w_{I}\overline{w}_{\bar{J}}\bigr]
\eea
where $T_{\alpha}$ are a combination of K\"ahler moduli and D7's Wilson lines. Indeed,
\be
T_\al + \overline{T}_\al\, =\, \frac{3}{2} \cK_\al + 6\ui\kappa_{4}^{2}\tau_{\uD 7}k^{2} \mathcal{C}_{\al}^{I\bar{J}} w_{I}\overline{w}_{\bar{J}}
\ee
where $\cK_\al$ control the the volume of the 4-cycles of the compactification. More precisely, if we express an unwarped Calabi-Yau volume in terms of the $v^\al$ defined in (\ref{eq:expandedkahler}),
\begin{equation}
  \mathcal{V}=\frac{1}{6}\mathcal{I}_{\alpha\beta\gamma} v^{\alpha}v^{\beta}v^{\gamma}
\end{equation}
then we have that, in general,
\be
\cK_\al\, =\, \mathcal{I}_{\alpha\beta\gamma} v^{\beta}v^{\gamma}
\ee
and in particular this expression applies for the K\"ahler moduli of $\bT^6$.

Expanding (\ref{KuwT6}) up to second order in the D7-brane Wilson lines $w^I$ we obtain that their unwarped K\"ahler metrics are given by
\be
\kappa_{4}^{2}\tau_{\uD 7}k^{2}
\sum_\al {3\ui
  \mathcal{C}_{\al}^{I\bar{J}} \over T_{\al}+\overline{T}_{\al}}\ w_Iw_{\bar{J}}
\ee
Comparing to our result (\ref{eq:warpedcykterm}), it is easy to see that a simple generalization that would reproduce the Wilson line warped metric is to replace
\be
T_\al + \overline{T}_\al\ \raw \  T^{\mathrm{w}}_{\alpha} + \overline{T}^{\mathrm{w}}_{\alpha}\, =\,
\frac{3}{2}  \mathcal{I}^{\mathrm{w}}_{\al\beta\gamma}v^{\beta}v^{\gamma} + 6\ui\kappa_{4}^{2}\tau_{\uD 7}k^{2} \mathcal{C}_{\al}^{I\bar{J}} w_{I}\overline{w}_{\bar{J}}
\label{rep}
\ee
in (\ref{KuwT6}). Here we have defined the warped intersection product\footnote{An alternative possibility would have been to set $\mathcal{I}^\text{w}_{\alpha\beta\gamma} = (\mathcal{V}_\text{w}/\mathcal{V}) \mathcal{I}_{\alpha\beta\gamma}$, although this would imply a very mild modification of the K\"ahler potential with respect to the unwarped case.}
\begin{equation}
  \mathcal{I}^{\mathrm{w}}_{\alpha\beta\gamma}
  =\int_{X_{6}}Z\, \omega_{\alpha}\wedge\omega_{\beta}\wedge\omega_{\gamma}
\end{equation}
that defines the warped volume as
\begin{equation}
  \mathcal{V}_{\mathrm{w}}=\frac{1}{6}\mathcal{I}^{\mathrm{w}}_{\alpha\beta\gamma}
  v^{\alpha}v^{\beta}v^{\gamma}
\end{equation}

One may then wonder whether this way of writing the warped K\"ahler potential is a particular feature of toroidal-like compactifications. A possible caveat is that the modification (\ref{rep}) is clearly different from  the modification of the gauge kinetic function (\ref{eq:gkfd7cy}) and that both quantities, $T^{\rm{w}}_\alpha$ and $f_{\text{D7}}$, should have a simple dependence on the K\"ahler moduli of the compactification.\footnote{Let us stress out that we are not identifying $T^{\rm{w}}_\alpha$ with the K\"ahler moduli of a warped compactification, but rather with the quantities that encode their appearance in the K\"ahler potential.} Indeed, the warp factor of the gauge kinetic function is integrated only over $\mathcal{S}_{4}$, while the warp factor in the definition of $T^{\mathrm{w}}_\alpha$ is integrated over the entire internal space. In fact, both definitions of warped volume can be put in the same form
\be
\text{Vol}^\text{w}_\xi(\cS_4)\, =\, - \frac12 \int_{X_6} \xi \wedge J \wedge J
\label{defw4cycle}
\ee
where $[\xi]$ is Poincar\'e dual to $[\cS_4]$, and $J\, =\, Z^{1/2} J^\CY$ is the warped K\"ahler form. Because $J^2$ is not closed, (\ref{defw4cycle}) depends on the representative $\xi \in [\xi]$. In particular, for  $T^\text{w}_\alpha$ $\xi$ is the harmonic representative, while for $f_{\text{D7}}$ $\xi$ should have $\delta$-function support on $\cS_4$.

Despite this discrepancy there is not necessarily a contradiction between (\ref{eq:gkfd7cy}) and our definition of $T^\text{w}_\alpha$. For instance, if one takes the definition of K\"ahler moduli given in \cite{Martucci09}, that in the present context translates into the shift $J \wedge J\raw J \wedge J + t^\alpha [\omega_\alpha]$, $[\omega_\alpha] \in H^{2,2}(X_6)$, we see that $T^\text{w}_\alpha$ and $f_{\text{D7}}$ have exactly the same dependence on $t^\alpha$, which suggest that they could differ by a holomorphic function of the compactification moduli. Indeed, for the case of a single K\"ahler modulus the results in \cite{bdkmmm06} (see also \cite{bhk}) show that one can express the warped volume of $\cS_4$ as
\be
{\cal V}_{\cS_4}^\text{w}\, =\, \int_{\cS_4} Z\,\ud \text{vol}_{\cS_4} \, =\, T^{\mathrm{w}}_{\alpha} + \overline{T}^{\mathrm{w}}_{\alpha} + [\varphi + \overline{\varphi}]
\label{holomf}
\ee
where $\varphi$ is a holomorphic function of D-brane position moduli. Hence, the real part of $\varphi$ is  precisely the difference between both choices of $\xi$ in (\ref{defw4cycle}). It would be interesting to try to extend (\ref{holomf}) to compactifications with several K\"ahler moduli.

In fact, compactifications with one K\"ahler modulus provide a further test to the above definition of warped K\"ahler potential. There, the unwarped K\"ahler potential reads \cite{jl04}
\begin{equation}
 -3\ln\bigl[T_{\Lambda}+\overline{T}_{\Lambda}
  -6\ui\kappa_{4}\tau_{\uD 7}k^{2}\mathcal{C}_{\Lambda}^{I\bar{J}}
  w_{I}\overline{w}_{\bar{J}}\bigr]
\end{equation}
where the single four-cycle $S_{\Lambda}$ is wrapped by the D7 brane.  According to our prescription (\ref{rep}), in the warped case this should be modified to
\begin{equation}
   -3\ln\bigl[T_{\Lambda}^{\mathrm{w}}+
  \overline{T}^{\mathrm{w}}_{\Lambda}
  -6\ui\kappa_{4}\tau_{\uD 7}k^{2}\mathcal{C}_{\Lambda}^{I\bar{J}}
  w_{I}\overline{w}_{\bar{J}}\bigr]
\end{equation}
and, in the absence of a D7 brane where $w_{I}=0$, this becomes
\begin{equation}
  -3\ln\bigl[T_{\Lambda}^{\mathrm{w}}+
  \overline{T}^{\mathrm{w}}_{\Lambda}\bigr]
\end{equation}
Note that this reproduces is the results of \cite{Frey:2008xw}.  Indeed, from our definition of
$T^{\text{w}}_{\alpha}$ we have that, in the absence of D7-branes,
\begin{equation}
  t_{\Lambda}^{\mathrm{w}}=\frac{3}{4}
  \mathcal{I}^{\left.\mathrm{w}\right.}_{\Lambda\Lambda\Lambda}
  \bigl(v^{\Lambda}\bigr)^{2}
\end{equation}
where $t_{\Lambda}^{\mathrm{w}}$ is the real part of $T_{\Lambda}^{\mathrm{w}}$. This real part  of the universal K\"ahler modulus can be identified as an $\R^{1,3}$-dependent shift $c$ in the warp factor \cite{Giddings:2005ff,stud,Frey:2008xw}\footnote{As explained in
\cite{Giddings:2005ff,stud,Frey:2008xw}, compensators are need to be added for consistency with the equations of motion for the closed string fluctuations. 
These are however unimportant for the discussion here since to quadratic order in fluctuations, the open string kinetic terms depend only on the background values of the closed string moduli.} 
\begin{equation}
  Z\left(x,y\right)=Z_{0}\bigl(y\bigr)+c\bigl(x\bigr)
\end{equation}
 Integrating this equation over $X_6$ gives an expression for the fluctuating warped volume\begin{equation}
  \label{eq:flucwarpvol}
  \mathcal{V}_{\mathrm{w}}\left(x\right)
  =\mathcal{V}_{\mathrm{w}}^{0}+c\bigl(x\bigr)\mathcal{V}
\end{equation}
As shown in \cite{Frey:2008xw}, the universal K\"ahler modulus is orthogonal
to the other metric fluctuations so we can freeze the value of
$\mathcal{V}$ to the fiducial value $\mathcal{V}^{0}$.
With this identification,
\begin{equation}
  \mathcal{I}^{\left.\mathrm{w}\right.}_{\alpha\beta\gamma}
  = \mathcal{I}^{\left.\mathrm{w_{0}}\right.}_{\alpha\beta\gamma}+c\, 
  \mathcal{I}_{\alpha\beta\gamma}
\end{equation}
where
\begin{equation}
  \mathcal{I}^{\left.\mathrm{w_{0}}\right.}_{\alpha\beta\gamma}
  =\int_{X_{6}}Z_{0}\,\omega_{\alpha}\wedge\omega_{\beta}\wedge\omega_{\gamma}
\end{equation}
While in general the warp factor may provide significant corrections to $\mathcal{I}_{\alpha\beta\gamma}$, %\footnote{For instance a vanishing $\mathcal{I}_{\alpha\beta\gamma}$ does not guarantee that $\mathcal{I}^{\mathrm{w}_{0}}_{\alpha\beta\gamma}$ vanishes.}  
 in the case of a single K\"ahler modulus $\Lambda$ the correction is simply a rescaling with the warped volume
\begin{equation}
  \mathcal{I}^{\left.\mathrm{w_{0}}\right.}_{\Lambda\Lambda\Lambda}
  =\mathcal{I}_{\Lambda\Lambda\Lambda}\frac{\mathcal{V}_{\mathrm{w}}^{0}}
  {\mathcal{V}^{0}}
\end{equation}
where $\mathcal{V}^{0}$ is again the fiducial volume of the unwarped Calabi-Yau.
This allows us to write
\begin{equation}
  t_{\Lambda}^{\mathrm{w}}=\bigl(c+\frac{\mathcal{V}_{\mathrm{w}}^{0}}{\alpha'^{3}}
  \bigr)\frac{3}{4}\mathcal{I}_{\Lambda\Lambda\Lambda}\bigl(v^{\Lambda}\bigr)^{2}
\end{equation}
so that the warping correction to the single K\"ahler modulus is an additive
shift proportional to
\begin{equation}
  \frac{\mathcal{V}_{\mathrm{w}}^{0}}{\mathcal{V}^{0}}
\end{equation}
And so, up to a multiplicative constant, we recover the result of~\cite{Frey:2008xw}, where all warping corrections to the K\"ahler potential for the universal K\"ahler modulus were summarized in an additive shift for the latter. We find it quite amusing that, at least in the case of a single K\"ahler modulus, such result can be reproduced by means of  a DBI analysis. It would be interesting to see if the same philosophy can be applied to compactifications with several K\"ahler moduli, as well as to K\"ahler potentials that involve K\"ahler moduli beyond the universal one. 

\subsection{A simple warped model}\label{model}

Let us now apply the above results to a model based on D7-branes which, besides a non-trivial warp factor, allows for semi-realistic features like 4D chiral fermions and Yukawa couplings. This will not only allow us to show the effects that warping can have on the 4D effective theory, but also to check that our results for the K\"ahler potential are compatible with the computation of physical quantities like Yukawa coupling. A simple way of constructing such model is to consider unmagnetized D7-branes in toroidal orbifolds. That is, we consider an internal manifold of the form $X_6 = \bT^6/\Gamma$, where $\Gamma$ is a discrete symmetry group of $\bT^6$, and place a stack of $N$ D7-branes wrapping a $\bT^4$ in the covering space. For trivial warp factor the phenomenological features of such models have been analyzed in \cite{imr98}. We would now like to see how 4D quantities change after introducing a warp factor. 

Let us then illustrate the warping effects by focusing in a particular toroidal model, namely the Pati-Salam $\mathbb{Z}_{4}$ toroidal orbifold model considered in \cite{Camara:2004jj}, Sec 9.1.  In
this model, the internal space is locally $X_{6}=\mathbf{T}^{6}/\mathbb{Z}_{4}$ where the $\mathbb{Z}_{4}$ action is
\begin{equation}
  \theta:\bigl(z_{1},z_{2},z_{3}\bigr)\mapsto
  \bigl(\ue^{2\pi\ui/4}z_{1},\ue^{2\pi\ui/4}z_{2},\ue^{\pi\ui}z_{3}\bigr)
\end{equation}
and the $\mathbf{T}^{6}$ has been factorized into three $\mathbf{T}_{i}^{2}$.  The gauge group and matter arise from a stack of eight D7-branes wrapping $(\bT^2)_1 \times (\bT^2)_2$ and located at an orbifold fixed point on the third torus. The orbifold action on the gauge degrees of freedom break the initial gauge group $U(8) \raw U(4) \times U(2)_L \times U(2)_R$, producing at the same time two quark/lepton generations $F_L^i = (4, \bar{2}, 1)$, $F_R^j = (\bar{4}, 1,2)$ $i,j =1,2$, a Higgs multiplet $H = (1,2, \bar{2})$, and Yukawa couplings $\ep_{ij} HF_L^iF_R^j$. The latter can be understood as arising from orbifolding and dimensionally reducing of the 8D SYM term
\beq
\int d^8 \xi \sqrt{g}\, \bar{\theta} \G^\al [A_\al, \theta]
\label{8DSYM}
\eeq
present in the initial U(8) D7-brane theory.\footnote{In fact, not all Yukawa couplings can be understood like this. In unwarped backgrounds without fluxes, a way to guess the missing Yukawas is to start from a 10D SYM action and reduce it to 8D in order to produce couplings beyond (\ref{8DSYM}), as in \cite{Conlon:2008qi}. We will however not discuss such approach, as (\ref{8DSYM}) will be enough for the purposes of this subsection.} 

When introducing the warp factor $Z$, the open string wavefunctions of this model will no longer be constant but develop a warp factor dependence following the analysis of Sec \ref{flat}. In particular, $F_{L,R}$ arise from (orbifolded) U(8) Wilson line multiplets, whereas $H$ arises from the transverse modulus + modulino. By Table \ref{table:warppowers}, we have that the warp factor dependence of their internal wavefunctions is given by
\begin{equation}
  H = (h, \psi_H)_{\text{4D}} \raw (Z^{0}, Z^{3/8}), \qquad F = (f, \psi_F)_{\text{4D}} \raw (Z^0,Z^{-1/8}).
\end{equation}
These wavefunctions must be inserted in the D7-brane fermionic action, where an analogous term to (\ref{8DSYM}) gives
\begin{equation}
  S^{\mathrm{Yuk}}_{\mathrm{\uD 7}}=\tau_{\uD 7}
  \int\ud^{8}\xi\sqrt{g}e^{\Phi_{0}} \ep_{ij} \bigl(\bar{\theta}_H  \G^{\bar{1}} A_{F_{L}}^i \theta_{F_R}^j + \bar{\theta}_H  \G^{\bar{2}} A_{F_R}^i \theta_{F_L}^j  + \text{h.c.} \bigr)
\end{equation}
and where both $\G$-matrices contain a factor of $Z^{-1/4}$. It is then easy to see that the full warp factor dependence cancels in the integral, performed upon dimensional reduction, and that one is left with an 4D effective action of the form
\begin{equation}
  S^{\mathrm{Yuk}}_{\uD 7}= \tau_{\uD 7} \frac{\alpha'^{6}}{\mathcal{V}_{\w}^{2}}  \ue^{\Phi_{0}} (\hat{g}^{1\bar{1}}_{\bT^4})^{1/2} \int_{\mathbb{R}^{1,3}}\ud^{4}x\,  f_L^{i}\bar{\psi}_H \psi^j_{F_R}  \ep_{ij} \int_{\mathbf{T}^{4}}\ud\hat{\vol}_{\mathbf{T}^{4}} \, W_{F_L}\eta_H^{\dagger}\eta_{F_R} \ + \ \dots
\end{equation}
where $s$ and $\eta$ are constant bosonic and fermionic internal wavefunctions, respectively, and where we have converted all quantities to the 4D Einstein frame. From Sec \ref{flat} we know that the normalization constants of such wavefunctions are
\bea
N_{\eta_H} & = & \left(e^{\Phi_0}\alpha'^{9/2}{\cal V}^{-3/2}_\w \int_{\bT^4} \ud\hat{\vol}_{\mathbf{T}^{4}} Z \right)^{-1/2}\\
N_{\eta_{F_R}} & = & \left(e^{\Phi_0}\alpha'^{9/2}{\cal V}^{-3/2}_\w \int_{\bT^4} \ud\hat{\vol}_{\mathbf{T}^{4}}  \right)^{-1/2}\\
N_{W_{F_L}} & = & \left(k^{2}\alpha'^{3}{\cal V}^{-1}_\w \hat{g}^{1\bar{1}}_{\bT^4}\int_{\bT^4} \ud\hat{\vol}_{\mathbf{T}^{4}} \right)^{-1/2}
\eea
and so, by imposing that our 4D fields are canonically normalized, we obtain the physical Yukawa coupling
\begin{equation}
  \label{eq:yuk}
  y_{HF_LF_R}\, =\,  \frac{\bigl(2\pi\bigr)^{3/2}k}
  {\left(\int_{\mathbf{T}^{4}}\ud\hat{\vol}_{\mathbf{T}^{4}}\, Z\right)^{1/2}}\, \sim\, g_{\uD 7}
\end{equation}
that should be compared to the standard supergravity formula 
\begin{equation}
  \label{eq:sugrayuk}
  y_{ijk}=\ue^{\mathcal{K}/2}\bigl(\mathcal{K}_{i\bar{i}}\mathcal{K}_{j\bar{j}}
  \mathcal{K}_{k\bar{k}}\bigr)^{-1/2}W_{ijk}
\end{equation}
and the results from subsection \ref{kahler}. Indeed, we see that by setting $W_{HF_LF_R}  =  1$ and using eqs.(\ref{kinmod}) and  (\ref{eq:flatwilsonkahler}), as well as $\cK =$ (\ref{Kal1}) + (\ref{KuwT6}), we can derive (\ref{eq:yuk}).

As emphasized in \cite{Giddings:2005ff,stud,Douglas:2008jx}, compensators are needed for consistency of the equations of motion for the closed string fluctuations, and thus the field space 
metrics for the closed string sector are in general highly complex. 
However, in comparing (\ref{eq:yuk}) and (\ref{eq:sugrayuk}), we do not 
need to evaluate derivatives of the K\"ahler potential $\mathcal{K}$ with respect to closed string moduli and so the issue of compensators do not concern us here.

In this particular model, the Higgs field propagates throughout
the worldvolume of the D7.  In contrast, in the Randall-Sundrum scenario the
Higgs is confined to or near the IR end of the geometry.  As
discussed
in section \ref{summary}, the 5D masses of the bulk fermions (except for the
gaugino) is a free parameter, though is related to the masses of the bulk
scalars.  The mass $m_{\Psi}=cK$ controls the profile of the fermion in the bulk,
with modes for $c>\frac{1}{2}$ being localized toward the IR and modes with
$c<\frac{1}{2}$ being localized towards the 
UV~\cite{Grossman:1999ra}.
This localization controls the overlap with the Higgs and hence the 4D Yukawa
couplings depend sensitively on $c$ so that this mechanism provides a
model of the fermion mass hierarchy.  However, the bosonic and fermionic actions
for D-branes do not have such mass terms.  Instead, the localization can be
controlled by
either using gauge instantons (as suggested in \cite{abv06}) or by localizing
the matter fermions on intersections of D7 branes (as used for example in
\cite{shamit}).

%\newpage

\section{Magnetized D7-branes}\label{sec:magnetized}

\subsection{Allowing a worldvolume flux}

As we have seen, D7-branes in warped backgrounds of the form (\ref{10metricE}) provide a wealth of gauge theories with warped internal wavefunctions. This is however far from being the most general possibility when producing such theories. Indeed, as discussed before the D7-brane action depends on a generalized field strength ${\cal F} = P[B] + 2\pi \al' F$ living on the D7-brane worldvolume $\R^{1,3} \times \cS_4$, which contains the 8D gauge boson degrees of freedom via the usual relation $F=dA$. Now, instead of consider a vanishing vev for ${\cal F}$ as in the previous section, one may allow a nontrivial vev for such worldvolume flux. Clearly this does not spoil 4D Poincar\'e invariance if we choose the indices of $\langle\cF\rangle$ to be along $\cS_4$ and, in fact, this is an essential ingredient to obtain 4D chiral fermions via D7-brane intersections. Finally, such ``magnetized'' D7-brane will be a stable BPS object if, in addition to demanding that $\cS_4$ is volume minimizing we impose that \cite{mmms99,gmm05}
\be
\cF\, =\, {*}_{\cS_4} \cF
\label{BPSmD7}
\ee
where here and henceforth we omit the brackets to refer to the vev of $\cF$. That is, magnetized D7-branes in warped backgrounds of the form (\ref{10metricE})  are BPS if $\cF$ is a self-dual 2-form of their internal dimensions $\cS_4$.\footnote{More precisely, $\cF = \pm *_{\cS_4} \cF$ if $2\, \ud \text{vol}_{\cS_4} = \mp P[J^2]$ (see footnote \ref{ft:sign}), and the choice taken in \cite{gmm05} was such that a BPS D7-brane should host an anti-self-dual flux $\cF$. Our conventions match those of \cite{ms05}, where the derivation of the D7 BPS conditions were also carried out for  more general supergravity backgrounds.}

It is easy to see that adding a non-trivial $\cF$ will change the zero mode equations for both fermions and bosons. In particular, the Einstein frame fermionic action is not longer of the form (\ref{D7fermE}), but rather (see \cite{mrvv05} and Appendix \ref{conv})
\begin{equation}
\label{mD7fermE}
  S_{\uD 7}^{\left.\mathrm{fer}\right.}\, =\, \tau_{\uD 7}
  \int\ud^{8}\xi\, \ue^{\Phi}\sqrt{\abs{\det\, M}}\,
  \bar{\Theta} P^{\uD7}_-(\cF)
\left( \G^\mu{\cal D}_\mu +   (\cM^{-1})^{ab}{\Gamma}_{a}\biggl(
 {\cal D}_{b}+\frac18\G_b{\cal O}\biggr)\right){\Theta}
\end{equation}
where as before $\mu$ stands for a $\R^{1,3}$ index and $a,b$ for indices in $\cS_4$.  The worldvolume flux dependence enters via the operators\footnote{The operator $\mathcal{M}$ corresponds to $\tilde{M}$ in \cite{mrvv05} and, while the definition here and in \cite{mrvv05} slightly differ, they are equivalent. For an expression of the fermionic action closer to that in \cite{mrvv05} see the appendix.}
\bes
\begin{align}
M\, =\,  & P[G] + e^{-\Phi/2} \cF\\
\cM \, =\,  & P[G] + e^{-\Phi/2} \cF \sig_3 \\
P^{\uD7}_\pm(\cF)\, =\, &  \frac{1}{2} \left(\Id \mp  \Gamma_{(8)}^\cF \otimes \sigma_2 \right)\\
 \Gamma_{(8)}^\cF \, =\, &  \Gamma_{(8)} \sqrt{\left|{\det P[G] \over \det M }\right|}\left(\Id -  e^{-\Phi/2}\slashed{\cF} \otimes \sig_3  + \frac32 e^{-\Phi} \slashed{\cF}^2\right)
 \label{G8m}
\end{align}
\ees
that clearly reduce to those in (\ref{D7fermE}) when taking $\cF \raw 0$.  Note that terms that do not appear with a tensor product implicitly act as the identity on the bispinor space.  Finally, one can show that $P^{\uD7}_\pm(\cF)$ are still projectors, and that (\ref{BPSmD7}) is equivalent to impose the usual BPS condition $P^{\uD7}_+(\cF) \ep = 0$, with  $\ep$ given by the Killing spinor (\ref{wkilling}) \cite{mmms99,gmm05,ms05}.

\subsection{Warped flat space}

Paralleling our previous discussion for unmagnetized D7-branes, let us first consider the case where our D7-brane wraps a conformally flat four-cycle $\cS_4 =\bT^4$ inside the warped internal manifold $X_6 = \bT^6$ which is also conformally flat, and so that the metric on the D7-brane worldvolume is of the form (\ref{flat10metricE}). Let us further simplify this situation by taking a factorizable setup where $\cS_4 = (\bT^2)_i \times (\bT^2)_j$ and
\bes
\label{factormD7}
\begin{align}
P[J]\, =& \ \ud{\text{vol}}_{(\bT^2)_i} + \ud {\text{vol}}_{(\bT^2)_j}\\
\cF\, =& \  b_i\, \ud \hat{\text{vol}}_{(\bT^2)_i} + b_j\, \ud \hat{\text{vol}}_{(\bT^2)_j}
\end{align}
\ees
where as before $\ud{\text{vol}}_{\bT^2} = Z^{1/2} \ud\hat{\text{vol}}_{\bT^2}$ stand for warped and unwarped volume elements. It is then easy to see that with the choice $\ud\text{vol}_{\cS_4} = - \ud \text{vol}_{(\bT^2)_i} \wedge \ud \text{vol}_{(\bT^2)_j} $ the BPS condition (\ref{BPSmD7}) is equivalent to $\cF \wedge P[J] = 0$, which is solved for $b = b_i = - b_j$. If in addition we consider a vanishing background B-field, then $\cF  = 2\pi \al' f$, where $f$ is a U(1) field strength of the form
\be
f\, =\, 2\pi  m_i\, {\ud \hat{\text{vol}}_{(\bT^2)_i} \over \hat{\text{vol}}_{(\bT^2)_i}} + 2\pi  m_j\, {\ud \hat{\text{vol}}_{(\bT^2)_j} \over \hat{\text{vol}}_{(\bT^2)_j}}
\ee
and where, because of Dirac's charge quantization, $m_i, m_j \in \mathbb{Z}$. The BPS conditions above then translate into the more familiar condition $m_i/ \hat{\text{vol}}_{(\bT^2)_i} + m_j/ \hat{\text{vol}}_{(\bT^2)_j} = 0$ used in the magnetized D7-brane literature.

\subsubsection{Fermions}\label{fermflatm}

Following the steps taken in subsection \ref{flatferm}, we have that the dilatino and gravitino operators  entering the fermionic action are again given by (\ref{flatD7op}). Hence, plugging them in (\ref{mD7fermE}) and taking the $\kappa$-fixing gauge (\ref{bispin1}), one finds a Dirac action of the form (\ref{flatD7fermE1}), where now
\bea
\nonumber
 \sqrt{{\det g_{\bT^4} \over \det M_{\bT^4}}}\, \slashed{D}^w& =&   {\slashed{\pa}}_4^{\text{ext}} + (M_{\bT^4}^{-1})^{ab}{\Gamma}_{a}\biggl( \pa_b - \frac18 \pa_b\ln Z\biggr) + \frac14  \Lam(-\cF)\G_{\text{Extra}} (M_{\bT^4}^{-1})^{ba}{\Gamma}_{a} \pa_b\ln Z \\
 \nonumber
 & - &  \frac12 \left(1 - \frac14 (M_{\bT^4}^{-1})^{ab}\G_a\G_b \right) \slashed{\pa}\ln Z \\
 & + &   \frac12 \Lam(-\cF) \G_{\text{Extra}} \left(1 - \frac14 (M_{\bT^4}^{-1})^{ba}\G_a\G_b \right) \slashed{\pa}\ln Z 
\label{wDiracmD7}
\eea
where
\be
\Lam(\cF)\, =\, \sqrt{{\det g_{\bT^4} \over \det M_{\bT^4}}} \left(\Id +  e^{-\Phi_0/2} \slashed{\cF}  + \frac32 e^{-\Phi_0} \slashed{\cF}^2\right) \quad  \quad \quad M_{\bT^4}\, =\, g_{\bT^4} + 2\pi \al' e^{-\Phi_0/2}f
\label{Lam}
\ee
and $g_{\bT^4} =Z^{1/2} \hat{g}_{\bT^4}$ stands for the warped $\bT^4$ metric.

Using now the factorized ansatz $\bT^4 =  (\bT^2)_i \times (\bT^2)_j$ and (\ref{factormD7}), it is easy to see that
\bes
\begin{align}
M_{\bT^4}\, = &\,
\left(
\begin{array}{cc}
M_{\bT^2_i} & 0 \\
0 & M_{\bT^2_j}
\end{array}
\right)\\
M_{\bT^2_i}\, = & \, 4\pi^2 \al'
\left[Z^{1/2}R_i^2
\left(
\begin{array}{cc}
1 & \re \tau_i \\ \re \tau_i & |\tau_i|^2
\end{array}
\right)
+
e^{-\Phi_0/2}
\left(
\begin{array}{cc}
0 & m_i \\ -m_i & 0
\end{array}
\right)
\right]
\end{align}
\ees
In terms of the complex coordinates $z^{m}=y^{m+3}+\tau_{m}y^{m+6}$ this reads
\begin{equation}
M_{\bT^2_i}\,=  \,
 \frac12 (4\pi^2 \al')\,
Z^{1/2}R_i^2
\left(
\begin{array}{cc}
0 & 1+ iB_i \\ 1- iB_i & 0
\end{array}
\right)\quad \text{with} \quad B_i \, =\, Z^{-1/2} e^{-\Phi_0/2} b_i
\label{cpxM}
\end{equation}
Then, also in this complex basis\footnote{Here $i,j$ denote particular
$\bT^2$'s and so there are no sums implicit in this kind of expressions.}
\be
\frac12 (M^{-1}_{\bT^4})^{ab}\G_a\G_b\, =\, {\Id - iB_i\G_{\bT^2_i} \over | 1+iB_i |^2} +  {\Id - iB_j\G_{\bT^2_j} \over | 1+iB_j |^2}
\ee
where $\G_{\bT^2_i} = - i\, \ud \slashed{\text{vol}}_{(\bT^2)_i}$ is the chirality matrix for $\bT^2_i$. Similarly, we have
\be
\Lam(\cF)\, =\,  {\Id + iB_i\G_{\bT^2_i} \over | 1+iB_i |} \cdot  {\Id + iB_j\G_{\bT^2_j} \over | 1+iB_j |}\, =\, e^{i\phi_i\G_{\bT^2_i}} \cdot e^{i\phi_j\G_{\bT^2_j}} 
\label{Lamflat}
\ee
where we have defined $\phi_i \equiv \text{arctan}\, B_i$. Notice that, unlike in the usual magnetized D-brane literature, $\phi_i$ is not a constant angle, having a non-trivial dependence on the warp factor. Finally we can express $\G_{\text{Extra}} = \ud \slashed{\text{vol}}_{\cS_4}= \G_{\bT^2_i} \G_{\bT^2_j}$.

We can now implement the dimensional reduction scheme of subsection \ref{flatferm}, taking again the ans\"atze (\ref{splitgaug1}) and (\ref{KKflatferm}). In order to find the eigenmodes of the Dirac operator, one first notices that given the setup above the first line of (\ref{wDiracmD7}) can be written as 
\be
{\slashed{\pa}}_4^{\text{ext}} + (M_{\bT^4}^{-1})^{ab}{\Gamma}_{a}\biggl[ \pa_b - \frac18 \pa_b\ln Z \left(1 + 2  \Lam(-\cF)\G_{\text{Extra}} \right) \biggr] 
\ee
In addition, considering the case where the worldvolume flux $\cF$ satisfies the BPS conditions $B_i = - B_j \iff \phi_i + \phi_j = 0$, it is easy to see that the second plus third lines of  (\ref{wDiracmD7}) vanish identically. Hence, we find a 6D internal eigenmode equation similar to (\ref{flatD7eigen1}) where the main differences come from the substitution $\hat{g}^{-1}_{\bT^4} \raw M^{-1}_{\bT^4}$ and the insertion of $\Lam(-\cF)$. In particular, the zero mode equation amounts to\footnote{The same discussion in Sec \ref{gaugefix} applies here as well.}
\be
\biggl[ \pa_b - \frac18 \pa_b\ln Z \left(1 + 2  \Lam(-\cF)\G_{\text{Extra}} \right) \biggr] \theta_{6D}^0\, =\, 0
\ee
whose solutions are
\bes
\label{flatmD7zero1}
\begin{align}
\label{flatmwilsonino1}
\theta_{6D}^0 = {Z^{-1/8} \over 1 + i B_i \G_{\bT_i^2}}\eta_- 
\quad & \text{for} \quad \G_{\text{Extra}}\, \eta_-\, =\, - \eta_- \quad \quad \text{Wilsonini}\\
\theta_{6D}^0 = {Z^{3/8}}\eta_+ 
\quad & \text{for} \quad \G_{\text{Extra}}\, \eta_+\, =\, \eta_+ \quad \quad \text{gaugino} + \text{modulino}
\label{flatmphotino1}
\end{align}
\ees
where $\eta_\pm$ are again constant 6D spinor modes with $\pm$ chirality in the D7-brane extra dimensions. In particular, for a D7-brane extended along $01234578$, we have that $\cS_4 = (\bT^2)_1 \times (\bT^2)_2 \subset (\bT^2)_1 \times (\bT^2)_2 \times (\bT^2)_3 = X_6$ and so the fermionic zero modes will have the following internal wavefunctions
\be
\theta_{6D}^{0,0}\, =\,
 Z^{3/8} \, \eta_{---}
\quad \quad
\theta_{6D}^{0,3}\, =\,
Z^{3/8} \, \eta_{++-}
\ee
and
\be
\label{eq:wilsoniniwavesm}
\theta_{6D}^{0,1}\, =\,
{Z^{-1/8} \over 1 - iB} \, \eta_{-++}
\quad \quad
\theta_{6D}^{0,2}\, =\, 
{Z^{-1/8} \over 1 + i B}\, \eta_{+-+}
\ee
where $B = B_1 = -B_2$, and again using the 6D fermionic basis defined in Appendix \ref{conv}. 

Notice that the new Wilsonini wavefunctions do not amount to a simple constant rescaling, as the `density of wordvolume flux' $B$ depends nontrivially on the warp factor. This dependence is however the one needed to cancel all warp factor dependence in the Wilsonini 4D kinetic terms. Indeed, by inserting (\ref{flatmwilsonino1}) into the $\kappa$-fixed fermionic action (\ref{flatD7fermE1}) we obtain again
\be
\label{ktermmwil}
S_{\uD 7}^{\left.\mathrm{fer}\right.}\, =\, \tau_{\uD 7} \,\ue^{\Phi_0}
\int_{\R^{1,3}} \ud^{4}x\, \bar{\theta}_{4D} \slashed{\pa}_{\R^{1,3}} \theta_{4D} \int_{\bT^4}  \ud\hat{\text{vol}}_{\bT^4}\, \eta_-^\dag \eta_- 
\ee
where we have taken into account the new volume factor appearing in the r.h.s of (\ref{wDiracmD7}), which in the BPS case reads
\be
\sqrt{{\det g_{\bT^4} \over \det M_{\bT^4}}}\, =\, \left| 1 + i B\right|^2
\ee
and where we are again expressing everything in terms of complex coordinates, as in (\ref{cpxM}). Regarding the gaugino and the modulino, the above factor does not cancel and so we have a kinetic term of the form
\be
\label{ktermmphot}
S_{\uD 7}^{\left.\mathrm{fer}\right.}\, =\, \tau_{\uD 7}\, \ue^{\Phi_0}
\int_{\R^{1,3}} \ud^{4}x\, \bar{\theta}_{4D} \slashed{\pa}_{\R^{1,3}} \theta_{4D} \int_{\bT^4}  \ud\hat{\text{vol}}_{\bT^4}\, |Z^{1/2} + i e^{-\Phi_0/2} b|^2 \eta_+^\dag \eta_+
\ee
that generalizes that obtained in (\ref{ktermphot}). As we will now see, such results can be rederived by analyzing the D7-brane bosonic wavefunctions.

\subsubsection{Bosons}\label{bosflatm}

In the presence of a world-volume flux, the 8D gauge boson $A_{\alpha}$ enters
into the D7-brane action through the field strength
$\mathcal{F}=P\left[B\right]+2\pi\alpha' f+2\pi\alpha' F$ where $f = \langle F \rangle$ is the
background field strength and $F=\ud A$.  The transverse oscillations again
enter through the pullback of the metric as in (\ref{pbmetric}).  In the case
of $B=0$ and constant dilaton $\Phi=\Phi_{0}$, the action for the
D7-brane up to quadratic in fluctuations order becomes
\begin{subequations}
\begin{equation}
  S_{\uD 7}^{\left.\mathrm{bos}\right.}
  =\bigl[S_{\uD 7}\bigr]_{0}
  + S_{\uD 7}^{\left.\mathrm{scal}\right.}
  + S_{\uD 7}^{\left.\mathrm{photon}\right.}
\end{equation}
where the action for the position moduli is
\begin{equation}
  S_{\uD 7}^{\left.\mathrm{scal}\right.}
  =-\bigl(8\pi^{3}k^{2}\bigr)^{-1}
  \int\ud^{8}\xi
  \sqrt{\absb{\det M}}\,
  \frac{1}{2}\ue^{\Phi_{0}}G_{ij}
  \bigl(M^{-1}\bigr)^{\left(\alpha\beta\right)}\partial_{\alpha}\sigma^{i}
  \partial_{\beta}\sigma^{i}
\end{equation}
and the action for the 8d gauge boson is
\begin{align}
  S_{\uD 7}^{\left.\mathrm{gauge}\right.}=
  -\frac{1}{2}\bigl(8\pi^{3}k^{2}\bigr)^{-1}
  \int\ud^{8}\xi
  \biggl\{&
  \sqrt{\absb{\det M}}
  \biggl[\frac{1}{2}\left(\bigl(M^{-1}\bigr)^{\left[\alpha\beta\right]}
  F_{\alpha\beta}\right)^{2}
  + \bigl(M^{-1}\bigr)^{\alpha\beta}\bigl(M^{-1}\bigr)^{\gamma\delta}
  F_{\alpha\delta}F_{\beta\gamma}\biggr] \notag \\
  & - \frac{1}{2}\biggl[ C_{4}^{\mathrm{int}}\epsilon^{\mu\nu\rho\sigma}
  F_{\mu\nu}F_{\rho\sigma}
  + C_{4}^{\mathrm{ext}}\epsilon^{abcd}F_{ab}F_{cd}\biggr] \notag \\
  & - \frac{1}{16}C_{0}\epsilon^{abcd}f_{ab}f_{cd}\epsilon^{\mu\nu\sigma\rho}
  F_{\mu\nu}F_{\sigma\rho}\, \biggr\}
\end{align}
\end{subequations}
where we have again used (\ref{eq:dettr}) and have separated the action between
a zero energy part and a part with derivatives.
In general, there are
three
more contributions to the action up to quadratic order including a term
that is linear in the field strength,
\begin{equation}
  \frac{1}{2}\bigl(8\pi^{3}k^{2}\bigr)^{-1}
  \int\ud^{8}\xi\,
  \bigl(\ue^{\Phi_{0}/2}
  Z^{-1}\sqrt{\absb{\det M_{\mathbf{T}^{4}}}}
  \bigl(M_{\mathbf{T}^{4}}^{-1}\bigr)^{\left[ab\right]}
  k F_{ab}+\frac{1}{2}\epsilon^{abcd}C_{4}^{\mathrm{ext}}f_{ab}F_{cd}\bigr)
\end{equation}
an interaction between the position moduli and the 8D gauge boson,
\begin{equation}
  \frac{1}{2}\bigl(8\pi^{3}k^{2}\bigr)^{-1}
  \int\ud^{8}\xi\, \biggl(
  \partial_{i}\bigl(\ue^{\Phi_{0}/2}Z^{-1}\sqrt{\absb{\det M_{\mathbf{T}^{4}}}}
  \bigl(M_{\mathbf{T}^{4}}^{-1}\bigr)^{\left[cd\right]}\bigr)
  + \frac{k}{2}\bigl(\partial_{i}C_{4}^{\mathrm{ext}}\bigr)\epsilon^{abcd}
  f_{ab}
  \biggr)
  F_{cd}\sigma^{i}
\end{equation}
and a potential term for the position moduli
\begin{equation}
  -\bigl(8\pi^{3}k^{2}\bigr)^{-1}\int\ud^{8}\xi\,
  \bigl(\frac{1}{k}\sigma_{i}\partial_{i} +
  \frac{1}{2}\sigma_{i}\sigma_{j}\partial_{i}\partial_{j}\bigr)
  \bigl(\ue^{\Phi_{0}}Z^{-1}\sqrt{\absb{\det M_{\mathbf{T}^{4}}}}
  -\frac{1}{8}C_{4}^{\mathrm{ext}}\epsilon^{abcd}
  f_{ab}f_{cd}
  \bigr)
\end{equation}
However, when the world-volume flux is self-dual, all three of these
contributions vanish up to surface terms.  This is most easily seen by
inserting the fluxes explicitly.

Expanding out the action for the position moduli,
\begin{equation}
  S_{\uD 7}^{\left.\mathrm{scal}\right.}
  =-\frac{1}{2}\bigl(8\pi^{3}k^{2}\bigr)^{-1}\ue^{\Phi_{0}}
  \int\ud^{8}\xi\, \sqrt{\absb{\det M}}
  \hat{g}_{ij}\biggl(Z\eta^{\mu\nu}\partial_{\mu}\sigma^{i}\partial_{\nu}\sigma^{j}
  + Z^{1/2}\bigl(M^{-1}\bigr)^{\left(ab\right)}
  \partial_{a}\sigma^{i}\partial_{b}\sigma^{j}\biggr)
\end{equation}
we obtain the 8D equation of motion
\begin{equation}
  \square_{\mathbb{R}^{1,3}}\sigma^{i}
  + \absb{\det M_{\mathbf{T}^{4}}}^{-1/2}
  \partial_{a}\bigl[Z^{-1/2}\sqrt{\absb{\det M_{\mathbf{T}^{4}}}}
  \bigl(M_{\mathbf{T}^{4}}^{-1}\bigr)^{\left(ab\right)}
  \partial_{b}\sigma^{i}\bigr]
  =0
\end{equation}
As in the unmagnetized case (\ref{kkmod}), performing a KK expansion gives
the eigenmode equation
\begin{equation}
  \partial_{a}\bigl[Z^{-1/2}\sqrt{\absb{\det M_{\mathbf{T}^{4}}}}
  \bigl(M_{\mathbf{T}^{4}}^{-1}\bigr)^{\left(ab\right)}
  \partial_{b}s_{\omega}^{i}\bigr]=
  -\sqrt{\absb{\det M_{\mathbf{T}^{4}}}}m_{\omega}^{2}s_{\omega}^{i}
\end{equation}
This depends on the warp factor and the magnetic flux, but for the massless
modes, the only well-defined solution is
$s_{0}^{i}=const$.  The resulting 4D kinetic term for the zero mode is
\begin{equation}
  S_{\uD 7}^{\left.\mathrm{scal}\right.}
  =-\frac{1}{2}\bigl(8\pi^{3}k^{2}\bigr)^{-1}
  \int_{\mathbb{R}^{1,3}}\ud^{4}x\, \hat{g}_{ij}
  \eta^{\mu\nu}\partial_{\mu}\zeta_{0}^{i}\partial_{\nu}
  \zeta_{0}^{j}
  \int_{\mathbf{T}^{4}}\ud\hat{\vol}_{\mathbf{T}^{4}}\, 
  \ue^{\Phi_{0}}\abs{Z^{1/2}+\ui\ue^{-\Phi_{0}/2}b}^{2}s_{0}^{i}s_{0}^{j}
\end{equation}
which again matches with kinetic term for the modulino (\ref{ktermmphot}).

Also as in the unmagnetized case, the action contains an interaction piece
between
the 4D photon $A_{\mu}$ and the 4D Wilson lines $A_{a}$ which, after integrating
by parts twice, is
\begin{equation}
  \bigl(8\pi^{3}k^{2}\bigr)^{-1}\int\ud^{8}\xi\,
  \partial_{a}\bigl(Z^{-1/2}\sqrt{\absb{\det M_{\mathbf{T}^{4}}}}
  \bigl(M_{\mathbf{T}^{4}}^{-1}\bigr)^{\left(ab\right)}\eta^{\mu\nu}
  A_{b}\partial_{\nu}A_{\mu}\bigr)
\end{equation}
In analogy with the unmagnetized case, this can be gauged away by considering
the class of $R_{\Xi}$ gauges with gauge-fixing term
\begin{equation}
  S_{\uD 7}^{\Xi} =
  \bigl(8\pi^{3}k^{2}\bigr)^{-1}
  \int\ud^{8}\xi\, \sqrt{\absb{\det M}}\mathcal{G}_{\Xi}\bigl(A\bigr)
\end{equation}
where we take
\begin{equation}
  \mathcal{G}_{\Xi}\bigl(A\bigr)=
  \frac{1}{2\Xi}
  \biggl[\partial^{\mu}A_{\mu}
  + \Xi Z^{1/2}\absb{\det M_{\mathbf{T}^{4}}}^{-1/2}
  \partial_{a}\bigl(\sqrt{\absb{\det M_{\mathbf{T}^{4}}}}
  \bigl(Z^{-1/2}M_{\mathbf{T}^{4}}^{-1}\bigr)^{\left(ab\right)}A_{b}\bigr)\biggr]^{2}
\end{equation}
The form of the gauge fixing is chosen so that the equations of
motion for $A_{\mu}$ decouple from the equations of motion for $A_{a}$
for any value of $\Xi$ and so that it reduces to
gauge-fixing term in the unmagnetized case (\ref{eq:gaugefixterm}).  For
$A_{\mu}$, the equation of motion in the $R_{\Xi}$ gauge is
\begin{equation}
  \square_{\mathbb{R}^{1,3}}A_{\nu}
  -\left(1-\frac{1}{\Xi}\right)\eta^{\mu\sigma}\partial_{\nu}\partial_{\mu}
  A_{\sigma}+\absb{\det M_{\mathbf{T}^{4}}}^{-1/2}
  \partial_{a}\bigl(Z^{-1/2}\sqrt{\absb{\det M_{\mathbf{T}^{4}}}}
  \bigl(M_{\mathbf{T}^{4}}^{-1}\bigr)^{\left(ab\right)}\partial_{b}A_{\nu}\bigr)=0
\end{equation}
while for $A_{a}$, the equation is
\begin{align}
  Z^{-1/2}&\sqrt{\absb{\det M_{\mathbf{T}^{4}}}}
  \bigl(M_{\mathbf{T}^{4}}^{-1}\bigr)^{\left(ab\right)}
  \square_{\mathbb{R}^{1,3}}A_{b} \notag \\
  &+ \partial_{b}\biggl[Z^{-1}\sqrt{\absb{\det M_{\mathbf{T}^{4}}}}
  \biggl(
  M_{\mathbf{T}^{4}}^{cbad}F_{cd}
  -\frac{1}{2}\bigl(M_{\mathbf{T}^{4}}^{-1}\bigr)^{\left[cd\right]}
  \bigl(M_{\mathbf{T}^{4}}^{-1}\bigr)^{\left[ab\right]}F_{cd}\biggr)\biggr] +
  \epsilon^{abcd}\partial_{b}\bigl(Z^{-1}F_{cd}\bigr)
  \notag \\
  & + \Xi\biggl[
  Z^{-1/2}\sqrt{\absb{\det M_{\mathbf{T}^{4}}}}
  \bigl(M_{\mathbf{T}^{4}}^{-1}\bigr)^{\left(ab\right)}
  \partial_{b}\bigl[
  \absb{\det M_{\mathbf{T}^{4}}}^{-1/2}
  \partial_{c}\bigl(Z^{-1/2}\sqrt{\absb{\det M_{\mathbf{T}^{4}}}}
  \bigl(M_{\mathbf{T}^{4}}^{-1}\bigr)^{\left(cd\right)}A_{d}\bigr)\biggr] =0
\end{align}
where we have defined
\begin{equation}
  M^{abcd}=\frac{1}{2}\bigl(M^{-1}\bigr)^{ab}\bigl(M^{-1}\bigr)^{cd}-
  \frac{1}{2}\bigl(M^{-1}\bigr)^{ac}\bigl(M^{-1}\bigr)^{bd}
\end{equation}
Note that the presence of warping and background world-volume flux together
has made the equation of motion rather complex, even in the case of flat
space.   With this gauge choice, the KK modes for the 4D gauge boson
satisfy
\begin{equation}
  \partial_{a}\bigl(\sqrt{\absb{\det M_{\mathbf{T}^{4}}}}Z^{-1/2}
  \bigl(M_{\mathbf{T}^{4}}^{-1}\bigr)^{\left(ab\right)}\partial_{b}a^{\omega}\bigr)
  =-\sqrt{\absb{\det M_{\mathbf{T}^{4}}}}m_{\omega}^{2}a^{\omega}
\end{equation}
so that the zero mode $a^{0}$ has a constant profile on the internal dimensions.
This gives a gauge kinetic function
\begin{equation}
  \label{eq:maggkf}
  f_{\uD 7}=\bigl(8\pi^{3}k^{2}\bigr)^{-1}
  \int_{\mathbf{T}^{4}}\frac{\ud\hat{\vol}_{\mathbf{T}^{4}}}
  {\sqrt{\hat{g}_{\mathbf{T}^{4}}}}\, \bigl(
  \abs{Z^{1/2}+\ui\ue^{-\Phi_{0}/2}b}^{2}
  +\ui \bigl(C_{4}^{\mathrm{int}}-C_{0}b^{2}\bigr)\left(\alpha^{0}\right)^{2}
\end{equation}
The real part matches the kinetic term for the gaugino (\ref{ktermphot})
and in the absence of warping agrees with that found in, e.g., \cite{lmrs04,Font:2004cx}.

The equation of motion for the Wilson lines simplify further in the
4D Lorenz gauge $\Xi=0$ though even then the equation of motion
is difficult to solve in general.  However, if we focus on the zero-modes which
satisfy
\begin{equation}
  \square_{\mathbb{R}^{1,3}}w_{a}^{0}=0
\end{equation}
then the equation of motion for the internal profiles becomes
\begin{equation}
  \partial_{b}\biggl[Z^{-1}\sqrt{\absb{\det M_{\mathbf{T}^{4}}}}
  \biggl(M_{\mathbf{T}^{4}}^{cbad}F_{cd}^{0}
  -\frac{1}{2}\bigl(M_{\mathbf{T}^{4}}^{-1}\bigr)^{\left[cd\right]}
  \bigl(M_{\mathbf{T}^{4}}^{-1}\bigr)^{\left[ab\right]}F_{cd}^{0}
  \biggr)\biggr] +
  \epsilon^{abcd}\partial_{b}\bigl(Z^{-1}F_{cd}^{0}\bigr)=0
\end{equation}
In the unmagnetized case, we deduced that the solution satisfied $F_{ab}^{0}=0$
and
this is clearly a solution in the magnetized case as well.  This again determines
the solution to be of the form $w_{a}^{0}=const.$ up to 
the residual gauge freedom $A_{a}\to A_{a} -
\partial_{a}\Lambda$ where $\partial_{\mu}\Lambda=0$.   This residual
freedom will not effect the 4D effective action,
\begin{equation}
  S_{\uD 7}^{\left.\mathrm{wl}\right.}=-\frac{1}{2}
  \bigl(8\pi^{3}k^{2}\bigr)^{-1}\int_{\mathbb{R}^{1,3}}\ud^{4}x\, 
  \eta^{\mu\nu}\partial_{\mu}w_{a}^{0}\partial_{\nu}w_{b}^{0}
  \int_{\mathbf{T}^{4}}\ud\hat{\vol}_{\mathbf{T}^{4}}\, 
  \abs{Z^{1/2}+\ui\ue^{-\Phi_{0}/2}b}^{2}
  Z^{-1/2}\bigl(M_{\mathbf{T^{4}}}^{-1}\bigr)^{\left(ab\right)}W_{a}^{0}W_{b}^{0}
\end{equation}
For $\Xi\neq 0$, there is an additional term in the equation of motion for
the internal wavefunction $W_{a}^{0}$ that depends on $\Xi$
\begin{equation}
  \Xi\biggl[
  Z^{-1/2}\sqrt{\absb{\det M_{\mathbf{T}^{4}}}}
  \bigl(M_{\mathbf{T}^{4}}^{-1}\bigr)^{\left(ab\right)}
  \partial_{b}\bigl[
  \absb{\det M_{\mathbf{T}^{4}}}^{-1/2}
  \partial_{c}\bigl(Z^{-1/2}\sqrt{\absb{\det M_{\mathbf{T}^{4}}}}
  \bigl(M_{\mathbf{T}^{4}}^{-1}\bigr)^{\left(cd\right)}A_{d}\bigr)\biggr]
\end{equation}
However, when the world-volume flux is self-dual or anti-self-dual, the
combination
\begin{equation}
  Z^{-1/2}\sqrt{\absb{\det M_{\mathbf{T}^{4}}}}
  \bigl(M_{\mathbf{T}^{4}}^{-1}\bigr)^{\left(cd\right)}
\end{equation}
is constant implying that $A_{a}=const.$ is still a solution for
arbitrary $\Xi$.  After complexifying the Wilson lines (\ref{eq:complexwl})
the kinetic term matches the kinetic term for the Wilsonini
(\ref{ktermmwil}) for any choice of $R_{\Xi}$ gauge.

\subsection{More general warped backgrounds}

Let us now consider magnetized D7-branes in more general warped backgrounds. Just as in the unmagnetized case, it proves useful to compute the D7-brane wavefunctions via an alternative choice of $\kappa$-fixing. Let us first do so for warped flat space. In this case, and before any $\kappa$-fixing, the operator in (\ref{mD7fermE}) between $\bar{\Theta}$ and $\Theta$ is given by
\be
\begin{array}{c}\vspace*{0.1cm}
P_-^{\uD 7} (\cF) \left[{\slashed{\pa}}_4^{\text{ext}} +  (\cM^{-1}_{\bT^4})^{ab}{\Gamma}_{a}\biggl( \pa_b + \pa_b\ln Z \left( \frac18 - \frac12 P_+^{O3}\right)\biggr)\right] \\
 -\, P_-^{\uD 7} (\cF) \left(1 - \frac14 (\cM^{-1}_{\bT^4})^{ab}\G_a\G_b \right) \slashed{\pa}\ln Z P_+^{O3}
\end{array}
\label{wDiracmD7full}
\ee
just like the last two lines of (\ref{wDiracmD7}), the second line of (\ref{wDiracmD7full}) vanishes when we impose the BPS condition on the worldvolume flux $\cF$. As a result, for BPS D7-branes such term can be discarded independently of the $\kappa$-fixing choice. Let us in particular take the choice $P_-^{\uD 7} (\cF) \,\Theta =0$, as in subsection \ref{altk}. This allows to remove $P_-^{\uD 7} (\cF)$ from (\ref{wDiracmD7full}), and so we find an fermionic action of the form (\ref{flatD7fermEalt}), with a Dirac operator
\be
\slashed{D}^w \, =\,  \sqrt{{\det M_{\bT^4} \over \det g_{\bT^4}}}\, \left[{\slashed{\pa}}_4^{\text{ext}} +  (\cM^{-1}_{\bT^4})^{ab}{\Gamma}_{a}\biggl( \pa_b + \pa_b\ln Z \left( \frac18 - \frac12 P_+^{O3}\right)\biggr)\right]
\label{wDiracmD7alt}
\ee
Hence, the main difference on $\slashed{D}^w$ with respect to the unmagnetized case (\ref{flatD7Dirac}) comes from substituting $g^{-1} \raw \cM^{-1}$. As $\cM^{-1}$ is obviously invertible, one would na\"ively say that the zero mode internal wavefunctions are the same as in the unmagnetized case. 

Note however that the $\kappa$-fixing condition $P_-^{\uD 7} (\cF)\, \Theta =0$ depends on $\cF$, and so will the set of 10D bispinors $\Theta$ that enter our fermionic action. Indeed, following \cite{bkop97} one can write
\be
 \Gamma_{(8)}^\cF \otimes \sig_2 \, =\, e^{-\frac i2 \left(\phi_i \G_{\bT^2_i} +  \phi_j \G_{\bT^2_j}\right)\otimes \sig_3} \left(  \Gamma_{(8)}\otimes \sig_2 \right) e^{\frac i2 \left(\phi_i \G_{\bT^2_i} +  \phi_j \G_{\bT^2_j}\right)\otimes \sig_3} 
\ee
where we have used the explicit form of $\Lam(\cF)$ in (\ref{Lamflat}). Hence, the bispinors surviving the projection $P_-^{\uD 7}(\cF)\, \Theta= 0$ are given by
\be
\Theta\, =\, e^{-\frac i2 \left(\phi_i \G_{\bT^2_i}  +  \phi_j \G_{\bT^2_j}\right)\otimes \sig_3} \, \Theta'\quad\quad \text{where}\quad \quad P_-^{\uD 7} \Theta'\, =\, 0
\label{rotatedbispin}
\ee
and where $P_-^{\uD 7}$ stands for the unmagnetized D7-projector (\ref{projD7F0}). We thus need to consider a basis of bispinors `rotated' with respect to the one used for unmagnetized D7-brane. As the rotation only acts on the internal D7-brane coordinates, one can still make the decomposition (\ref{decomp46alt}), with the 4D spinor $\theta_{4D}$ intact and the 6D bispinor $\Theta_{6D}$ rotated as in (\ref{rotatedbispin}). In particular, if we impose the BPS condition $\phi_i + \phi_j = 0$, $\Theta_{6D}$ takes the form
\bes
\label{ansmD7alt}
\begin{align}
\label{answilsoninoalt}
\Theta_{6D,-} = &\, \frac{\psi_-}{\sqrt{2}} 
e^{-i \phi_i \G_{\bT^2_i}\otimes \sig_3}
\left(
\begin{array}{c}
\eta_{-}\\ i\eta_{-}
\end{array}
\right) 
\quad  \text{for} \quad \G_{\text{Extra}} \eta_-\, =\, - \eta_-\\%\quad \quad \text{Wilsonini}\\
\Theta_{6D,+} = &\,  \frac{\psi_+}{\sqrt{2}}
\left(
\begin{array}{c}
i\eta_{+}\\ \eta_{+}
\end{array}
\right)
\quad \quad \quad\quad \ \quad\text{for} \quad 
\G_{\text{Extra}} \eta_+\, =\, \eta_+%\quad \quad \text{gaugino + modulino}
\label{ansphotinoalt}
\end{align}
\ees
and so the bispinors $\Theta_{6D,+}$ with positive extra-dimensional chirality are exactly those of the unmagnetized case, while those of negative chirality $\Theta_{6D,-}$ are rotated by a (warping dependent) phase. 

From the above, it is easy to see that the zero modes coming from $\Theta_{6D,+}$ have as wavefunction $\psi^0_+ = Z^{3/8}$, just like in the unmagnetized case. On the other hand, plugging (\ref{answilsoninoalt}) into (\ref{wDiracmD7alt}) we obtain a zero mode equation quite similar to that found Wilsonini in subsection \ref{fermflatm}, and so we find that $\psi_-^0 = Z^{-1/8} |1 + i B_i|^{-1}$. As a result, the zero mode wavefunctions are given by

\bes
\label{flatmD7zeroalt}
\begin{align}
\label{flatwilsoninomalt}
\Theta_{6D,-}^0 = &\, {Z^{-1/8}/\sqrt{2} \over 1 + i B_i \G_{\bT^2_i}\otimes\sig_3}
\left(
\begin{array}{c}
\eta_{-}\\ i\eta_{-}
\end{array}
\right) 
\quad  \text{for} \quad \G_{\text{Extra}} \eta_-\, =\, - \eta_- \quad \quad \text{Wilsonini}\\
\Theta_{6D,+}^0 = &\,  \frac{Z^{3/8}}{\sqrt{2}}
\left(
\begin{array}{c}
i\eta_{+}\\ \eta_{+}
\end{array}
\right)
\quad \quad \quad \ \quad \quad\text{for} \quad 
\G_{\text{Extra}} \eta_+\, =\, \eta_+\quad \quad \text{gaugino + modulino}
\label{flatphotinomalt}
\end{align}
\ees
where, via matching of the 4D kinetic functions, we have identified the fermionic 4D zero modes that they correspond to. Note that again the Wilsonini have an extra warp factor dependence with respect to the unmagnetized case, which is contained in $B_i$.

On can then proceed to generalize the above computation to the case of a D7-brane in a warped Calabi-Yau. Imposing the $\kappa$-fixing choice $P_-^{\uD 7} (\cF) \Theta = 0$ and the BPS condition $*_{\cS_4}\cF = \cF$, the Dirac operator reads
\be
\slashed{D}^w \, =\,  \sqrt{{\det M_{\bT^4} \over \det g_{\bT^4}}}\, \left[{\slashed{\pa}}_4^{\text{ext}} +  (\cM^{-1}_{\cS_4})^{ab}{\Gamma}_{a}\biggl( \nabla_b^\CY + \pa_b\ln Z \left( \frac18 - \frac12 P_+^{O3}\right)\biggr)\right]
\label{wCYDiracmD7alt}
\ee
where we have removed the term coming from the second line of (\ref{wDiracmD7full}), using the fact that it vanishes for a BPS worldvolume flux $\cF$.\footnote{Indeed, even if we are no longer in flat space, there is locally always a choice of worldvolume vielbein where \cite{mrvv05}
\bes
\begin{align}\nonumber
&\frac12 (M^{-1}_{\cS_4})^{ab}\G_a\G_b\, = \, {\Id - iB_i \sig_3^i \over | 1+iB_i |^2} +  {\Id - iB_j\sig_3^j \over | 1+iB_j |^2} \\ \nonumber
&\Lam(\cF)\, = \, e^{i\left( \phi_i\sig_3^i + \phi_j\sig_3^j\right)} 
\end{align}
\ees
where $\sig_3^1 \equiv \sig_3 \otimes \Id_2 \otimes \Id_2$, $\sig_3^2 \equiv \Id_2 \otimes \sig_3 \otimes \Id_2$ and $\sig_3^3 \equiv \Id_2 \otimes \Id_2 \otimes  \sig_3$ act on the 6D spinor basis (\ref{spinorbasis:ap}). In this basis $*_{\cS_4}\cF = \cF$ is equivalent to $\phi_i + \phi_j =0$, and so all the algebraic manipulations carried out for flat space also apply. In particular, the second line of (\ref{wDiracmD7full}) identically vanishes.}

In addition to the Dirac operator, one needs to know how the worldvolume fermions satisfying $P_-^{\uD 7} (\cF) \Theta = 0$ look like. From our discussion above we know that this $\kappa$-fixing choice selects bispinors of the form 
\be
\Theta\, =\, 
\left(
\begin{array}{cc}
\Lam(-\cF)^{1/2} \\
& \Lam(\cF)^{1/2}
\end{array}
\right)\, \Theta' \quad\quad 
\text{with}\quad \quad P_-^{\uD 7} \Theta'\, =\, 0
\label{rotCY}
\ee
where again $P_-^{\uD 7}$ stands for (\ref{projD7F0}). In general, the rotation $\Lam(\cF)$ will be an element of  $Spin(4) = SU(2)_1 \times SU(2)_2$. If we identify $SU(2)_1$ with the $SU(2)$ inside the holonomy group $U(2)$ of $\cS_4$, then following \cite{Conlon:2008qi} we can classify our fermionic modes in terms of $Spin(4)$ representations as
\be
\begin{array}{lcr}
P_-^{O3} \Theta'\, =\, 0 \quad \quad & \Theta' \quad \text{transforms as} & \quad \quad (\bf{1, 2})\\
P_+^{O3} \Theta'\, =\, 0 \quad \quad & \Theta' \quad \text{transforms as} & \quad \quad (\bf{2 , 1})\\
\end{array}
\ee
In addition, if we impose the BPS condition $*_{\cS_4} \cF = \cF$ then $\Lam(\cF) \in SU(2)_1$, and so bispinors projected out by $P_-^{O3}$ are left invariant by the rotation in (\ref{rotCY}). In particular, this applies to the bispinor (\ref{gauginoCY}), that describes the D7-brane gaugino for the unwarped Calabi-Yau case. As discussed in section \ref{warpcy}, this same fermionic wavefunction will be a solution of the unmagnetized, warped Dirac operator (\ref{CYD7Dirac}) if we multiply it by $Z^{3/8}$. Finally, since (\ref{gauginoCY}) satisfies $P_-^{\uD 7} \Theta\, =\, 0$ and (\ref{CYD7Dirac}) and (\ref{wCYDiracmD7alt}) imply the same zero mode equation, it follows that the wavefunction of the D7-brane gaugino is also of the form
\beq
\Theta\, = \, Z^{3/8}\left[\theta_{4D} \otimes 
\frac{1}{\sqrt{2}}
\left(
\begin{array}{c}
i \eta^\CY_- \\  \eta^\CY_-
\end{array}
\right) - i 
B_4^* \theta_{4D}^* \otimes 
\frac{1}{\sqrt{2}}
\left(
\begin{array}{c}
\eta^\CY_+ \\ i\eta^\CY_+
\end{array}
\right)\right]
\label{gauginomCY}
\eeq
as already pointed out in \cite{lmmt08}.

On the other hand, bispinors of the form (\ref{wilsoninoCY}) are projected out by $P_+^{O3}$ and so are non-trivially rotated by $\Lam(\mp\cF)$ even assuming the BPS condition for $\cF$. One can then see that the corresponding zero modes, which correspond to the D7-brane Wilsonini, should have as wavefunction
\be
\Theta\, = \, Z^{-1/8}\frac{1}{4}(\cM^{-1}_{\cS_4})^{ab}\G_a\G_b \left[
B_4^* \theta_{4D}^* \otimes 
\frac{1}{\sqrt{2}}
\left(
\begin{array}{c}
i\eta_W \\ \eta_W
\end{array}
\right)
-i
\theta_{4D} \otimes 
\frac{B_6}{\sqrt{2}}
\left(
\begin{array}{c}
\eta_W^* \\  i \eta_W^*
\end{array}
\right) 
\right]
\label{wilsoninomCY}
\ee
which is the obvious generalization of the warped flat space solution (\ref{flatwilsoninomalt}). Again, the warp factor dependence of this solution is contained in both $Z^{-1/8}$ and in $\cM^{-1}_{\cS_4}$, and both cancel out with $\sqrt{\det \, M_{\cS_4}/\det\, g_{\cS_4}}$ when computing the Wilsonini 4D kinetic term.

Finally, one may consider fermionic wavefunctions of the form (\ref{modulinoCY}), also invariant under the rotation (\ref{rotCY}), and whose zero modes give rise to D7-brane modulini. The analogy with flat space, suggests that to any zero mode of the unwarped case a factor of $Z^{-3/8}$ should be added to obtain the warped zero mode. Let us however point out that, by the results of \cite{gghssy05,km06} one would expect that many of these would-be moduli and modulini are lifted due to the presence of the worldvolume flux $\cF$ and to global properties of $\cS_4$. Thus, the question of which are the zero mode profile of modulini is a tricky one even in the unwarped case, and so we will refrain from analyzing them in detail.

\subsection{Warped K\"ahler metrics}

Let us now proceed to compute the warped K\"ahler metrics for open strings on magnetized D7-branes, following the same approach taken in Sec \ref{kahler} for unmagnetized D7-branes. One first realizes that 
the gauge kinetic function is given by
\begin{equation}
  f_{\uD 7}=  \bigl(8\pi^{3}k^{2}\bigr)^{-1}
  \int_{\mathcal{S}_{4}}
  \frac{\ud\hat{\vol}_{\mathcal{S}_{4}}}
  {\sqrt{\hat{g}_{\mathcal{S}_{4}}}}
  \bigl(\sqrt{\absb{\det M_{\cS_4}}}
  - i(C_{4}^{\mathrm{int}} + C_0\, f \wedge f)\bigr)
\end{equation}
where again $f = \langle \cF\rangle$. This can be written as a holomorphic function by using the BPS condition
\begin{equation}
  \ud\hat{\vol}_{\cS_4}
  \sqrt{\absb{\det M_{\cS_4}}}=
  \frac{1}{2}\bigl(-P\left[J\wedge J\right] +
    \ue^{-\Phi_{0}}\mathcal{F}\wedge\mathcal{F}\bigr)
\end{equation}
and the identity (\ref{holomf}). Note that $J = Z^{1/2}J^\CY$ is the warped K\"ahler form, and that the only dependence of $f_{\uD 7}$ in the warp factor is contained in $J^2$. Hence, the extra piece in $f_{\uD 7}$ that comes from the magnetic flux is precisely as in the unwarped case.

Regarding the position modulus and modulino, they again combine into an $\mathcal{N}=1$
supermultiplet. In the toroidal case, assuming the setup of (\ref{factormD7}) and the BPS condition $b = b_i = - b_j$, we have a the K\"ahler metric of the form
\begin{equation}
  \kappa_{4}^{2}\mathcal{K}_{\zeta\bar{\zeta}}=\frac{k^{2}}{\mathcal{V}_{\mathrm{w}}}
  \int_{\mathbf{T}^{4}}\ud\hat{\vol}_{\mathbf{T}^{4}}\ue^{\Phi_{0}}
  \abs{Z^{1/2}+\ui\ue^{-\Phi_{0}/2}b}^{2}s_{0}s_{0}^{\ast}
  \left(\hat{g}_{\mathbf{T}^{4}}\right)_{k\bar{k}}
\end{equation}
that can be read from the corresponding kinetic term. Note that
\begin{equation}
 \ue^{\Phi_{0}} \abs{Z^{1/2}+\ui\ue^{-\Phi_{0}/2}b}^{2} = \ue^{\Phi_{0}}Z + b^{2}
\end{equation}
and so we again have a warp-factor independent extra term. In order to find out how this generalizes to D7-branes in warped Calabi-Yau backgrounds, let us first recall the results for the unwarped Calabi-Yau. Following \cite{jl05}, one can see that the presence of the magnetic flux $\cF$ modifies the kinetic term (\ref{ktermmod}) to
\begin{equation}
  \tau_{\uD 7}\int_{\mathbb{R}^{1,3}}\ui\mathcal{L}_{A\bar{B}}
  \left(\ue^{\Phi_0}+4G_{ab}{\cal B}^{a}{\cal B}^{b}-\frac{v^\Lambda}{\cal V}Q_{\tilde{f}}\right)
  \ud\zeta^{A}\wedge\ast_{4}\ud\bar{\zeta}^{\bar{B}}
  \label{ktermmodmag}
\end{equation}
Here the background world-volume flux has been split as
\be
f\, =\, f_{X_6} + \tilde{f}\, =\, f_{X_6}^a P[\omega_a] + \tilde{f}
\ee
where $\omega_a$ is a basis of (1,1)-forms of $X_6$\footnote{More precisely, as the analysis of \cite{jl05} takes place in the context of orientifold compactifications, $\omega_a \in H^{(1,1)}_-(X_6,\R)$, that is to those (1,1)-forms that are odd under the orientifold involution.} to be pulled-back into the D7-brane 4-cycle $\cS_4$, and $\tilde{f}$ is the component of $f$ that cannot be seen as a pull-back. One then defines
\begin{equation}
 {\cal B}^a\, =\, b^a - k f_{X_6}^a\quad \quad B\, =\, b^a\, \omega_a
\end{equation}
where $B$ is the bulk B-field as well as
\begin{equation}
  G_{ab}\, =\, \frac{1}{4\mathcal{V}}\int_{X_{6}}\omega_{a}\wedge\ast_{6}\omega_{b}
\end{equation}
where $\mathcal{V}$ is the volume of the unwarped Calabi-Yau, and
\be
Q_{\tilde{f}}\, =\, k^2 \int_{\cS_4} \tilde{f} \wedge \tilde{f}
\ee
Finally, recall that $v^{\Lambda}$ is defined by (\ref{eq:expandedkahler}), $\omega_\Lambda$ corresponding to the Calabi-Yau harmonic 2-form Poincar\'e dual to $\cS_4$. Then, from the explicit computation of the kinetic term in the toroidal case, it is easy to see that the natural generalization of (\ref{ktermmodmag}) to warped compactifications is
\begin{equation}
  \tau_{\uD 7}\int_{\mathbb{R}^{1,3}}\left(\ui\mathcal{L}^{\mathrm{w}}_{A\bar{B}}
  \ue^{\Phi_0}+\ui\tilde{\mathcal{L}}^{\mathrm{w}}_{A\bar{B}}
  G_{ab}{\cal B}^{a}{\cal B}^{b} -\frac{v^\Lambda}{\cal V_\w}Q_{\tilde{f}} \right)
  \ud\zeta^{A}\wedge\ast_{4}\ud\bar{\zeta}^{\bar{B}}
\end{equation}
in agreement with the (string frame) K\"ahler metric derived in \cite{Martucci06}. As before, we have that
\begin{equation}
  \mathcal{L}_{A\bar{B}}^{\mathrm{w}}=
  \frac{\int_{\mathcal{S}_{4}} Z\, m_{A}\wedge
    m_{\bar{B}}}
  {\int_{X_{6}} Z\, \Omega^{\mathrm{CY}}\wedge\bar{\Omega}^{\mathrm{CY}}}
\end{equation}
while we have also defined
\begin{equation}
  \tilde{\mathcal{L}}^{\mathrm{w}}_{A\bar{B}}=
  \frac{\int_{\mathcal{S}_{4}} \, m_{A}\wedge
    m_{\bar{B}}}
  {\int_{X_{6}} Z\, \Omega^{\mathrm{CY}}\wedge\bar{\Omega}^{\mathrm{CY}}}
\end{equation}
Note that both terms involve the warped internal volume which comes from
moving to the 4D Einstein frame while the first term has an additional power of
the warp factor in the integral over the internal profiles, as we found in the
toroidal case.

Finally, the Wilson lines and Wilsonini also combine into
$\mathcal{N}=1$ chiral supermultiplets.  For the factorizable torus, the kinetic
term for the complexified Wilson lines defined in (\ref{eq:complexwl}) is
\begin{equation}
  S_{\uD 7}^{\left.\mathrm{wl}\right.}
  =-\frac{k^{2}}{\kappa_{4}^{2}\mathcal{V}_{\mathrm{w}}}
  \int_{\mathbb{R}^{1,3}}\ud^{4}x
  \hat{g}_{\mathbf{T}^{4}}^{a\bar{b}}\eta^{\mu\nu}
  \partial_{\mu}w_{a}\partial_{\nu}w^{\ast}_{\bar{b}}
  \int_{\mathbf{\mathbf{T}^{4}}}\ud\hat{\vol}_{\mathbf{T}^{4}}W_{a}^{\left(0\right)}
  W_{\bar{b}}^{\ast\left(0\right)}
\end{equation}
The presence of the magnetic flux cancels out, as found for the Wilsonini in
(\ref{ktermmwil}) and in the warped Calabi-Yau case.  This gives the K\"ahler metric for the Wilson
supermultiplets
\begin{equation}
  \kappa_{4}^{2}\mathcal{K}_{a\bar{b}}
  =\frac{k^{2}}{\mathcal{V}_{\mathrm{w}}}
  \int_{\mathbf{T}^{4}}\ud\hat{\vol}_{\mathbf{T}^{4}}
  W_{a}W_{\bar{b}}^{\ast\left(0\right)}\hat{g}_{\mathbf{T}^{4}}^{a\bar{b}}
\end{equation}
We thus find that kinetic term for the Wilsonini is then unchanged with the addition of magnetic
flux, and so the kinetic terms are the same as those found in Sec \ref{kahler}.

\section{Conclusions and Outlook}\label{sec:conclusions}

In this paper we have analyzed the wavefunctions for open string degrees of freedom in warped compactifications. In particular, we have focused on type IIB supergravity backgrounds with O3/O7-planes, and explicitly computed the zero mode wavefunctions for open strings with both ends on a probe D7-brane. Such analysis has been performed for both the bosonic and fermionic D7-brane degrees of freedom, in the case of warped flat space, warped Calabi-Yau and warped F-theory backgrounds, and finally in the case of D7-branes with and without internal worldvolume fluxes.

One clear motivation to carry out such computation is the fact that models of D7-branes in warped backgrounds provide a string theory realization of the Randall-Sundrum scenario. In particular, they reproduce the basic features of 
%those
5D WED models where gauge bosons and chiral fermions are allowed to propagate in the bulk. On the other hand, since by considering D7-branes we are embedding such WED scenarios in a UV complete theory, one may naturally wonder if new features may also arise. Indeed, string theory/supergravity contains a sector of RR antisymmetric fields which is not present in the RS 5D construction, and whose field strengths are required to be non-trivial in warped backgrounds by consistency of the equations of motion. We found that such background RR fluxes couple non-trivially to the fermionic wavefunctions, leading to qualitatively different behavior depending on their extra-dimensional chirality. We have shown that these different behaviors are not accidental, but are necessary in order to provide a sensible description of SUSY or spontaneously broken SUSY 4D theories upon dimensional reduction, and in particular to produce models where the kinetic terms for bosons and fermions can be understood in terms of a 4D K\"ahler potential.

In fact, computing the open string K\"ahler potential turns out to be a very fruitful excercise since, as we have shown, it suggests a general method of extracting the closed string K\"ahler potential from (an often simpler) open string computation. Indeed, from this point of view the open strings serve as probes of the background geometry, as the consistency of their couplings to the closed string degrees of freedom enable us to use their K\"ahler metrics to deduce their closed string counterparts. We have shown that  this simple procedure reproduces the recently derived closed string results of \cite{stud,Frey:2008xw}, which were obtained in a highly complicated way. Moreover, we expect our open-closed string method to be useful in probing the structure of 
%other closed string K\"ahler potentials beyond those ones.
K\"ahler potentials in more general cases.

Returning to the WED perspective, the present work can be viewed as an initial step in the studies of the Warped String Standard Model. Such studies should involve the computation of phenomenologically relevant quantities like Yukawa couplings and flavor mixing. Even if we have illustrated such kind of computations in a very simple class of models, namely D7-branes at singularities,   our results are also relevant for more realistic constructions like those in \cite{ms04}, that involve backgrounds fluxes and magnetized intersecting D7-branes. Note, however, that the chiral sector in this latter kind of constructions arises from the intersection of D7-branes, for which a worldvolume action is still lacking. It would then be very interesting to extend our analysis to describe the degrees of freedom at the intersection of D7-branes in the presence of bulk fluxes.

Finally, let us point out that we have focussed our discussions to supersymmetric backgrounds for the sake of simplicity, but that our analysis is applicable to non-supersymmetric models as well. In such non-SUSY models, warping provides an alternative mechanism of generating the electroweak hierarchy \cite{RS}, which by way of the gauge/gravity duality can be understood as a dual description of technicolor theories. The above wavefunctions and their overlaps allows us to compute via a weakly coupled theory interactions in the strongly coupled dual, and may then offer insights into technicolor model building. Hence, other than realizing the Standard Model, constructing chiral gauge theories in warped backgrounds may also help in understanding the physics of strongly coupled hidden sectors, an element in many SUSY  breaking scenarios. For instance, recent work \cite{shamit} has shown that the strongly coupled hidden sector in general gauge mediation \cite{Meade:2008wd} can be holographically described in terms of the dual warped geometries. The open string wavefunctions obtained here can thus play an important role in determining the soft terms in such supersymmetry breaking scenarios.

%We hope to return to these interesting directions in the future.

\paragraph{}

\section*{Acknowledgments}
We would like to thank F. Benini, P.G. C\'amara, A. Dymarsky, H. Jockers, S. Kachru, L. Martucci, P. Ouyang, D. Simi\'c, and G. Torroba for helpful discussions.
The work of PM and GS  was supported in part by NSF CAREER Award No. PHY-0348093, DOE grant DE-FG-02-95ER40896, a Research Innovation Award and a Cottrell Scholar Award from Research
Corporation, a Vilas Associate Award from the University of Wisconsin, and a John Simon Guggenheim Memorial Foundation Fellowship.
GS thanks the theory division at CERN for hospitality during the course of this work.
PM and GS also thank the Stanford Institute for Theoretical Physics and SLAC for hospitality and support while this work was written.

\newpage

\appendix

\section{Conventions}\label{conv}

\subsection{Bulk supergravity action}

The bosonic sector of type IIB supergravity consists of the metric
$G_{MN}$, 2-form $B_{MN}$ and dilaton $\Phi$ in the NS-NS sector and the
$p$-form potentials $C_{0}$, $C_{2}$, and $C_{4}$ in the R-R sector.  The
string frame action for these fields is
\begin{subequations}
\begin{align}
  S_{\mathrm{IIB}}=&S_{\mathrm{NS}}+S_{\mathrm{R}}+S_{\mathrm{CS}} \\
  S_{\mathrm{NS}}=&
  \frac{1}{2\kappa_{10}^{2}}\int\ud^{10}x\, \ue^{-2\Phi}
  \sqrt{\abs{\det G}}
  \biggl\{\mathcal{R} + 4\partial_{M}\Phi\partial_{M}\Phi
  -\frac{1}{2} H_{3}^{\, 2}\biggr\} \\
  S_{\mathrm{R}}=&-\frac{1}{4\kappa_{10}^{2}}
  \int\ud^{10}x\sqrt{\abs{\det G}}
  \biggl\{F_{1}^{\, 2}+ F_{3}^{\, 2}+\frac{1}{2} F_{5}^{\, 2}\biggr\}
  \\
  S_{\mathrm{CS}}=&-\frac{1}{4\kappa_{10}^{2}}
  \int C_{4}\wedge H_{3}\wedge F_{3}
\end{align}
\end{subequations}
where $2\kappa_{10}^2 = (2\pi)^7 \al'^{\, 4}$ and
\begin{subequations}
\begin{align}
  F_{1}=&\ud C \\
  F_{3}=&\ud C_{2}-H_{3} \\
  F_{5}=&\ud C_{4}-\frac{1}{2}C_{2}\wedge H_{3}+\frac{1}{2}B_{2}\wedge F_{3}
\end{align}
\end{subequations}
and $H_{3}=\ud B_{2}$.  Here for any $p$-form $\omega$ we define $\omega^2 = \omega \cdot \omega$, where $\cdot$ is given by
\be
\omega_{p}\cdot\chi_{p}\, =\, \frac{1}{p!}\omega_{M_1\ldots M_p}\chi^{M_1\ldots M_p}
\label{cdot}
\ee
Finally, $\mathcal{R}$ is the Ricci scalar built from the metric $G$.

\subsection{D-brane fermionic action}

The fermionic action for a single D$p$-brane, up to quadratic order in the fermions and in the string frame, was computed in \cite{mms03}. I was shown in \cite{mrvv05} that one can express it as
\begin{equation}
\label{ap:Dpfermst}
  S_{\uD p}^{\mathrm{fer}}\, =\, \tau_{\uD p}
  \int\ud^{p+1}\xi\, \ue^{-\Phi}\sqrt{\abs{\det\, \bigl(P[G] + {\cal F}\bigr)}}\,
 \, \bar{\Theta} P^{Dp}_-({\cal F})
  \biggl(\bigl({\mathcal{M}}^{-1}\bigr)^{\alpha\beta} {\Gamma}_{\beta}
  {\cal D}_{\al}-\frac{1}{2}{\cal O}\biggr)\,{\Theta}
\end{equation}
where $\tau_{\uD p}^{-1}=\left(2\pi\right)^{p}\alpha'^{\frac{p+1}{2}}$ is the
tension of the $\uD p$-brane, $P[\dots]$ indicates a pull-back into the D$p$-brane worldvolume, and $\Theta$ is a 10D Majorana-Weyl bispinor,
\begin{equation}
\label{ap:bispinor}
  \Theta = \begin{pmatrix} \theta_{1} \\ \theta_{2}\end{pmatrix}
\end{equation}
with $\theta_{1},\theta_{2}$ 10D MW spinors.  Gamma matrices act on such bispinor as
\begin{equation}
  \Gamma_{M}\Theta =\begin{pmatrix}\Gamma_{M}\theta_{1} \\ \Gamma_{M}\theta_{2}
  \end{pmatrix}
\end{equation}
This action involves the generalized field strength $\mathcal{F}=P[B]+2\pi\alpha' F$ (where $F$ is the world-volume field strength of the U(1) gauge theory) through several quantities. An obvious one is the integration measure $\det(P[G]+\cF)$ that substitutes the more conventional volume element. A more crucial quantity for the analysis of Sec \ref{sec:magnetized} is $\mathcal{M}_{\alpha\beta}=G_{\alpha\beta}+\mathcal{F}_{\alpha\beta} \Gamma_{\left(10\right)}\otimes\sigma_{3}$, that encodes the D-brane world-volume natural metric in the presence of a non-trivial $\cF$. Finally, $\cF$ also appears in the projection operators
\begin{equation}
  P_{\pm}^{\uD p}=\frac{1}{2}\bigl(\mathbb{I}\pm\Gamma_{\uD p}\bigr)
  \label{projDp:ap}
\end{equation}
where $\Gamma_{\uD p}$ can be written as \cite{kallosh97}
\begin{equation}
  \Gamma_{\uD p}=\begin{pmatrix} 0 & \breve{\Gamma}_{\uD p}^{-1} \\
    \breve{\Gamma}_{\uD p} & 0\end{pmatrix}
\end{equation}
with
\begin{equation}
  \breve{\Gamma}_{\uD p}=
  i^{\left(p-2\right)\left(p-3\right)}
  \Gamma_{\uD p}^{(0)}
  \frac{\sqrt{\abs{\det P[G]}}}
  {\sqrt{\abs{\det\left(P[G]+\mathcal{F}\right)}}}
  \sum_{q}\frac{\Gamma^{\alpha_{1}\ldots\alpha_{2q}}}{q!2^{q}}
  \mathcal{F}_{\alpha_{1}\alpha_{2}}\cdots
  \mathcal{F}_{\alpha_{2q-1}}\mathcal{F}_{\alpha_{2q}}
\end{equation}
and
\be
\G_{\uD p}^{(0)}\, =\, \frac{\ep^{\al_1\dots \al_{p+1}}\G_{\al_1\dots \al_{p+1}}}{(p+1)!\sqrt{|\det P[G]|}}
\label{Gamma0:ap}
\ee
Then, for $p =2k+1$,
\be
i^{\left(p-2\right)\left(p-3\right)}\G_{\uD p}^{(0)}\, =\, i^{(p -1)/2} \G_{(p+1)}
\ee
with $\G_{(p+1)}$ as defined in footnote \ref{chiralf}. Hence, for D3 and D7-branes with $\cF=0$ we have that
\begin{equation}
  \Gamma_{\uD 3}=\begin{pmatrix} 0 & - i {\Gamma}_{(4)} \\
    i {\Gamma}_{(4)} & 0\end{pmatrix}
     = \G_{(4)} \otimes \sigma_2 \quad \text{and} \quad  \Gamma_{\uD 7} = - \G_{(8)} \otimes \sigma_2
\end{equation}
so that eqs.(\ref{projD7F0}) and (\ref{projD3}) follow from (\ref{projDp:ap}). 

The operators ${\cal O}$ and ${\cal D}_\al$ are defined from the dilatino and gravitino SUSY variations
\begin{subequations}
\label{ap:SUSYst}
\begin{align}
\delta\psi_M\,  = & \, {\cal D}_M \epsilon\, = \, \left[\nabla_M+\frac14(\slashed{H}_3)_M\sigma_3+\frac{1}{16}e^\Phi
\left(\begin{array}{cc} 0 & \slashed{F} \\ -\sigma(\slashed{F})& 0 \end{array}\right)\Gamma_M\Gamma_{(10)}\right]\epsilon \\
\delta\lambda\, =&\, {\cal O} \epsilon \, =\, \left[ \slashed{\partial}\Phi+\frac12\slashed{H}_3\sigma_3+\frac{1}{16}e^{\Phi}\Gamma^M  \left(\begin{array}{cc} 0 & \slashed{F} \\ -\sigma(\slashed{F})& 0 \end{array}\right) \Gamma_M\Gamma_{(10)}\right] \epsilon
\end{align}
\end{subequations}
where
\begin{equation}
  \label{ap:slashnot}
  \slashed{F}_{p}=\frac{1}{p!}F_{M_{1}\cdots M_{p}}\Gamma^{M_{1}\cdots M_{p}}
\end{equation}
indicates a contraction over bulk indices and $\sigma$ indicates that the order of indices in the contraction is reversed,
\begin{equation}
  \sigma\bigl(\slashed{F}_{p}\bigr)
  =\frac{1}{p!}F_{M_{1}\cdots M_{p}}\Gamma^{M_{p}\cdots M_{1}}
\end{equation}
In type IIB theory one then has that
\begin{subequations}
\label{ap:SUSYst2}
\begin{align}
 {\cal D}_M \, = &\, \nabla_M+\frac14(\slashed{H}_3)_M\sigma_3+ \frac{1}{8}e^\Phi \left(\slashed{F}_1 i \sigma_2  + \slashed{F}_3\sigma_1 + \slashed{F}^{\text{int}}_5 i \sigma_2 \right)
\Gamma_M\\
{\cal O} \, = & \, \slashed{\partial}\Phi+\frac12\slashed{H}_3\sigma_3- e^{\Phi}  \left(\slashed{F}_1 i \sigma_2 + \frac{1}{2}\slashed{F}_3 \sigma_1\right)\end{align}
\end{subequations}

For converting (\ref{ap:Dpfermst}) to the Einstein frame we have to do the following fermion redefinitions 
\be
\begin{array}{rcl}
\Theta^E & = & e^{-\Phi/8} \Theta\\
{\cal O}^E & = & e^{\Phi/8} {\cal O} \\
{\cal D}_M^E & = & e^{-\Phi/8} \left( {\cal D} - \frac{1}{8} \Gamma_M {\cal O} \right)
\end{array}
\ee
After which we obtain
\begin{eqnarray}
\nonumber
  S_{\uD p}^{\mathrm{fer}}&  = &  \tau_{\uD p}
  \int\ud^{p+1}\xi\, \ue^{\left( \frac{p-3}{4}\right)\Phi}\sqrt{\abs{\det\, \bigl(G + {\cal F}\bigr)}}\,
 \, \bar{\Theta}^E P^{\uD p}_-({\cal F})
  \biggl(\bigl({\mathcal{M}}^{-1}\bigr)^{\alpha\beta} {\Gamma}_{\beta}
  \left({\cal D}_{\al}^E + \frac{1}{8} \Gamma_\al {\cal O}^E\right)-\frac{1}{2}{\cal O}^E\biggr)\,{\Theta}^E\\
  \nonumber
  & = &  \tau_{\uD p}
  \int\ud^{p+1}\xi\, \ue^{\left( \frac{p-3}{4}\right)\Phi}\sqrt{\abs{\det\, \bigl(G + {\cal F}\bigr)}}\,
 \, \bar{\Theta}^E P^{\uD p}_-({\cal F})
  \biggl(\Gamma^\mu {\cal D}_\mu^E  + \bigl({\mathcal{M}}^{-1}\bigr)^{mn} {\Gamma}_{n}
  \left({\cal D}_{m}^E + \frac{1}{8} \Gamma_m {\cal O}^E\right)\biggr)\,{\Theta}^E
  \label{ap:DpfermE}
\end{eqnarray}
where in the second line we have taken into account that we are reducing to 4D, and where the $\Gamma$'s and ${\cal M}$ are converted to the Einstein frame. In the unmagnetized case ${\cal F} = 0$ we have 
\begin{equation}
  S_{\uD p}^{\mathrm{fer}}  =   \tau_{\uD p}
  \int\ud^{p+1}\xi\, \ue^{\left( \frac{p-3}{4}\right)\Phi}\abs{\det\, \bigl(P[G]
\bigr)}^{\frac{1}{2}}\,
 \, \bar{\Theta}^E P^{\uD p}_-
  \biggl({\Gamma}^{\al}{\cal D}_{\al}^E  + \frac{p-3}{8} {\cal O}^E\biggr)\,{\Theta}^E
  \label{ap:DpfermE2}
\end{equation}
matching (\ref{D7fermE}) for the case $p=7$. Finally, the gravitino and dilatino operators in the Einstein frame are
\begin{subequations}
\label{ap:SUSYE}
\begin{align}
 {\cal D}_M^E \, = &\, \nabla_M+\frac18  e^{\Phi/2} \left({\frak G}_3^+ \Gamma_M + \frac12\Gamma_M {\frak G}_3^+ \right)+ \frac{1}{4}\left( e^\Phi (F_1)_M   + \frac{1}{2}\slashed{F}^{\text{int}}_5 \Gamma_M \right)  i \sigma_2\\
{\cal O}^E \, = & \, \slashed{\partial}\Phi- \frac12 e^{\Phi/2} {\frak G}_3^- - e^{\Phi} \slashed{F}_1 i \sigma_2
\end{align}
\end{subequations}
where we have defined ${\frak G}_3^\pm \equiv \slashed{F}_3\sigma_1\pm e^{-\Phi}\slashed{H}_3\sigma_3$.

\subsection{Fermion conventions}

In order to describe explicitly fermionic wavefunctions we take the following representation for $\G$-matrices in flat 10D space
\be
\G^{\ul{\mu}}\, = \, \ga^\mu \otimes \Id_2 \otimes \Id_2 \otimes \Id_2 \quad \quad  \quad \G^{\ul{m}}\, =\, \ga_{(4)} \otimes \tilde{\ga}^{m-3}
\label{ulG:ap}
\ee
where $\mu = 0, \dots, 3$, labels the 4D Minkowski coordinates, whose gamma matrices are
\be
\ga^0\, =\, 
\left(
\begin{array}{cc}
 0 & -\Id_2 \\ \Id_2 & 0
\end{array}
\right)
\quad
\ga^i\, =\, 
\left(
\begin{array}{cc}
 0 & \sig_i \\ \sig_i & 0
\end{array}
\right)
\ee
$m = 4, \dots, 9$ labels the extra $\R^6$ coordinates
\be
\begin{array}{lll}
\tilde{\ga}^{1}\, = \, \sig_1 \otimes  \Id_2 \otimes \Id_2 & \quad \quad &  \tilde{\ga}^{4}\, = \,  \sig_2 \otimes  \Id_2 \otimes \Id_2 \\
\tilde{\ga}^{2}\, = \, \sig_3 \otimes  \sig_1 \otimes \Id_2 & \quad \quad &  \tilde{\ga}^{5}\, = \,  \sig_3 \otimes  \sig_2 \otimes \Id_2 \\
\tilde{\ga}^{3}\, = \,  \sig_3 \otimes \sig_3 \otimes \sig_1 & \quad \quad &  \tilde{\ga}^{6}\, = \,  \sig_3 \otimes \sig_3 \otimes \sig_2
\end{array}
\ee
and $\sig_i$ indicate the usual Pauli matrices. The 4D chirality operator is then given by
\be
\G_{(4)}\, = \, \ga_{(4)} \otimes  \Id_2 \otimes \Id_2 \otimes \Id_2 
\ee
where $\ga_{(4)} = i \ga^0\ga^1\ga^2\ga^3$, and the 10D chirality operator by
\be
\G_{(10)}\, = \, \ga_{(4)} \otimes \ga_{(6)}\, =\, \left(
\begin{array}{cc}
 \Id_2 & 0 \\ 0 & -\Id_2
\end{array}
\right) \otimes  \sig_3 \otimes \sig_3 \otimes \sig_3
\ee
with $\ga_{(6)} = -i \tilde{\ga}^1\tilde{\ga}^2\tilde{\ga}^3\tilde{\ga}^4\tilde{\ga}^5\tilde{\ga}^6$. Finally, in this choice of representation a Majorana matrix is given by
\be
\label{ap:Maj}
B\, =\, \G^{\ul{2}}\G^{\ul{7}}\G^{\ul{8}}\G^{\ul{9}}\, =\, 
\left(
\begin{array}{cc}
 0 & \sig_2 \\ -\sig_2 & 0
\end{array}
\right)
\otimes \sig_2 \otimes i\sig_1 \otimes \sig_2 \,= \, B_4 \otimes B_6
\ee 
which indeed satisfies the conditions $BB^* = \Id$ and $B\, \G^{\ul{M}} B^* = \G^{\ul{M}*}$. Notice that the 4D and 6D Majorana matrices $B_4 \equiv \ga^2 \ga_{(4)}$ and $B_6 \equiv \tilde{\ga}^4 \tilde{\ga}^5 \tilde{\ga}^6$ satisfy analogous conditions $B_4B_4^* = B_6B_6^* = \Id$ and $B_4\, \ga^{\mu} B_4^* = \ga^{\mu*}$, $B_6\, \ga^{m} B_6^* = -  \ga^{m*}$.

In the text we mainly work with 10D Majorana-Weyl spinors, meaning those spinors $\theta$ satisfying $\theta = \G_{(10)} \theta = B^*\theta^*$. In the conventions above this means that we have spinors of the form
\bes
\label{basisMW}
\begin{align}
\theta^0\, =\, 
\psi^0 \,
\left(
\begin{array}{c}
0 \\ \xi_- 
\end{array}
\right) \otimes \eta_{---} 
- i (\psi^0)^*\,
\left(
\begin{array}{c}
\sig_2\xi_-^* \\ 0
\end{array}
\right) \otimes
\eta_{+++}\\
\theta^1\, =\, 
\psi^1 \,
\left(
\begin{array}{c}
0 \\ \xi_- 
\end{array}
\right) \otimes \eta_{-++} 
+ i (\psi^1)^*\,
\left(
\begin{array}{c}
\sig_2\xi_-^* \\ 0
\end{array}
\right) \otimes
\eta_{+--}\\
\theta^2\, =\, 
\psi^2 \,
\left(
\begin{array}{c}
0 \\ \xi_- 
\end{array}
\right) \otimes \eta_{+-+} 
- i (\psi^2)^*\,
\left(
\begin{array}{c}
\sig_2\xi_-^* \\ 0
\end{array}
\right) \otimes
\eta_{-+-}\\
\theta^3\, =\, 
\psi^3 \,
\left(
\begin{array}{c}
0 \\ \xi_- 
\end{array}
\right) \otimes \eta_{++-} 
+ i (\psi^3)^*\,
\left(
\begin{array}{c}
\sig_2\xi_-^* \\ 0
\end{array}
\right) \otimes
\eta_{--+}
\end{align}
\ees
where $\psi^j$ is the spinor wavefunction, $(0\  \xi_-)^t$ is a 4D spinor of negative chirality and $\eta_{\epsilon_1\epsilon_2\epsilon_3}$ is a basis of 6D spinors of such that
\be
\eta_{---}\, =\,
\left(
\begin{array}{c}
0\\1
\end{array}
\right) \otimes
\left(
\begin{array}{c}
0\\1
\end{array}
\right) \otimes
\left(
\begin{array}{c}
0\\1
\end{array}
\right)
\quad \quad
\eta_{+++}\, =\,
\left(
\begin{array}{c}
1\\0
\end{array}
\right) \otimes
\left(
\begin{array}{c}
1\\0
\end{array}
\right) \otimes
\left(
\begin{array}{c}
1\\0
\end{array}
\right)
\label{spinorbasis:ap}
\ee
etc. Note that these basis elements are eigenstates of the 6D chirality operator $\ga_{(6)}$, with eigenvalues $\epsilon_1\epsilon_2\epsilon_3$.

In fact, that enters into the fermionic D7-brane action is a bispinor $\Theta$ of the form (\ref{bispin}), where each of $\theta_1$, $\theta_2$ is given by (\ref{basisMW}) or a linear combinations thereof. Both components of the bispinor are however not independent, but rather related by the choice of $\kappa$-fixing. Indeed, note that the fermionic action (\ref{ap:Dpfermst}) is invariant under the transformation $\Theta \rightarrow \Theta + P_-^{\uD p} \kappa$, with $\kappa$ an arbitrary 10D MW bispinor. This means that half of the degrees of freedom in $\Theta$ are not physical and can be gauged away. In practice, this amounts to impose on $\Theta = P_-^{\uD p} \Theta + P_+^{\uD p} \Theta$ a condition that fixes $P_-^{\uD p} \Theta$.

Let us for instance consider a D7-brane with ${\cal F}=0$. Taking the $\kappa$-gauge $P_-^{\uD 7} \Theta = 0$, we have that
\be
\Theta\,=\,
\left(
\begin{array}{c}
\theta_1 \\ \theta_2 
\end{array}
\right)
\, =\,
i\left(
\begin{array}{c}
\G_{(8)} \theta_2\\
-\G_{(8)} \theta_1
\end{array}
\right)
\, =\,
\left(
\begin{array}{c}
\theta\\
-i \G_{(8)} \theta
\end{array}
\right)
\ee
where $\theta$ is a spinor of the form (\ref{basisMW}). If in addition the D7-brane spans the coordinates $01234578$ with positive orientation, then the 8D chirality operator is $\G_{(8)} = -i \G^{\ul{01234578}}$, and so the wavefunctions $\psi_i^j$ of both spinors are related as
\be
\psi_2^0\, =\, - i \psi_1^0\quad \quad \psi_2^1\, =\, i \psi_1^1\quad \quad \psi_2^2\, =\, i \psi_1^2\quad \quad \psi_2^3\, =\, -i \psi_1^3
\ee
so that there are only four independent spinors wavefunctions after imposing this constraint. 
If we now define the projectors
\be
P_\pm^{\uD 3}\, =\, \frac12 \left(\Id \pm \G_{(4)} \otimes \sigma_2 \right)  \quad \quad P_\pm^{O3}\, =\, \frac12 \left(\Id \pm \G_{(6)} \otimes \sigma_2 \right) 
\label{ap:proj}
\ee
with $\G_{(6)} = \Id_4 \otimes \gamma_{(6)}$, then we see that two bispinors satisfy $P_+^{O3} \Theta = P_+^{\uD 3} \Theta =0$, namely
\be
\Theta^1\, =\, 
\left(
\begin{array}{c}
\theta^1\\
-i \G_{(8)} \theta^1
\end{array}
\right)
\quad \text{and} \quad
\Theta^2\, =\, 
\left(
\begin{array}{c}
\theta^2\\
-i \G_{(8)} \theta^2
\end{array}
\right)
\ee
and two satisfy $P_-^{O3} \Theta = P_-^{D3} \Theta =0$
\be
\Theta^0\, =\, 
\left(
\begin{array}{c}
\theta^0\\
-i \G_{(8)} \theta^0
\end{array}
\right)
\quad \text{and} \quad
\Theta^3\, =\, 
\left(
\begin{array}{c}
\theta^3\\
-i \G_{(8)} \theta^3
\end{array}
\right)
\ee

Finally, let us recall that to dimensionally reduce a D7-brane fermionic action, one has to simultaneously diagonalize two Dirac operators: $\slashed{\pa}_4$ and $\slashed{D}^{w}$, built from $\G^{\ul{\mu}}$ and $\G^{\ul{m}}$, respectively. However, as these two set of $\G$-matrices do not commute, nor will $\slashed{\pa}_4$ and $\slashed{D}^{w}$, and so we need instead to construct these Dirac operators from the alternative $\G$-matrices
\be
\tilde{\G}^{\ul{\mu}}\, = \, \G_{(4)} \G^{\ul{\mu}}\\, = \, \G_{(4)}\ga^\mu \otimes \Id_2 \otimes \Id_2 \otimes \Id_2 \quad \quad  \quad  \tilde{\G}^{\ul{m}}\, =\, \G_{(4)} \G^{\ul{m}}\, =\, \Id_4 \otimes \tilde{\ga}^{m-3}
\ee
%
%following the common practice in the literature.

\end{document}